\documentclass[12pt,a4paper]{article} 
\usepackage{times}
\usepackage{epsfig}
\usepackage{pstricks}

\let\xepsfbox\epsfbox
\def\epsfbox#1{\scalebox{1.2}{\xepsfbox{#1}}}

\evensidemargin \oddsidemargin 
\makeatletter
 
\DeclareSymbolFont{boldmath}{OML}{cmm}{b}{it}
\DeclareSymbolFontAlphabet{\mathb}{boldmath}
\DeclareMathAlphabet{\Bbb}{U}{msb}{m}{n}
\DeclareMathAlphabet{\euf}{U}{euf}{b}{n}
\DeclareMathSymbol{\balpha}{0}{boldmath}{"0B}
\DeclareMathSymbol{\bbeta}{0}{boldmath}{"0C}
\DeclareMathSymbol{\bgamma}{0}{boldmath}{"0D}
\DeclareMathSymbol{\bdelta}{0}{boldmath}{"0E}
\DeclareMathSymbol{\bepsilon}{0}{boldmath}{"0F}
\DeclareMathSymbol{\bzeta}{0}{boldmath}{"10}
\DeclareMathSymbol{\bfeta}{0}{boldmath}{"11}
\DeclareMathSymbol{\btheta}{0}{boldmath}{"12}
\DeclareMathSymbol{\biota}{0}{boldmath}{"13}
\DeclareMathSymbol{\bkappa}{0}{boldmath}{"14}
\DeclareMathSymbol{\blambda}{0}{boldmath}{"15}
\DeclareMathSymbol{\bmu}{0}{boldmath}{"16}
\DeclareMathSymbol{\bnu}{0}{boldmath}{"17}
\DeclareMathSymbol{\bxi}{0}{boldmath}{"18}
\DeclareMathSymbol{\bpi}{0}{boldmath}{"19}
\DeclareMathSymbol{\brho}{0}{boldmath}{"1A}
\DeclareMathSymbol{\bsigma}{0}{boldmath}{"1B}
\DeclareMathSymbol{\btau}{0}{boldmath}{"1C}
\DeclareMathSymbol{\bupsilon}{0}{boldmath}{"1D}
\DeclareMathSymbol{\bphi}{0}{boldmath}{"1E}
\DeclareMathSymbol{\bchi}{0}{boldmath}{"1F}
\DeclareMathSymbol{\bpsi}{0}{boldmath}{"20}
\DeclareMathSymbol{\bomega}{0}{boldmath}{"21}
\DeclareMathSymbol{\beps}{0}{boldmath}{"22}
\DeclareMathSymbol{\bthet}{0}{boldmath}{"23}
\DeclareMathSymbol{\bomeg}{0}{boldmath}{"24}    
\DeclareMathSymbol{\bvphi}{0}{boldmath}{"27}

\newcommand{\nwl}{\nonumber\\}
\newcommand{\txt}[1]{\quad\hbox{#1}\quad}

\@addtoreset{equation}{section}

\def\xcaption#1{\begin{quote}
               \def\normalsize{\small}\small
               \caption{#1}\global\edef\@currentlabel{\@currentlabel}
               \end{quote}\hrule}

\newcommand{\sref}[1]{section~\ref{#1}} 
\newcommand{\fref}[1]{figure~\ref{#1}} 

\newcommand{\eref}[1]{(\ref{#1})}

\newcommand{\ssty}{\scriptstyle}

\def\^{\hspace{-.3em}^} 
\def\_{\hspace{-.1em}_}

\newcommand\follows{\quad\Rightarrow\quad}
\newcommand\equivalent{\qquad\Leftrightarrow\qquad}

\newcommand{\ft}[2]{{\textstyle{{#1}\over{#2}}}}
\newcommand{\intl}[2]{\int_{\makebox[0pt][l]{$\ssty #1$}}
                          ^{\makebox[0pt][l]{$\ssty #2$}}\,\,\,}
\newcommand{\inti}[1]{\int_{\makebox[0pt][l]{$\ssty #1$}}\,\,\,}
\newcommand{\sumi}[1]{\sum_{\makebox[0pt][c]{$\ssty #1$}}\,}
\newcommand{\prodi}[1]{\prod_{\makebox[0pt][c]{$\ssty #1$}}\,}

\newcommand{\Pexp}{\mathop{\mathcal{P}\mathrm{exp}}}

\newcommand{\expo}[1]{\mathrm{e}^{#1}} 

\newcommand{\comm}[2]{[#1,#2]} 
\newcommand{\pois}[2]{\bigl\{#1,#2\bigr\}}

\newcommand{\del}{\partial} 
 
\newcommand{\dd}{\mathrm{d}}

\newcommand{\ii}{\mathrm{i}} 
\newcommand{\Tr}{\mathrm{Tr}} 
\newcommand{\Trr}[1]{\Tr(#1)}
\newcommand{\Trrr}[1]{\Tr\bigl(#1\bigr)}

\def\inv{\@ifnextchar_{\iinv}{^{-1}}}

\def\iinv_#1{\makebox[0pt][l]{$_{#1}$}
             \makebox[0pt][l]{\raisebox{.3ex}{$^{\,-1}$}}
             \hphantom{{}^{\,-1}_{#1}}}

\newcommand{\grpSL}{\mathsf{SL}} 
\newcommand{\grpSO}{\mathsf{SO}} 
\newcommand{\algsl}{\euf{sl}}

\newcommand{\ZZ}{\Bbb{Z}} 
\newcommand{\RR}{\Bbb{R}} 
\newcommand{\SP}{\mathsf{S}}
\newcommand{\tng}{\mathsf{T}}

\newcommand{\N}{\mathcal{N}} 
\newcommand{\M}{\mathcal{M}} 
\newcommand{\U}{\mathcal{U}}

\newcommand{\bnd}{\mathcal{B}}

\newcommand{\con}{\mathcal{C}}
\newcommand{\mul}{\zeta}

\newcommand{\pcon}{\mathb{Z}}      \newcommand{\pconv}{Z}

\newcommand{\econ}{\mathcal{Z}}
\newcommand{\emul}{\omega}

\newcommand{\jcon}{\mathcal{J}}
\newcommand{\jmul}{\epsilon}

\newcommand{\state}{\mathit{\Phi}}

\newcommand{\sym}{\mathit{\Omega}} 
\newcommand{\pot}{\mathit{\Theta}}

\newcommand{\gam}{\bgamma} 

\newcommand{\one}{\mathbf{1}}

\newcommand{\ee}{\mathb{e}} 
\newcommand{\om}{\bomega}

\newcommand{\FF}{\mathb{F}} 
\newcommand{\TT}{\mathb{T}} 

\newcommand{\norm}{\mathb{n}}

\newcommand{\fx}{\mathb{f}} 
\newcommand{\gx}{\mathb{g}}

\newcommand{\blpar}{\mathb{h}}    
\newcommand{\blgen}{\bchi}        
\newcommand{\btpar}{\mathb{v}}    

\newcommand{\hol}{\mathb{u}}  \newcommand{\hols}{u}
\newcommand{\mom}{\mathb{p}}  \newcommand{\momv}{p}
\newcommand{\pos}{\mathb{x}}  \newcommand{\posv}{x}
\newcommand{\dis}{\mathb{z}}  \newcommand{\disv}{z}

\newcommand{\chol}{\mathb{g}}  
\newcommand{\cmom}{\mathb{q}}  \newcommand{\cmomv}{q}
\newcommand{\cang}{\mathb{f}}

\newcommand{\vv}{\mathb{v}}
\newcommand{\ww}{\mathb{w}}
\newcommand{\uu}{\mathb{u}}
\newcommand{\pp}{\mathb{p}}

\newcommand{\cuts}{\mathit{\Gamma}}
\newcommand{\newcuts}{\mathit{\tilde\Gamma}}
\newcommand{\cut}{\lambda}

\newcommand{\acut}{\lambda}
\newcommand{\bcut}{\kappa}

\newcommand{\ecut}{\eta}
\newcommand{\fcut}{\varrho}

\newcommand{\pols}{\mathit{\Pi}}
\newcommand{\newpols}{\mathit{\tilde\Pi}}
\newcommand{\pol}{\Delta}

\newcommand{\prt}{\pi}

\newcommand{\nprt}{n}

\newcommand{\ncut}{\ell}
\newcommand{\npol}{\wp}
\newcommand{\ninf}{k}

\newcommand{\genus}{g}

\newcommand{\cdir}{\phi}
\newcommand{\bdir}{\theta}

\newcommand{\tang}{\mathb{J}}
\newcommand{\tmom}{\mathb{P}}

\newcommand{\atau}{\tau}
\newcommand{\adir}{\alpha}

\newcommand{\eps}{\varepsilon}

\newcommand{\p}{\varphi}

\newcommand{\lag}{\mathcal{L}}
\newcommand{\ham}{\mathcal{H}}
\newcommand{\kin}{\mathcal{K}}
\newcommand{\Mpl}{M_\mathrm{Pl}}
\newcommand{\lpl}{\ell_\mathrm{Pl}}

\newcommand{\psp}{\mathcal{P}}
\newcommand{\esp}{\mathcal{Q}}
\newcommand{\ksp}{\mathcal{S}}

\newcommand{\tpar}{\Delta T}
\newcommand{\rpar}{\Delta\cdir}

\makeatother

\begin{document}

\setcounter{page}{0}
\thispagestyle{empty}

\begin{flushright}
MZ-TH/00-44\\
gr-qc/0103084
\end{flushright}

\begin{center}
  \LARGE \textsc{The Phase Space Structure of\\
  Multi Particle Models in 2+1 Gravity}
\end{center}
 
\vspace*{8mm}

\begin{center}
  \textbf{Hans-J\"urgen Matschull}\\[2ex]
  Institut f\"ur Physik, Johannes Gutenberg-Universit\"at\\
  55099 Mainz, Germany\\
  matschul@thep.physik.uni-mainz.de
\end{center}

\vspace*{10mm}

\begin{center}
  March 2001
\end{center}

\vspace*{10mm}

\begin{abstract}
  What can we learn about quantum gravity from a simple toy model,
  without actually quantizing it?  The toy model consists of a finite
  number of point particles, coupled to three dimensional Einstein
  gravity. It has finitely many physical degrees of freedom.  These
  are basically the relative positions of the particles in spacetime
  and the conjugate momenta. The resulting reduced phase space is
  derived from Einstein gravity as a topological field theory. The
  crucial point is thereby that we do not make any a priori
  assumptions about this phase space, except that the dynamics of the
  gravitational field is defined by the Einstein Hilbert action. This
  already leads to some interesting features of the reduced phase
  space, such as a non-commutative structure of spacetime when the
  model is quantized.
\end{abstract}

\newpage

\section*{Outline and summary}
It is widely believed that in a quantum theory of gravity, the concept
of a smooth spacetime manifold has to be replaced by something new,
perhaps a kind of \emph{non-commutative} spacetime, whatever this
precisely means. In a non-commutative spacetime the position of a
particle in space at a moment of time is not a well defined point on a
manifold. There is some kind of uncertainty relation, which makes it
impossible for the particle to be localized. The spacetime itself
obtains a foamy structure at small length and time scales. It is no
longer possible to measure, for example, the relative position of two
objects with arbitrary precision. The length scale at which this is
expected to happen is the \emph{Planck scale}, which is about
$\lpl\approx 10^{-33}$cm.

The purpose of the present article is to study a toy model, which is
based on Einstein gravity in three spacetime dimensions. There are no
local excitations of the gravitational field, and in particular no
gravitational waves is a three dimensional spacetime
\cite{starus,carlip}. But nevertheless it is possible to define
non-trivial dynamical systems. The special feature of these toy models
is that they have finitely many physical degrees of freedom, while at
the level of Einstein gravity as a field theory they share all those
feature with the full theory in higher dimensions which are interesting
from the point of view of quantum gravity. These features are, in
particular, the diffeomorphism group of the spacetime manifold, which
appears as a gauge group, and the presence of a dimensionful coupling
constant, which is Newton's constant $G$.

There are two classes of such toy models. The \emph{topological}
models are those that do not include matter. Physical degrees of
freedom only arise if the universe has a non-trivial topology
\cite{witten,matrev}. The \emph{particle} models contain pointlike
particles as matter sources. The physical degrees of freedom are then
basically the relative positions and momenta of the particles in
spacetime \cite{djh,welrh}. There are many interesting and unexpected
features of such particle models, some of them even related to time
machines and black hole physics \cite{gott,steif,matbtz,holmat}. In
the context of quantum gravity, the basic idea behind the models is to
use a pointlike particle to \emph{probe} the structure of the
spacetime at small length scales \cite{hooft1,hooft2,matwel}. If
quantum gravity sets some principle limitation on the possible length
scales that can be resolved, then this should be seen when the toy
model is quantized.

In three spacetime dimensions, Newton's constant $G$ is an inverse
mass or energy, in units where the velocity of light is equal to one.
Its inverse defines the \emph{Planck mass} $\Mpl$, which is a
classical constant in the sense that $\hbar$ is not involved in the
definition. So, there is a natural energy scale in the classical
theory of general relativity. Another special feature of three
dimensional gravity is that it admits pointlike matter sources, which
are not hidden in black holes. The gravitational field of a point
particle with mass $m$ is simply a cone with a deficit angle of $8\pi
Gm$ \cite{starus,djh}. The spacetime in the neighbourhood of the world
line is the direct product of this cone with a real line.

Due to this conical structure of the spacetime, there is a
gravitational interaction between particles, even though there are no
local gravitational forces acting on the particles. Each world line is
just a timelike or lightlike geodesics, thus a straight line in an
otherwise flat spacetime. However due to the non-trivial global
behaviour of a geodesic on a cone, the particles are effectively
attracted by each other when they pass each other. But actually, we
are here not interested in these \emph{dynamical} features the
particles. There are various methods to solve the classical equations
of motion, and to obtain an overview of all possible spacetimes
\cite{djh,welrh,bcv}.

What we are rather interested in are the \emph{kinematical} features
of the particle models. By this we mean, for example, the phase space
structures, and in particular the way gravity influences the
symplectic structure and the Poisson brackets. It is this what
distinguishes gravity from other typical interactions that can be
defined between point particles. It turns out that gravity not only
enters the Hamiltonian, by adding some kind of interaction potential
or whatsoever, but also the symplectic structure. When the model is
quantized, this has an immediate influence on the commutation
relations between certain operators, long before the actual dynamics
is imposed by the Hamiltonian. So, we expect the principle effects of
quantum gravity at show up at this kinematical level.

It is therefore not of much help to know the classical solutions to
the equations of motion, hence in this case Einstein's equations in a
three dimensional spacetime. Instead, we have to define a proper
action functional for the particle model, and derive from this the
phase space structures. As we are dealing with a toy model that has no
counterpart in the real world, we have to be very careful with this
definition, if we want to learn something about quantum gravity and
the way it influences the small scale structure of spacetime. In
particular, we should not make any a priori assumptions, which might
be motivated by the expected results. The \emph{only} assumption that
we are going to make is that the kinematical and dynamical features of
the gravitational field are completely specified by the Einstein
Hilbert action.

The particle model is in fact well defined as a dynamical system at
the level of Einstein gravity as a field theory on a fixed spacetime
manifold. We shall show this in \sref{reduc}. At this level, the
definition of the particle model has more or less become standard
\cite{bcv,ms,cms}. However, there are then various different ways to
proceed. The model is defined as a field theory, where the basic
variables are the metric or the dreibein and the spin connection on a
given spacetime. So it has an infinite dimensional configuration
space, though only finitely many physical degrees of freedom. One can
use the ADM formulation of general relativity \cite{mtw}, to set up a
Hamiltonian framework. But then one still has an infinite dimensional
phase space.

To this one can apply the usual phase space reduction methods. There
are constraints and associated gauge symmetries, and these can be
solved and divided out. What remains is a finite dimensional reduced
phase space, which is basically spanned by the position and momentum
coordinates of the particles. A possible way to derive this reduced
phase space, which is motivated by the conventional methods for gauge
field theories, is to impose a gauge condition directly on the basic
fields, hence the metric. In the ADM formulation of general
relativity, this means that the foliation of the spacetime is chosen
in a particular way, introducing a globally defined absolute time and
space. The gauge fixed metric on the space at a moment of time is then
specified by finitely many variables, and roughly speaking they define
the positions and velocities of the particles in space at that moment
of time \cite{welrh,bcv,ms,cms}.

So far, this approach is completely straightforward. However, it turns
out to be quite hard to return to physics after this mathematical
framework has been set up. The reason is the following. By
construction, the reduced phase space is covered by a global canonical
coordinate chart. One has a Poisson bracket which is diagonal in the
basic variables, and the Hamiltonian is some function of these
variables. Formally, a state can be specified by fixing $\nprt$
points on a plane and their velocities, where $\nprt$ is the number of
particles. However, this plane and the points on the plane have not
very much to do with the real space in which the particles are moving.
The real space at a moment of time is a conical surface with $\nprt$
tips, which is embedded into a locally flat spacetime with $\nprt$
conical singularities.

The actual geometry of this conical space is encoded in the locations
and velocities of the points on the plane in a somewhat implicit way.
Consider for example the following question. Given two particles, what
is the distance of these particles in spacetime, at a fixed moment of
time? The spacetime is locally flat, and thus a geodesic connecting
two particles is just a straight line. There might be several such
geodesics in a conical spacetime, but then we can choose one and ask
for its length. Expressing this length as a phase space function is an
interesting question, because it is the corresponding quantum operator
which tells us something about the spacetime structure when the model
is quantized. But unfortunately, in the gauge fixed formulation this
simple quantity is a very complicated function of the basic phase
space variables.

To know the geodesic distance between two particles, we first have to
know the spacetime metric in terms of the coordinates used, and to
know this we have to take into account all particles and their
relative motion. Hence, a very simple object, the flat spacetime
metric, is encoded in a complicated way. The idea of this article is
therefore to present an alternative phase space reduction, which leads
to a different set of basic phase space variables. The price that we
have to pay is that these coordinates do not longer provide a global
chart on the reduced phase space. Instead, it will be covered by an
atlas of finitely many local coordinate charts. But on the other hand,
these coordinates will have an immediate geometric and thus physical
interpretation.

To explain this, it is convenient first to consider the limit $G\to0$,
where the gravitational interaction is switched off. In this case, the
particle model reduces to a special-relativistic free particle model.
The particles are moving freely in a flat three dimensional Minkowski
space. The phase space of this model is spanned by the \emph{absolute}
positions and momenta of the particles with respect to the embedding
Minkowski space, which serves as a reference frame. But it is also
possible to introduce \emph{relative} coordinates and momenta, and to
define a phase space spanned by them. This will be shown in
\sref{free}. The idea is, in a sense, to use the free particle model
as a much simpler and well known toy model for our real toy model, and
to show the analogy to the ADM, or Hamiltonian framework of general
relativity.

We can think of the flat Minkowski space in which the particles are
living as a spacetime manifold, which is foliated by a family of
spacelike slices. They are labeled by a kind of ADM time coordinate,
and the state at a moment of time is specified by the geometry of the
respective slice. A typical such slice is shown in \fref{fspc}. Its
geometry is, on the other hand, specified by a set of relative
position and momentum coordinates for the particles. They refer to the
links of a \emph{triangulation}. The time evolution of this geometry
is provided by the mass shell constraints for the particles, in
analogy of the Hamiltonian constraints of general relativity.

So far, this is just a strange way to describe a very simply system on
$\nprt$ free relativistic point particles. However, the remarkable
feature of this description is that we only have to make a few
modifications, or \emph{deformations}, in order to describe the
geometry of the spacetime of the interacting system. This will be the
subject of \sref{inter}, and the resulting phase space structures will
be discussed in \sref{phase}. In a sense, we can \emph{switch on} the
gravitational interaction in a continuous way, using Newton's constant
$G$ as a deformation parameter. First we have to cut the polygons of
the triangulation apart, as indicated in \fref{fpol}, then they are
deformed, as shown in \fref{ipol}, and finally we can glue them
together again. The result is a spacelike surface which is no longer
globally embedded into Minkowski space.

Instead, it is a conical surface with $\nprt$ tips. This is the space
at a moment of time with the gravitational interaction between the
particles switched on. The crucial point is thereby that the link
variables, hence those phase space variables that define the geometry
of the individual polygons, are still the same as before. In
particular, they still represent the same physical quantities, for
example the relative position of two particles in spacetime. This is
very different to the gauge fixed method, where the phase space
variables are also the usual position and momentum coordinates of the
free particles in the limit $G\to0$. But when gravity is switched on,
the physical interpretation of the variables is lost.

The only problem is that it is now no longer possible to use these
geometric variables as global coordinates on the phase space. There
are some principle obstacles, and therefore it is not possible to have
at the same time a global coordinate chart on the phase space, and a
straightforward geometric interpretation of the coordinates
\cite{matcs}. So, we have to cover the phase space by an atlas of
local coordinate charts. But as long as we are interested in the local
features of the phase space, and the symplectic structure is a local
object, we can stick to one of the charts and need not care about the
global structure of the phase space. 

It is at this point, where the perhaps most interesting feature arise,
namely the \emph{non-commutative} structure of spacetime. Let us
briefly explain what happens. For a pair of free particles in flat
Minkowski space, we can introduce a vector $z^a$ ($a=0,1,2$),
representing the relative position of the particles in spacetime.
Clearly, the Poisson brackets for the components of this vector are
zero. Such a vector can still be defined when gravity is switched on,
and it still has the same physical interpretation as a relative
position of two particles in spacetime.  However, now it turns out
that the Poisson brackets and therefore the resulting quantum
commutators are given by
\begin{displaymath}
  \pois{z^a}{z^b} = 8\pi G \,  \eps^{abc} \, z_c  \follows
  \comm{z^a}{z^b} = 8\pi \ii \,  \lpl \, \eps^{abc} \, z_c ,
\end{displaymath}
where $\lpl=G\hbar$ is again the \emph{Planck length}, and
$\eps^{abc}$ is the Levi Civita tensor. It defines the structure
constants for the three dimensional Lorentz group. So, we have in this
case a well defined notion of what is meant by a non-commutative
spacetime. The position coordinates of the particle no longer commute.
And obviously this non-commutative structure disappears in the limit
$G\to0$.

The same quantum commutator has also been found for a much simpler
single particle model \cite{matwel}. The only difference was that
there the vector $z^a$ did not represent the relative position of two
particles, but the absolute position of the particle with respect to a
reference frame. Unfortunately, there was some ambiguity in the
definition of the reference frame, and we had to make some additional
assumptions, beyond the basic one about the Einstein Hilbert action.
One could therefore argue a little bit about the result, and the
suggestion was that these problems could be overcome when going over
from a single to a multi particle model.

And in fact, we shall here see that the non-commutative structure
arises from the phase space reduction applied to the Einstein Hilbert
action without any further assumption. The reason why the multi
particle model is, in a sense, better defined than a single particle
model has to do with the asymptotic structure of a three dimensional
spacetime. To set up a proper Hamiltonian framework of general
relativity, one has to impose a fall off condition on the metric at
infinity. This is typically some kind of asymptotical flatness
condition. Moreover, one has to introduce a \emph{reference frame} at
spatial infinity, which can be interpreted as the rest frame of some
external observer. 

The phase space and the Hamiltonian are well defined only if such a
reference frame is included into the configuration space of the model
\cite{nico}. The Killing symmetries of the asymptotic metric are then
of special interest. They represent the possible translations and
rotations of the universe with respect to the reference frame. In the
Hamiltonian framework, these are the rigid symmetries of the model,
and the associated conserved charges represent the total momentum and
angular momentum of the universe. Now, the special feature of Einstein
gravity in three dimensions is the conical structure of spacetime,
which also manifests itself at spatial infinity. The universe is not
asymptotically flat, but asymptotically conical.

The symmetry group of a cone, however, is smaller than the symmetry
group of a flat Minkowski space. The only symmetries are time
translations and spatial rotations, whereas boosts and spatial
translations are not allowed. This restricted symmetry group is that
of the \emph{centre of mass frame} of a system of point particles. It
is therefore appropriate to choose the reference frame so that it
coincides with the centre of mass frame. But on the other hand, this
does not make sense for a single particle model, because there is then
no physical degree of freedom left. And this was the reason for the
various difficulties and ambiguities to arise, which were related to
the definition of the reference frame.

All these ambiguities disappear when we define the multi particle
model so that the reference frame coincides with the centre of mass
frame of the universe. In \sref{reduc} we shall show this explicitly,
defining the model in the first order formalism of general relativity,
using the dreibein and the spin connection. We'll see that the
required boundary terms to be added to the Einstein Hilbert action,
defining the reference frame at spatial infinity, are uniquely fixed
without making any further assumptions, in contrast to the single
particle model where this was not the case. But in all other aspect
the multi particle model will be a straightforward generalization of
the single particle model.

In particular, the coupling of the particles to the gravitational
field will be defined in the same way, by formally converting the
matter degrees of freedom into topological degrees of freedom, and
introducing generalized, group valued momenta of the particles. As all
this is explained in detail in \cite{matwel}. So, we shall here only
briefly sketch this. The actual subject of \sref{reduc} is then to
perform the phase space reduction, and to derive all those phase space
structures which we have to define without further motivation in
\sref{phase}. The logical order of the article is actually so that
\sref{reduc} comes first, then \sref{inter} describes the model from
the spacetime point of view, \sref{phase} takes the phase space point
of view, and finally in \sref{free} we take the limit $G\to0$.

There are then various ways to proceed, but this is not any more the
subject of this article. It actually serves a rather technical
purpose, namely to set up a proper classical Hamiltonian framework for
the particle model, which is as general as possible, but also as
simple as possible, in the sense that no further assumption enters the
model beyond the basic one mentioned above about the Einstein Hilbert
action. The results can therefore serve as a starting point for a
further analysis after some simplifications are made. For example, the
simplest case is the \emph{Kepler system} with $\nprt=2$ particles.
This can be solve completely both at the classical level, deriving the
trajectories in the phase space, and at the quantum level, solving the
Schr\"odinger, or actually a generalized Klein Gordon equation
\cite{loumat,matlou}.

And for this particular example, we find indeed some interesting
features of quantum gravity. It is no longer possible to localize the
particles in space, or to bring the particles closer to each other
than a certain distance, which is of the order of the Planck length.
All this can be seen very explicitly and clearly when the model is
quantized canonically. It is even possible to derive the wave
functions explicitly and see how the energy eigenstates of look like,
when expressed in terms of physical variables. Hence the states can
directly be interpreted as probability amplitudes, and we see
explicitly how quantum gravity restrict the ability to probe small
length scales. 

Another way to proceed further might be to consider the low coupling
limit, in which case it could be possible to solve and quantize the
$\nprt$ particle system exactly, or to consider a closed universe with
a few particles as a new kind of cosmological toy model. But whatever
one does in this direction, \emph{solving} always means more then just
writing down the solutions to the equations of motion or the quantum
states in terms of \emph{some} variables. This tells us almost nothing
about the physics. To learn something about quantum gravity one has to
give the solutions, hence the classical trajectories or the quantum
states, in terms of \emph{physical} variables. And for this purpose
the Hamiltonian framework presented here might be useful.

\section{Free particles}
\label{free}
Consider a very simple system of $\nprt\ge2$ free relativistic point
particles, living in a three dimensional, flat Minkowski space. As a
vector space, we identify this with the spinor representation
$\algsl(2)$ of the three dimensional Lorentz algebra.  The position
variables $\pos_\prt=\posv_\prt^a\gam_a$ and the momentum vectors
$\mom_\prt=\momv_\prt^a\gam_a$ of the particles are represented as
traceless $2\times2$ matrices, with $\gam_a$ ($a=0,1,2$) being the
orthonormal basis of the usual gamma matrices \eref{gamma}. A
collection of definitions and formulas regarding the vector and matrix
notation can be found in the appendix.  The index $\prt$ is used to
label the individual particles.

We have a $6\nprt$ dimensional phase space which is spanned by the
variables $\pos_\prt$ and $\mom_\prt$. The symplectic potential and
the resulting Poisson brackets are 
\begin{equation}
  \label{rpp-pot-pois}
  \pot = \sumi{\prt}
    \ft12\Trr{\mom_\prt \, \dd\pos_\prt} \follows
  \pois{\momv_\prt^a}{\posv_\prt^b} = \eta^{ab}.
\end{equation}
Note that we sometimes switch between the matrix and vector notation,
and use whatever is more convenient. The Hamiltonian is a linear
combination of the mass shell constraints $\con_\prt$, with Lagrange
multipliers $\mul_\prt\in\RR$ as coefficients,
\begin{equation}
  \label{rpp-ham}
  \ham = \sumi{\prt} \mul_\prt \, \con_\prt, \qquad 
   \con_\prt = \ft14\Trr{\mom_\prt\^2} + \ft12 m_\prt\^2 
     \approx 0.
\end{equation}
The physical phase space is the subspace defined by the mass shell
constraints and the positive energy conditions,
\begin{equation}
  \label{rpp-erg}
   \ft12\Trr{\mom_\prt\^2} = - m_\prt\^2, \qquad  
  \momv_\prt^0 = \ft12\Trr{\mom_\prt \gam^0} > 0 .
\end{equation}
The mass shell constraints are first class constraints, and the
associated gauge symmetries are the reparameterizations of the world
lines as functions of a common, unphysical time parameter $t$. The
gauge freedom corresponds to the freedom to choose the multipliers in
the time evolution equations,
\begin{equation}
  \label{rpp-evolve}
  \dot \mom_\prt = \pois\ham{\mom_\prt} = 0, \qquad
  \dot \pos_\prt = \pois\ham{\pos_\prt} = \mul_\prt \, \mom_\prt.
\end{equation}
The dot denotes the derivative with respect to $t$. The free particle
system is a simple toy model for the ADM formulation of general
relativity. The coordinate $t$ on the world lines is the globally
defined, but unphysical ADM time. The mass shell constraints
$\con_\prt$ are the Hamiltonian constraints. The multipliers
$\mul_\prt$ represent the lapse function. And the gauge symmetries,
the reparameterizations of the world lines, are the spacetime
diffeomorphisms. To make this analogy even closer, we impose the
following two restrictions on the phase space variables. The first is
a gauge restriction. It is analogous to the restriction to spacelike
foliations in general relativity. The second restriction is a kind of
asymptotical flatness condition, although at the moment this is not
immediately obvious.

The gauge restriction is as follows. At each moment of ADM time $t$,
the particles must be located on a common spacelike surface in the
embedding Minkowski space. In other words, the parameterization of the
world lines is induced by a foliation of Minkowski space by a family
of spacelike slices. For simplicity, let us further assume that the
states where the particles are at the same point in space are excluded
from the phase space. As long as we do not study the dynamics, that is
the classical trajectories, this is not going to be a problem. The
gauge condition is then requires all relative position vectors to be
spacelike. Hence,
\begin{equation}
  \label{spacelike}
  \pos_{\prt_2}-\pos_{\prt_1} 
     \txt{is spacelike for all $\prt_1,\prt_2$.}
\end{equation}
With this restriction, the phase space is no longer a vector space.
But it is still a $6\nprt$ dimensional manifold, because the set of
all spacelike vectors is an open subset of Minkowski space. The number
of independent gauge symmetries in not affected by this restriction.

The second restriction has to do with the rigid symmetries. The
positions $\pos_\prt$ and momenta $\mom_\prt$ are defined with respect
to a \emph{reference frame}. We can think of it as the rest frame of
some external observer. The rigid symmetries are the translations and
Lorentz rotations of the particles with respect to this reference
frame. The associated Noether charges are the components of the total
momentum and angular momentum vector,
\begin{equation}
  \label{rpp-tot}
  \tmom = \sumi{\prt} \mom_\prt , \qquad 
  \tang = \ft12 \sumi{\prt} \comm{\mom_\prt}{\pos_\prt}.
\end{equation}
The reference frame coincides with the \emph{centre of mass} frame if
\begin{equation}
  \label{rpp-com}
  \tmom = M \, \gam_0 , \qquad 
  \tang = S \, \gam_0 , \qquad M,S\in\RR, \quad M>0.
\end{equation}
This is the second restriction that we impose on the phase space
variables. In the centre of mass frame, there exists an absolute time
and an absolute space. The symmetry group is restricted to the two
dimensional subgroup of time translations and spatial rotations. A
plane orthogonal to the $\gam_0$-axis represents an instant of
absolute time. The $\gam_0$-axis is the world line of a
\emph{fictitious} centre of mass particle. The total energy $M$ is the
mass, and the total angular momentum $S$ is the spin of this particle.
These are the conserved charges associated with the remaining
symmetries.

Except for one very special situation, a centre of mass frame always
exists. If all particles are massless, and if they move with the
velocity of light into the same spatial direction, then the total
momentum is lightlike. These states are excluded. The dimension of the
restricted phase space $6\nprt-4$, since only the four spatial
components of the equations \eref{rpp-com} are restrictions on the
phase space variables. The time components are the definitions of the
charges $M$ and $S$. The gauge symmetries are not affected. We still
have the freedom to reparameterize the world lines independently,
within the range that is allowed by \eref{spacelike}.

From the phase space point of view, the restriction to the centre of
mass frame provides two pairs of second class constraints, whereas the
mass shell constraints are still first class. For a proper Hamiltonian
description of the restricted system, we either have to replace the
Poisson brackets with Dirac brackets, or we have to eliminate the
second class constraints, by going over to a new set of independent
phase space variables. This is what we are going to do now. We are
looking for a convenient way to parameterize the solutions to the
equations $\tmom=M\gam_0$ and $\tang=S\gam_0$, which can later be
generalized to describe the interacting particles. 

\subsubsection*{Triangulations}
\begin{figure}[t]
  \begin{center}
    \epsfbox{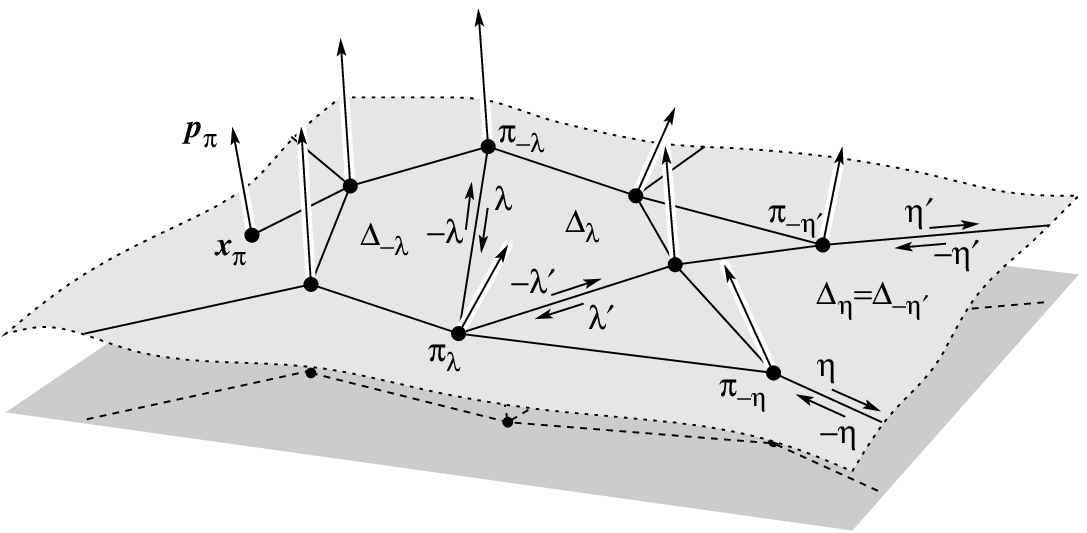}
    \xcaption{The triangulated ADM surface of the free particle
    system. The particles are located at the positions $\pos_\prt$ in
    the embedding Minkowski space, which defines the centre of mass
    frame and also the reference frame. The momentum vectors
    $\mom_\prt$ are the tangent vectors of the world lines attached to
    these points. The surface is cut along the links $\cut$ of a graph
    $\cuts$, and divided into a collection $\pols$ of simply connected
    polygons $\pol$.}
    \label{fspc}
  \end{center}
\end{figure}

Consider a foliation of the embedding Minkowski space by a family of
spacelike slices, labeled by an ADM time coordinate $t$, and so that
the parameterization of the world lines is induced by this foliation.
A typical slice is shown in \fref{fspc}. Let us call it the \emph{ADM
surface} at a moment of time $t$. Note that we have to keep the ADM
time $t$ apart from the absolute time defined by the
$\gam_0$-coordinate in the centre of mass frame. The ADM surface need
not be an spatial plane in Minkowski space. For the time being, it can
be any spacelike surface. The positions $\pos_\prt$ of the particles
at the given moment of time are the intersections of the world lines
with the ADM surface, and the momenta $\mom_\prt$ are the tangent
vectors of the world lines attached to these points.

A \emph{triangulation} is defined by a collection $\cuts$ of
\emph{oriented links}. Every link $\cut\in\cuts$ represents a geodesic
in Minkowski space. It either connects two particles or extends from a
particle to infinity. The particles are thus the \emph{vertices} of a
\emph{graph}, and there is an additional, special vertex at infinity.
The graph must be connected, hence there must be at least one link
attached to every vertex, and it must be possible to go from every
particle to every other along the links. Furthermore, the graph must
be spacelike, which means that the ADM surface can be smoothly
deformed, so that all the links are contained in the surface. And
finally, let us also require that all the links extending to infinity
are \emph{spatial half lines}. A spatial half line is a geodesic which
is orthogonal to the $\gam_0$-axis, and extends from some point to
infinity.

If this is the case, then the graph $\cuts$ divides the ADM surface
into a collection $\pols$ of simply connected \emph{polygons}. To
define the various relations between the vertices, the links, and the
polygons, we introduce the following notation. With each link
$\cut\in\cuts$, we also have the \emph{reversed} link $-\cut\in\cuts$.
It represents the same geodesic with opposite orientation. We
distinguish between \emph{internal} and \emph{external} links. An
internal link $\cut$ \emph{begins} at a particle $\prt_{-\cut}$, and
it \emph{ends} at a particle $\prt_\cut$. For the reversed link
$-\cut$, these two particles are interchanged, which explains the
notation. The subset of all internal links is called
$\cuts_0\subset\cuts$.

It is sometimes useful to assign a preferred orientation to each
internal link. For this purpose, we split the set
$\cuts_0=\cuts_+\cup\cuts_-$ into two disjoint subsets, so that
$\cut\in\cuts_+$ implies $-\cut\in\cuts_-$ and vice versa. Since a
priori there is no such preferred orientation, this decomposition is
arbitrary. Whenever we use it we have to make sure that the results
are independent of the particular decomposition. The external links
already have a preferred orientation. We define
$\cuts_\infty\subset\cuts$ to be the set of all external links
oriented towards infinity, and $\cuts_{-\infty}$ is the set of all
reversed external links oriented towards the particles. Again, it is
so that $\ecut\in\cuts_\infty$ implies $-\ecut\in\cuts_{-\infty}$
and vice versa.

An external link $\ecut\in\cuts_\infty$ begins at some particle
$\prt_{-\ecut}$, but it does not end at any particle, and vice versa
for the reversed external link $-\ecut\in\cuts_{-\infty}$. We use the
symbol $\cut$ to denote a link in general or an internal link, whereas
the symbol $\ecut$ always denotes an external link. A useful identity
is the following decomposition of the graph $\cuts$ into four disjoint
subsets,
\begin{equation}
  \label{graph-union}
  \cuts = \cuts_+ \cup \cuts_- \cup \cuts_\infty \cup \cuts_{-\infty},
  \qquad \cuts_0 = \cuts_+ \cup \cuts_- .
\end{equation}
If $\prt$ is some vertex, then
$\cuts_\prt=\{\cut\,|\,\prt_\cut=\prt\}$ is the set of all links
\emph{ending} at $\prt$, and
$\cuts_{-\prt}=\{\cut\,|\,\prt_{-\cut}=\prt\}$ is the set of all links
\emph{beginning} at $\prt$. For the special vertex at infinity, this
agrees with the definition of $\cuts_\infty$ and $\cuts_{-\infty}$
above. Since every link begins and ends at some vertex, including the
one at infinity, we have the following alternative disjoint
decomposition of the graph,
\begin{equation}
  \label{graph-prt}
  \cuts = \cuts_\infty \cup \bigcup_\prt \cuts_\prt
        = \cuts_{-\infty} \cup \bigcup_\prt \cuts_{-\prt}. 
\end{equation}
It is useful to assign a cyclic ordering to each of these sets. Let
$\prt$ be a particle, and consider a loop surrounding this particle in
clockwise direction. The ordering in which the links cross this loop
defines the cyclic ordering of the set $\cuts_\prt$. If
$\cut\in\cuts_\prt$ is a link ending at $\prt$, then the next
following link in clockwise direction is called the \emph{successor}
of $\cut$, which is denoted by $\cut'\in\cuts_\prt$.

If there is only one link attached of a vertex, then we have
$\cut'=\cut$, otherwise this obviously defines a cyclic ordering.  The
same applies to the set $\cuts_\infty$ of external links, but for
consistency this is ordered in counter clockwise direction. Thus every
external link $\ecut\in\cuts_\infty$ has a unique successor
$\ecut'\in\cuts_\infty$, which is the next external link that we cross
when we walk around a circle at infinity in counter clockwise
direction. Since every link is contained in exactly one of the subsets
$\cuts_\prt$ or $\cuts_\infty$, every link also has a unique
successor, which is an element of the same subset.

Each pair of links $\cut$ and $-\cut$ represents the common boundary
of two polygons $\pol_\cut$ and $\pol_{-\cut}$. We define $\pol_\cut$
to be the polygon that lies \emph{to the left} of $\cut$, and
consequently $\pol_{-\cut}$ lies \emph{to the right} of $\cut$. For
each polygon $\pol\in\pols$, the set
$\cuts_\pol=\{\cut\,|\,\pol_\cut=\pol\}$ represents the boundary of
the polygon, traversed in counter clockwise direction. A link
$\cut\in\cuts_\pol$ is called an \emph{edge} of the polygon $\pol$.
Since every oriented link is an edge of exactly one polygon, we have
yet another disjoint decomposition of the graph, namely
\begin{equation}
  \label{graph-pol}
  \cuts = \bigcup_{\pol\in\pols} \cuts_\pol.
\end{equation}
There is relation between the cyclically ordered sets $\cuts_\prt$ and
$\cuts_\infty$, and the sets $\cuts_\pol$ of edges. If
$\cut,\cut'\in\cuts_\prt$ are two successive links ending at a vertex
$\prt$, which might also be the one at infinity, then
$\cut,-\cut'\in\cuts_\pol$ are two successive edges of some polygon
$\pol$. For $\cut,\cut'\in\cuts_\prt$ we always have
$\pol_\cut=\pol_{-\cut'}$.

We distinguish between \emph{compact} polygons $\pol\in\pols_0$, and
\emph{non-compact} polygons $\pol\in\pols_\infty$, so that
$\pols=\pols_0\cup\pols_\infty$. A compact polygon $\pol\in\pols_0$ is
bounded by at least three internal edges
$\cut\in\cuts_\pol\subset\cuts_0$. A non-compact polygon has exactly
one external edge $\ecut\in\cuts_\pol\cap\cuts_\infty$ oriented
towards infinity, exactly one reversed external edge
$-\ecut'\in\cuts_\pol\cap\cuts_{-\infty}$, and one or more internal
edges $\cut\in\cuts_\pol\cap\cuts_0$. And the links
$\ecut,\ecut'\in\cuts_\infty$ are thereby always two successive
external links. Consequently, there are as many non-compact polygons
as there are external links.

There is also a relation between the number of compact polygons and
the number of internal links. It follows from Euler's polyhedron
formula, and it involves the number of particles. Let
$\ncut_0=|\cuts_+|=|\cuts_-|$ be the number of internal links, so
that each link is only counted with one orientation, and let
$\ncut_\infty=|\cuts_\infty|=|\cuts_{-\infty}|$ be the number of
external links. Similarly, $\npol_0=|\pols_0|$ is the number of
compact polygons, and $\npol_\infty=|\pols_\infty|$ is the number of
non-compact polygons. Then we have the following relations,
\begin{equation}
  \label{rpp-R2-top}
  \ncut_0 - \npol_0 = \nprt - 1 , \qquad 
  \ncut_\infty = \npol_\infty.
\end{equation}
They basically define the topology of the surface which is
triangulated. We may release these conditions later on when we
consider the coupled system, considering more general spatial
topologies. But for the free particles the ADM surface always has the
topology of $\RR^2$.

\subsubsection*{Relative momenta}
Let us now come back to the phase space of the particles. The primary
problem was to solve the second class constraints \eref{rpp-com}. Let
us first consider the equation $\tmom=M\gam_0$, which only involves
the momenta of the particles. To find the general solution, we
introduce a \emph{relative momentum} vector $\cmom_\cut\in\algsl(2)$
assigned to every link $\cut\in\cuts$, so that
\begin{equation}
  \label{rpp-cmom-mom}
  \mom_\prt = \sumi{\cut\in\cuts_\prt} \cmom_\cut, \qquad
   \cmom_{-\cut} = -\cmom_\cut. 
\end{equation}
The momentum $\mom_\prt$ of a particle $\prt$ is the sum of the
relative momenta $\cmom_\cut$ of all links ending at $\prt$. It is
useful to think of the vectors $\cmom_\cut$ as currents flowing
through the links. The second equation then states that the current
flowing through the reversed link is the reversed current. And the
first equation states that the momentum of a particle $\prt$ is the
total current that flows out of the graph at the vertex $\prt$. Using
this picture, and the fact that the graph $\cuts$ is connected, it is
easy to show that, for a given set of momenta $\mom_\prt$, it is
always possible to find a suitable set of relative momenta
$\cmom_\cut$, so that \eref{rpp-cmom-mom} holds.

However, there is no one-to-one relation between the momenta
$\mom_\prt$ and the relative momenta $\cmom_\cut$. We are free to add
an extra current, which only flows along the boundary of one of the
polygons, but never enters or leaves the graph at any vertex. This has
no influence on the particles. More explicitly, let
$\blgen_\pol\in\algsl(2)$ be a Minkowski vector for each polygon
$\pol\in\pols$, and consider the following transformation of the
relative momenta,
\begin{equation}
  \label{rpp-cmom-lor}
  \cmom_\cut \mapsto \cmom_\cut 
      + \blgen_{\pol_{-\cut}} - \blgen_{\pol_{\cut}}. 
\end{equation}
Inserting this into \eref{rpp-cmom-mom}, we find 
\begin{equation}
  \label{rpp-mom-lor}
  \mom_\prt \mapsto \mom_\prt 
    + \sumi{\cut\in\cuts_\prt} 
        (\blgen_{\pol_{-\cut}} - \blgen_{\pol_\cut})
    = \mom_\prt 
    + \sumi{\cut\in\cuts_\prt} \blgen_{\pol_{-\cut'}} 
    - \sumi{\cut\in\cuts_\prt} \blgen_{\pol_\cut} = \mom_\prt.
\end{equation}
For the first equality, we split the sum into two, and then we
replaced the index $\cut\in\cuts_\prt$ in the first sum by its
successor $\cut'\in\cuts_\prt$. Thus we shifted the summation index
cyclically. But now, remember that for two successive links
$\cut,\cut'\in\cuts_\prt$, we always have $\pol_\cut=\pol_{-\cut'}$.
Therefore the two sums are equal and cancel. We find that the momentum
vectors are invariant under any transformation of the form
\eref{rpp-cmom-lor}.

Before considering this in more detail, let us show how the link
variables can be used to solve the constraint equation
$\tmom=M\gam_0$. The total momentum vector can be written as
\begin{equation}
  \label{rpp-tmom-link}
  \tmom = \sumi{\prt} \mom_\prt 
        = \sumi{\prt} \sumi{\cut\in\cuts_\prt} \cmom_\cut
        = \sumi{\cut\in\cuts_+} \cmom_\cut
        + \sumi{\cut\in\cuts_-} \cmom_\cut
        + \sumi{\ecut\in\cuts_{-\infty}} \cmom_\ecut
        =  - \sumi{\ecut\in\cuts_\infty} \cmom_\ecut.
\end{equation}
The first equality is the definition of $\tmom$, and the second
follows from \eref{rpp-cmom-mom}. To derive the third equality, we
used \eref{graph-union} and \eref{graph-prt}, which implies
\begin{equation}
  \label{graph-sum}
  \bigcup_{\prt} \cuts_\prt = 
   \cuts_+ \cup \cuts_- \cup \cuts_{-\infty}. 
\end{equation}
Finally, the last equality in \eref{rpp-tmom-link} follows from the
fact that $\cmom_{-\cut}=-\cmom_\cut$ for all links $\cut\in\cuts$.
The sums over $\cuts_+$ and $\cuts_-$ cancel, and the sum over
$\cuts_{-\infty}$ can be written as a sum over $\cuts_\infty$.  Now,
consider the equation $\tmom=M\gam_0$. This is obviously satisfied if
we choose the relative momentum vectors for the external links to be
proportional to $\gam_0$, hence
\begin{equation}
  \label{rpp-ecut-cmom}
  \cmom_{\ecut} =  - M_\ecut \, \gam_0  ,\qquad
  \cmom_{-\ecut} = M_\ecut \, \gam_0, \qquad 
    \ecut \in\cuts_\infty. 
\end{equation}
The variable $M_\ecut$ is called the \emph{energy} of the external
link $\ecut\in\cuts_\infty$. The total energy is then the sum of all
energies assigned to the external links, 
\begin{equation}
  \label{rpp-tmom-M}
  \tmom = \sumi{\ecut\in\cuts_\infty} M_\ecut \, \gam_0
   \follows M = \sumi{\ecut\in\cuts_\infty} M_\ecut .
\end{equation}
So, what we found is a partly redundant but complete parameterization
of the momentum vectors $\mom_\prt$ of the particles, which are
subject to the equation $\tmom=M\gam_0$. The independent parameters
are the relative momentum vectors $\cmom_\cut$ for $\cut\in\cuts_+$,
and the energies $M_\ecut$ for $\ecut\in\cuts_\infty$.

The redundancy transformations are still given by \eref{rpp-cmom-lor}.
But there is now a certain restriction on the parameters
$\blgen_\pol$. The external relative momenta $\cmom_\ecut$ for
$\ecut\in\cuts_\infty$ must be proportional to $\gam_0$, before and
after the transformation. This is the case if and only if
$\blgen_\pol=-\emul_\pol\gam_0$ for all non-compact polygons
$\pol\in\pols_\infty$, where $\emul_\pol\in\RR$. This implies that the
energies transform as
\begin{equation}
  \label{rpp-M-lor}
  M_\ecut \mapsto M_\ecut 
      + \emul_{\pol_{-\ecut}} - \emul_{\pol_\ecut}. 
\end{equation}
As a cross check, let us count the number of independent redundancies.
There are three components of the vector $\blgen_\pol$ for each
compact polygon $\pol\in\pols_0$, and one real number $\emul_\pol$ for
each non-compact polygon $\pol\in\pols_\infty$. Thus all together we
have $3\npol_0+\npol_\infty$ parameters. However, there is actually
one degree of freedom less. If we set $\blgen_\pol=-\emul\gam_0$ for
all polygons $\pol\in\pols$, and $\emul_\pol=\emul$ for all
non-compact polygons $\pol\in\pols_\infty$, where $\emul\in\RR$ is
some fixed number, then the transformations \eref{rpp-cmom-lor} and
\eref{rpp-M-lor} are void. Therefore, the number of independent
redundancies is
\begin{equation}
  \label{rpp-mom-redun}
  3\npol_0 + \npol_\infty - 1 = 
  (3\ncut_0 + \ncut_\infty) - (3 \nprt - 2). 
\end{equation}
The equality follows from \eref{rpp-R2-top}. The result is what we
have to expect. On the right hand side, we have the difference between
the number of independent link variables, and the original number of
momentum variables. There are $3\ncut_0$ independent components of the
relative momenta $\cmom_\cut$ for $\cut\in\cuts_+$, and $\ncut_\infty$
independent energies $M_\ecut$ for $\ecut\in\cuts_\infty$. Originally,
we had $3\nprt$ independent components of the momentum vectors
$\mom_\prt$, and we imposed two constraints, the spatial components of
the equations $\tmom=M\gam_0$.

\subsubsection*{Relative positions}
\begin{figure}[t]
  \begin{center}
    \epsfbox{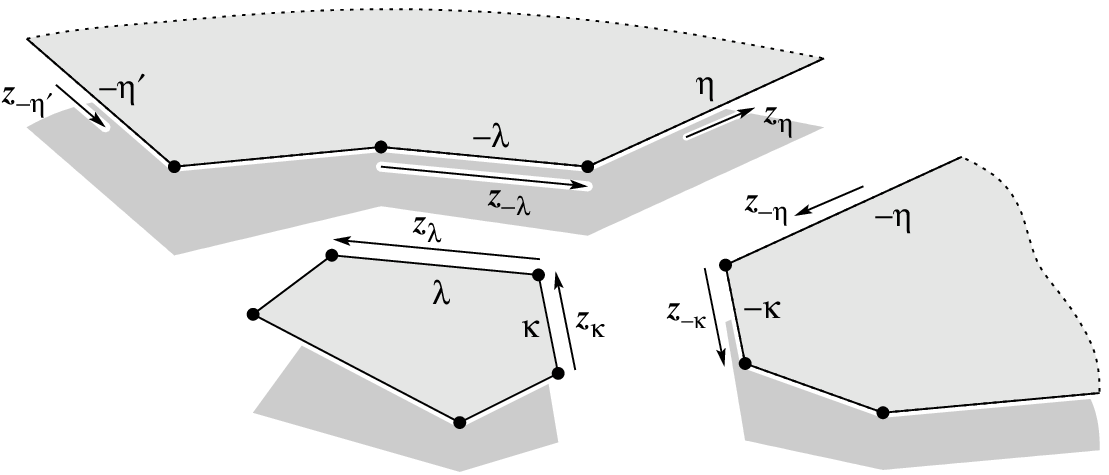}
    \xcaption{The geometry of the individual polygons is determined,
    up to smooth deformations, by the relative position vectors
    $\dis_\cut$ for $\cut\in\cuts_0$, and the spatial unit vectors
    $\dis_\ecut$ for $\ecut\in\cuts_{\pm\infty}$. The ADM surface in
    \fref{fspc} is reconstructed by gluing the polygons together along
    the edges. The relation $\dis_{-\cut}=-\dis_\cut$ ensures that
    they always fit together.}
    \label{fpol}
  \end{center}
\end{figure}

The positions $\pos_\prt$ of the particles are subject to the
constraint $\tang=S\gam_0$. They can now be parameterized in a
similar, but in a sense dual way. For each link $\cut\in\cuts$, we
introduce yet another vectors $\dis_\cut\in\algsl(2)$. For an internal
link, it represents the \emph{relative position} of the two particles
connected by the link,
\begin{equation}
  \label{rpp-dis}
  \dis_\cut = \pos_{\prt_\cut} - \pos_{\prt_{-\cut}}
  \follows \dis_{-\cut} = -\dis_\cut, 
   \qquad \cut\in\cuts_0.
\end{equation}
For an external link, it is a spacelike unit vector, which specifies
the \emph{direction} of the link in the embedding Minkowski space.
Remember that we required the external links to be spatial half lines.
The vector $\dis_\ecut$ for $\ecut\in\cuts_\infty$ is therefore a
spacelike unit vector which is orthogonal to the $\gam_0$-axis, and
the same applies to the vector $\dis_{-\ecut}$, which points into the
opposite direction. Using the definition \eref{gamma-rot} we can write
\begin{equation}
  \label{rpp-ecut-dis}
  \dis_\ecut = \gam(\cdir_\ecut) , \qquad 
  \dis_{-\ecut} = - \gam(\cdir_\ecut), \qquad
   \ecut\in\cuts_\infty,
\end{equation}
where $\cdir_\ecut$ is the angular direction of the external link
$\ecut$. In \fref{fpol} we have cut the ADM surface apart, and
embedded the individual polygons into some auxiliary Minkowski space,
to visualize the definition of the vectors $\dis_\cut$ for internal
and external links.

Finally, there is yet another link variable that can be assigned to
every external link $\ecut\in\cuts_\infty$. Consider a \emph{clock}
which is located at the far end of the link, and which shows that
absolute time in the centre of mass frame. The time shown on this
clock is then equal to the $\gam_0$-coordinate of the particle
$\prt_{-\ecut}$ to which the link is attached, thus 
\begin{equation}
  \label{rpp-T}
  T_\ecut = \posv^0_{\prt_{-\ecut}} , \qquad \ecut\in\cuts_\infty.
\end{equation}
So, we have now assigned a vector $\dis_\cut$ to each internal link
$\cut\in\cuts_0$, and two scalars $\cdir_\ecut$ and $T_\ecut$ to each
external link. But these variables are not all independent. It follows
from the definition \eref{rpp-dis} that, for every compact polygon,
the relative position vectors representing its edges add up to zero,
\begin{equation}
  \label{rpp-icond}
  \sumi{\cut\in\cuts_\pol} \dis_\cut = 0 , \qquad
  \pol\in\pols_0.
\end{equation}
For non-compact polygons, there is a similar relation. Let
$\ecut,\ecut'\in\cuts_\infty$ be two successive external links, and
let $\pol=\pol_\ecut=\pol_{-\ecut'}$ be the non-compact polygon in
between. There is then the following relation between the clocks and
the relative position vectors, which follows from \eref{rpp-dis} and
\eref{rpp-T}, 
\begin{equation}
  \label{rpp-econd}
    T_{\ecut'} - T_{\ecut} 
     +  \sumi{\cut\in\cuts_\pol\cap\cuts_0} \disv^0_\cut = 0 , 
    \qquad \pol\in\pols_\infty. 
\end{equation}
Hence, the difference between the absolute time on the two external
edges is equal to the minus the sum of the the time components of the
internal edges. 

The statement is now the following. Given the vectors $\dis_\cut$ for
all $\cut\in\cuts$, and the clocks $T_\ecut$ for all
$\ecut\in\cuts_\infty$, and provided that the \emph{consistency
conditions} \eref{rpp-icond} and \eref{rpp-econd} are satisfied, then
the positions $\pos_\prt$ of the particles are uniquely determined by
the system of linear equations \eref{rpp-dis} and \eref{rpp-T}, and
the constraint equation $\tang=S\gam_0$. The proof is not very
complicated. Given the vectors $\dis_\cut$ for all $\cut\in\cuts_0$,
can already derive the relative position of any two particles. We just
have to walk along the internal links from one particle to the other,
and add up the relative position vectors. The relation
\eref{rpp-icond} ensures that the result is independent of the path
that we took. 

So, we already know the absolute positions up to an overall
translation $\pos_\prt\mapsto\pos_\prt+\btpar$, where
$\btpar\in\algsl(2)$ is some unknown Minkowski vector. Now, consider
the equation $\tang=S\gam_0$. Let us express the angular momentum
vector $\tang$ in terms of the link variables. Using the definition
\eref{rpp-tot}, and inserting the representation \eref{rpp-cmom-mom}
of the momentum vectors, we get
\begin{equation}
  \label{rpp-J-link-1}
  \tang = \ft12\sumi\prt \comm{\mom_\prt}{\pos_\prt} 
        = \ft12\sumi\prt \sumi{\cut\in\cuts_\prt}
                         \comm{\cmom_\cut}{\pos_{\prt_\cut}}.
\end{equation}
Here we used that $\prt=\prt_\cut$ for all $\cut\in\cuts_\prt$. It
follows from \eref{graph-sum} that this can be written as
\begin{equation}
  \label{rpp-J-link-2}
  \tang = 
      \ft12\sumi{\cut\in\cuts_+} \comm{\cmom_\cut}{\pos_{\prt_\cut}}
    + \ft12\sumi{\cut\in\cuts_-} \comm{\cmom_\cut}{\pos_{\prt_\cut}}
    + \ft12\sumi{\ecut\in\cuts_{-\infty}} 
                               \comm{\cmom_\ecut}{\pos_{\prt_\ecut}}.
\end{equation}
The first two sums can be combined, using $\cmom_{-\cut}=-\cmom_\cut$,
and then we can insert the definition \eref{rpp-dis} of $\dis_\cut$.
To simplify the last sum, we use the definition \eref{rpp-ecut-cmom}
of $M_\ecut$. The result is
\begin{equation}
  \label{rpp-J-link-3}
  \tang = 
      \ft12\sumi{\cut\in\cuts_+} \comm{\cmom_\cut}{\dis_\cut}
    + \ft12\sumi{\cut\in\cuts_\infty} 
        M_\ecut  \, \comm{\gam_0}{\pos_{\prt_{-\ecut}}}.
\end{equation}
Now, consider an overall translation
$\pos_\prt\mapsto\pos_\prt+\btpar$. The first term is obviously
invariant, and so the angular momentum vector transforms as
\begin{equation}
  \label{rpp-tang-trans}
  \tang \mapsto \tang + \ft12 M \, \comm{\gam_0}{\btpar}.
\end{equation}
This implies that the spatial components of the unknown vector
$\btpar$ are determined by the equation $\tang=S\gam_0$. But then we
still have the freedom to perform a time translation
$\pos_\prt\mapsto\pos_\prt+v\gam_0$. To fix the remaining degree of
freedom, we need to know the absolute time coordinate of at least one
particle. This additional information is provided by the clocks
$T_\ecut$. It is sufficient to know one of them, but it does not
matter which one. The consistency condition \eref{rpp-econd} ensures
that we always get the same results.

So, all together we find that the link variables provide sufficient
information to reconstruct the original position and momentum
variables $\pos_\prt$ and $\mom_\prt$ of the particles, which are
subject to the constraints $\tmom=M\gam_0$ and $\tang=S\gam_0$. The
charges $M$ and $S$ can then also be expressed in terms of the link
variables. According to \eref{rpp-tmom-M} and \eref{rpp-J-link-3} we
have
\begin{equation}
  \label{rpp-M-S}
  M = \sumi{\ecut\in\cuts_\infty} M_\ecut, \qquad 
  S = \sumi{\cut\in\cuts_+} L_\cut, \txt{where}
  L_\cut = \ft14  \Trr{ \, \comm{\cmom_\cut}{\dis_\cut} \, \gam^0}.
\end{equation}
The total energy is, in a sense, distributed over the external links,
and the total angular momentum is distributed over the internal links.
Note that $L_{-\cut}=L_\cut$, which implies that the sum is
independent of the decomposition $\cuts_0=\cuts_+\cup\cuts_-$.

But still, the parameterization is not unique. We still have the
redundancy transformation \eref{rpp-cmom-lor} and \eref{rpp-M-lor},
acting on the relative momenta and energies. Moreover, we never used
the unit vectors $\dis_{\pm\ecut}$ for $\ecut\in\cuts_\infty$ when we
derived the absolute positions $\pos_\prt$ of the particles. We are
therefore free to perform the following additional redundancy
transformations, 
\begin{equation}
  \label{rpp-tws-cdir}
  \cdir_\ecut \mapsto \cdir_\ecut + \jmul_\ecut, \qquad 
    \ecut\in\cuts_\infty,
\end{equation}
where $\jmul_\ecut\in\RR$ are free parameters. This is the freedom
that we also have when we introduce the triangulation in \fref{fspc}.
The directions of the external links are not fixed, so within a
certain range we can rotated them. This requires a smooth deformation
of the ADM surface, but it does not affect the positions or the
momenta of the particles. 

Taking this into account, we have the following total number of
redundancies,
\begin{equation}
  \label{rpp-redun}
  3\npol_0 + \npol_\infty + \ncut_\infty - 1.  
\end{equation}
On the other hand, we also have the consistency conditions to be
satisfied by the link variables. There are $3\npol_0$ components of
the vector equations \eref{rpp-icond}, and $\npol_\infty$ scalar
equations \eref{rpp-econd}. But these are not all independent. Let us
add the $\gam_0$-components of all equations \eref{rpp-icond}, and all
equations \eref{rpp-econd}. Then the clocks drop out, and what remains
is the sum of the $\gam_0$-components of all vectors $\dis_\cut$ for
$\cut\in\cuts_0$. But this sum is zero, because for each link
$\cut\in\cuts_0$, there is also a reversed link $-\cut\in\cuts_0$. So,
the total number of consistency conditions is
$3\npol_0+\npol_\infty-1$.

The idea is to consider these consistency conditions as first class
constraints, and the redundancy transformations as the associated
gauge symmetries. To treat the rotations \eref{rpp-tws-cdir} of the
external links in the same way, we shall add an auxiliary
\emph{angular momentum} $L_\ecut$ for $\ecut\in\cuts_\infty$ to the
link variables, which is canonically conjugate to the direction
$\cdir_\ecut$, and impose the constraint $L_\ecut=0$, which then
generates the gauge transformation \eref{rpp-tws-cdir}. With this
extension of the phase space variables, the number of consistency
conditions, or constraints, becomes equal to the number of
redundancies, or gauge symmetries \eref{rpp-redun}.

\subsubsection*{The extended phase space}
Let us summarize all this in the following definition of an
\emph{extended phase space} $\esp$, for a system of $\nprt$ free point
particles in the centre of mass frame. Since there are finitely many
possible ways to define a graph $\cuts$, the extended phase space
$\esp$ consists of a finite number of disconnected components
$\esp_\cuts$.  There is one component for each possible graph $\cuts$
with $\nprt$ particles. We shall discuss these global features of the
phase space $\esp$ in a moment. Let us first consider a fixed graph
$\cuts$, and define the component $\esp_\cuts$. It is spanned by the
following \emph{link variables},
\begin{itemize}
\item[-] a \emph{relative momentum vector} $\cmom_\cut\in\algsl(2)$
  and a \emph{relative position vector} $\dis_\cut\in\algsl(2)$ for
  every internal link $\cut\in\cuts_0$, and
\item[-] an \emph{energy} $M_\ecut$, a \emph{clock} $T_\ecut$, an
  \emph{angular momentum} $L_\ecut$, and a \emph{direction}
  $\cdir_\ecut$ for every external link.  
\end{itemize}
The vectors are not all independent. For the internal links we have
\begin{equation}
  \label{rpp-cut-inv}
    \cmom_{-\cut} = - \cmom_\cut, \qquad 
    \dis_{-\cut} = - \dis_\cut, \qquad \cut\in\cuts.
\end{equation}
The independent phase space variables are thus $\cmom_\cut$ and
$\dis_\cut$ for $\cut\in\cuts_+$, where $\cuts_0=\cuts_+\cup\cuts_-$
can be any possible decomposition. The relation \eref{rpp-cut-inv} is
also valid for external links, if we define the associated vectors to
be
\begin{equation}
  \label{rpp-ecut-vec}
  \cmom_\ecut = - \cmom_{-\ecut} = - M_\ecut \, \gam_0 , \qquad
  \dis_\ecut = - \dis_{-\ecut} = \gam(\cdir_\ecut) , \qquad
   \ecut\in\cuts_\infty.
\end{equation}
The symplectic structure on $\esp$ is defined by the following
canonical symplectic potential,
\begin{equation}
  \label{rpp-pot-ext}
  \pot = \sumi{\ecut\in\cuts_\infty}  
          ( T_\ecut \, \dd M_\ecut 
                  +  L_\ecut \, \dd \cdir_\ecut ) 
       - \ft12 \sumi{\cut\in\cuts_+} 
             \Trr{\dd \cmom_\cut \, \dis_\cut}.
\end{equation}
Of course, there is no particular motivation for this at the moment.
And in fact, we must show that this coincides with the original
symplectic structure \eref{rpp-pot-pois}. Otherwise the definition
does not make sense. We still want to consider the same dynamical
system. The proof will be given further below, so let us for the
moment assume that the definition is correct. We can then read off the
following non-vanishing Poisson brackets. For each external link, we
have two canonical pairs
\begin{equation}
  \label{rpp-pois-ecut}
  \pois{T_\ecut}{M_\ecut} = 1, \qquad 
  \pois{L_\ecut}{\cdir_\ecut} = 1,   \qquad \ecut\in\cuts_\infty.
\end{equation}
For each internal link $\cut\in\cuts_0$, the vectors $\cmom_\cut$ and
$\dis_\cut$ are canonically conjugate to each other. However, we have
to take into account the relations \eref{rpp-cut-inv}, stating that
the vectors for $\cut$ and $-\cut$ are not independent. What we find
is
\begin{equation}
  \label{rpp-pois-cut}
  \pois{\cmomv_\cut^a}{\disv_\cut^b} =  \eta^{ab} \follows 
  \pois{\cmomv_{-\cut}^a}{\disv_\cut^b} = 
  \pois{\cmomv_\cut^a}{\disv_{-\cut}^b} = - \eta^{ab} ,
     \qquad \cut\in\cuts_0.
\end{equation}
All other brackets are zero, in particular those between variables
assigned to different links. We should also note that the expression
\eref{rpp-pot-ext} for the symplectic potential is independent of the
decomposition $\cuts_0=\cuts_+\cup\cuts_-$. We have to sum over all
internal links with some preferred orientation, but it does not matter
how this orientation is chosen. If we replace $\cut$ with $-\cut$,
both $\cmom_\cut$ and $\dis_\cut$ change their sign, and thus the
corresponding contribution to $\pot$ is invariant. The same holds for
the Poisson brackets \eref{rpp-pois-cut}.

\subsubsection*{Kinematical constraints}
How is this extended phase space $\esp$ related to the original phase
space of the free particles, restricted to the centre of mass frame?
We have seen that the link variables are subject to some consistency
conditions. The idea is to impose them as \emph{kinematical
constraints} on the extended phase space $\esp$. Thus, for each
compact polygon, there is a vector valued constraint \eref{rpp-icond},
\begin{equation}
  \label{rpp-icon}
  \pcon_\pol = \sumi{\cut\in\pol} \dis_\cut \approx 0, 
    \qquad \pol \in\pols_0.  
\end{equation}
For each non-compact polygon, there is a scalar constraint
\eref{rpp-econd},
\begin{equation}
  \label{rpp-econ}
  \econ_\pol = T_{\ecut'} - T_\ecut 
     +  \sumi{\cut\in\cuts_\pol\cap\cuts_0} \disv^0_\cut \approx 0, 
    \qquad \pol \in\pols_\infty. 
\end{equation}
And finally, for each external link we impose a constraint which
eliminates the artificially introduced angular momentum $L_\ecut$,
\begin{equation}
  \label{rpp-jcon}
  \jcon_\ecut = L_\ecut \approx 0, \qquad \ecut\in\cuts_\infty. 
\end{equation}
The subspace $\ksp_\cuts\subset\esp_\cuts$ defined by these
constraints is called the \emph{kinematical phase space}. 

It is then straightforward to check that the kinematical constraints
are first class, and that the associated gauge symmetries are the
previously considered redundancy transformations. It is convenient
first to define a general linear combination. We introduce a set of
Lagrange multipliers, a vector $\blgen_\pol\in\algsl(2)$ for each
compact polygon $\pol\in\pols_0$, a scalar $\emul_\pol\in\RR$ for each
non-compact polygon $\pol\in\pols_\infty$, and another scalar
$\jmul_\ecut\in\RR$ for each external link $\ecut\in\cuts_\infty$.
Then we define
\begin{equation}
  \label{rpp-kin}
  \kin = \ft12 \sumi{\pol\in\pols_0} \Trr{\blgen_\pol\pcon_\pol} 
        + \sumi{\pol\in\pols_\infty} \emul_\pol \, \econ_\pol
        + \sumi{\ecut\in\cuts_\infty} \jmul_\ecut \, \jcon_\ecut.
\end{equation}
Using the definition of the constraints, and rearranging the
summation, this can also be written as
\begin{equation}
  \label{rpp-kin-sum}
  \kin = \ft12 \sumi{\cut\in\cuts_0} 
         \Trrr{\blgen_{\pol_\cut} \dis_\cut} 
       + \sumi{\ecut\in\cuts_\infty} 
         (\emul_{\pol_{-\ecut}} - \emul_{\pol_{\ecut}} ) \, T_\ecut
       + \sumi{\ecut\in\cuts_\infty} 
         \jmul_\ecut \, L_\ecut.
\end{equation}
Here we set $\blgen_\pol=-\emul_\pol\gam_0$ for $\pol\in\pols_\infty$,
so that the first sum also contains the contributions from the
non-compact polygons. It is then very easy to derive the following
non-vanishing Poisson brackets,
\begin{eqnarray}
  \label{rpp-kin-pois}
  \pois{\kin}{\cmom_\cut} &=& 
  \blgen_{\pol_{-\cut}} - \blgen_{\pol_{\cut}}, 
         \hspace{8em}     \qquad \cut\in\cuts_0,\nwl  
  \pois{\kin}{M_\ecut} &=& 
  \emul_{\pol_{-\ecut}} - \emul_{\pol_{\ecut}} , \qquad 
  \pois{\kin}{\cdir_\ecut} = \jmul_\ecut, 
                           \qquad \ecut\in\cuts_\infty.
\end{eqnarray}
These are obviously the infinitesimal generators of the redundancy
transformations \eref{rpp-cmom-lor}, \eref{rpp-M-lor}, and
\eref{rpp-tws-cdir}. We can also see that the constraints are not all
independent. Setting $\jmul_\ecut=0$, $\emul_\pol=\emul$, and
$\blgen_\pol=-\emul\,\gam_0$ for some fixed $\emul\in\RR$, we find
that the linear combination \eref{rpp-kin-sum} vanishes identically,
again because the sum over all vectors $\dis_\cut$ for
$\cut\in\cuts_0$ is zero. We have the following relation between the
constraints,
\begin{equation}
  \label{rpp-kin-sum-zero}
     \sumi{\pol\in\pols_0} \pconv_\pol^0 
   + \sumi{\pol\in\pols_\infty} \econ_\pol = 0.
\end{equation}
Everything fits together consistently. We can also recover the
original phase space spanned by the absolute positions $\pos_\prt$ and
momentum vectors $\mom_\prt$ of the particles. It is the quotient
space $\ksp_\cuts/{\sim}$, where two states
$\state_1,\state_2\in\ksp_\cuts$ are equivalent,
$\state_1\sim\state_2$, if they can be transformed into each other by
a redundancy transformation generated by the constraints.

We can now also check that the symplectic potential \eref{rpp-pot-ext}
coincide with the original one on the kinematical subspace
$\ksp_\cuts$. We start from the original expression
\eref{rpp-pot-pois}, and insert the definition \eref{rpp-cmom-mom} of
the momentum vectors $\mom_\prt$,
\begin{equation}
  \pot = \ft12\sumi\prt \Trr{\mom_\prt \, \dd\pos_\prt} 
       = \ft12\sumi\prt \sumi{\cut\in\cuts_\prt}
                        \Trr{\cmom_\cut \, \dd\pos_{\prt_\cut}}. 
\end{equation}
Again, we can rearrange the summation using the identity
\eref{graph-sum}, which gives
\begin{equation}
  \pot = \ft12\sumi{\cut\in\cuts_+} 
           \Trr{\cmom_\cut \, \dd\pos_{\prt_\cut}}
       + \ft12\sumi{\cut\in\cuts_-} 
           \Trr{\cmom_\cut \, \dd\pos_{\prt_\cut}}
       + \ft12\sumi{\ecut\in\cuts_{-\infty}} 
           \Trr{\cmom_\ecut \, \dd\pos_{\prt_\ecut}}.
\end{equation}
Then we replace the index $\cut$ in the second sum by $-\cut$, so that
this also becomes a sum over $\cuts_+$. And finally we insert the
definition of the relative position vectors $\dis_\cut$, the energies
$M_\ecut$, and the clocks $T_\ecut$. The result is
\begin{equation}
  \label{rpp-pot-rel}
  \pot = \ft12\sumi{\cut\in\cuts_+} \Trr{\cmom_\cut \, \dd\dis_\cut}
       - \sumi{\ecut\in\cuts_\infty}  M_\ecut \, \dd  T_\ecut.
\end{equation}
Up to a total derivative, which can always be added to the symplectic
potential, this is equal to \eref{rpp-pot-ext} on the kinematical
constraint surface $\ksp_\cuts\subset\esp_\cuts$, because there we
have $L_\ecut=0$. 

\subsubsection*{Mass shell constraints}
In addition to the kinematical constraints, we also have to impose the
\emph{dynamical}, or mass shell constraints, providing the actual time
evolution with respect to the ADM time $t$. They are still given by
\eref{rpp-ham}, and the Hamiltonian is a general linear combination
thereof,
\begin{equation}
  \label{rpp-mss-rel}
  \ham = \sumi\prt \mul_\prt \, \con_\prt , \qquad 
  \con_\prt = \ft14 \Trr{\mom_\prt\^2} + \ft12 m_\prt\^2 \approx 0. 
\end{equation}
The subspace $\psp_\cuts\subset\ksp_\cuts$ defined by the mass shell
constraints is called the \emph{physical phase space}. 

To derive the time evolution equations, we have to express the momenta
$\mom_\prt$ as functions of the relative momenta $\cmom_\cut$ for
$\cut\in\cuts_0$, and the energies $M_\ecut$ for
$\ecut\in\cuts_\infty$. Consequently, the mass shell constraints have
non-vanishing brackets with the relative position vectors $\dis_\cut$
for $\cut\in\cuts_0$, and the clocks $T_\ecut$ for
$\ecut\in\cuts_\infty$. A straightforward calculation yields the
following brackets of a momentum vector $\mom_\prt$ with a relative
position vector $\dis_\cut$,
\begin{equation}
  \label{rpp-pois-dis}
  \pois{\momv_\prt^a}{\disv_\acut^b} 
  = \sumi{\bcut\in\cuts_\prt} \pois{\cmomv_\bcut^a}{\disv_\acut^b} 
  = \cases{ \phantom{-} \eta^{ab} & if $\acut\in\cuts_\prt$, \cr 
                     -  \eta^{ab} & if $\acut\in\cuts_{-\prt}$, \cr
            \phantom{-}0 & otherwise.}
\end{equation}
And for the clocks we finds
\begin{equation}
  \label{rpp-pois-T}
  \pois{\momv_\prt^a}{T_\ecut} 
  = \sumi{\bcut\in\cuts_\prt} \pois{\cmomv_\bcut^a}{T_\ecut}
  = - \sumi{\bcut\in\cuts_\prt\cap\cuts_{-\infty}} 
             \pois{M_\bcut}{T_\ecut} \, \eta^{a0} 
  = \cases{ \eta^{a0} & if $\ecut\in\cuts_{-\prt}$, \cr 
            \phantom{-} 0 & otherwise. }
\end{equation}
Finally, the brackets with the mass shell constraints are
\begin{equation}
  \label{rpp-pois-mss}
  \pois{\con_\prt}{\dis_\cut} 
     = \cases{  \phantom{-} \mom_\prt & if $\prt=\prt_\cut$, \cr 
                         -  \mom_\prt & if $\prt=\prt_{-\cut}$, \cr
                \phantom{-}  0       & otherwise, }
  \qquad
  \pois{ \con_\prt }{T_\ecut} 
    = \cases{  \momv_\prt^0  & if $\prt=\prt_{-\ecut}$ \cr
                0            & otherwise. }
\end{equation}
Using all this, we can easily derive the following brackets with
$\ham$, providing the time evolution of the link variables with
respect to the ADM time $t$,
\begin{equation}
  \label{rpp-evolve-rel}
  \dot\dis_\cut = \pois{\ham}{\dis_\cut}  
                =  \mul_{\prt_{\cut}} \, \mom_{\prt_{\cut}} 
                 - \mul_{\prt_{-\cut}} \, \mom_{\prt_{-\cut}}, \qquad 
  \dot T_\ecut = \pois{\ham}{T_\ecut} 
               = \mul_{\prt_{-\ecut}} \, \momv^0_{\prt_{-\ecut}}. 
\end{equation}
This is exactly how we expect the relative position vectors
$\dis_\cut$ and the clocks $T_\ecut$ to behave when the particles are
moving along their world lines. We should also note that this is
consistent with the definitions \eref{rpp-dis} of $\dis_\cut$ and
\eref{rpp-T} of $T_\ecut$, and the original time evolution equations
\eref{rpp-evolve}.

\subsubsection*{Symmetries}
Let us also consider the rigid symmetries of the system at the level
of the extended phase space $\esp$. We have seen that the charges are
given by
\begin{equation}
  \label{rpp-charges}
  M = \sumi{\ecut\in\cuts_\infty} M_\ecut, \qquad
  S = \sumi{\cut\in\cuts_+} L_\cut,
\end{equation}
where $M_\ecut$ are the energies assigned to the external links, and
$L_\cut$ are the angular momenta of the internal links, which were
given by \eref{rpp-M-S}. The rigid symmetry generated by $M$ is very
simple. The only non-vanishing brackets are
\begin{equation}
  \label{rpp-pois-M}
  \pois{M}{T_\ecut} = - 1 \follows 
   T_\ecut \mapsto T_\ecut - \tpar ,
\end{equation}
where $\tpar$ is some real parameter. The symmetry only acts on the
clocks, which are all turned backwards by the same amount. In
\fref{fspc}, this corresponds to a translation of the ADM surface
backwards in time, or to a forward time translation of the reference
frame with respect to the particles. The total energy $M$ is thus the
generator of the absolute time evolution in the centre of mass frame.
This must be distinguished from the ADM time evolution with respect to
the coordinate time $t$, which is generated by the weakly vanishing,
unphysical Hamiltonian $\ham$.

To find the symmetry generated by $S$, let us first evaluate the
brackets of the link variables with $L_\cut$. A straightforward
calculation shows that the only non-vanishing brackets are
\begin{equation}
  \label{rpp-pois-L-cut}
  \pois{L_\cut}{\dis_{\pm\cut}} = \ft12\comm{\gam_0}{\dis_{\pm\cut}},
   \qquad
  \pois{L_\cut}{\cmom_{\pm\cut}} = \ft12\comm{\gam_0}{\cmom_{\pm\cut}},
   \qquad \cut\in\cuts_0.
\end{equation}
This is the generator of a counter clockwise rotation about the
$\gam_0$-axis, which acts on the vectors associated with the links
$\cut$ and $-\cut$. If we sum over all internal links
$\cut\in\cuts_+$, then the transformation acts in the same way on all
internal links,
\begin{equation}
  \label{rpp-pois-S}
  \pois{S}{\dis_{\cut}} = \ft12\comm{\gam_0}{\dis_{\cut}},
   \qquad
  \pois{S}{\cmom_{\cut}} = \ft12\comm{\gam_0}{\cmom_{\cut}},
   \qquad \cut\in\cuts_0.
\end{equation}
The resulting transformation is a rotation \eref{rotation} about the
$\gam_0$-axis,
\begin{equation}
  \label{rpp-rot-S}
  \cmom_\cut \mapsto 
   \expo{\rpar \gam_0/2} \cmom_\cut \, 
       \expo{-\rpar \gam_0/2}, \qquad
  \dis_\cut \mapsto 
   \expo{\rpar\gam_0/2} \dis_\cut \,
       \expo{-\rpar\gam_0/2}, \qquad
   \cut\in\cuts_0, 
\end{equation}
where $\rpar$ is the angle of rotation. But now, we have the following
problem. Consider the action of this transformation on the polygons in
\fref{fpol}. All internal links are rotated by an angle $\rpar$ in
counter clockwise direction. But the directions of the external links,
specified by the unit vectors $\dis_{\pm\ecut}$, are unchanged. As a
consequence, there is a certain restriction on the parameter $\rpar$,
because for too large angles the non-compact polygons are twisted in
such a way that they can no longer be realized as spacelike surfaces.

This can be avoided as follows. We redefine the total angular
momentum, including also the angular momenta $L_\ecut$ of the external
links,
\begin{equation}
  \label{rpp-J}
  J = \sumi{\cut\in\cuts_+} L_\cut
    + \sumi{\ecut\in\cuts_\infty} L_\ecut. 
\end{equation}
Due to the constraints \eref{rpp-jcon}, the charges $S$ and $J$ are in
fact weakly equal, thus $S\approx J$. The associated rigid symmetries
are therefore equal up to a kinematical gauge symmetry. For $J$, we
have the additional non-vanishing brackets
\begin{equation}
  \label{rpp-pois-J}
  \pois{J}{\cdir_\ecut} = 1 \follows
  \pois{J}{\dis_{\pm\ecut}} = \ft12\comm{\gam_0}{\dis_{\pm\ecut}},
  \qquad  \ecut\in\cuts_\infty.
\end{equation}
The second equation follows from the definition \eref{rpp-ecut-vec} of
the unit vectors $\dis_{\pm\ecut}$. Hence, if we replace $S$ by $J$,
then the transformation \eref{rpp-rot-S} also applies to the external
links. For the relative momenta $\cmom_{\pm\ecut}$ this is trivial,
because they are proportional to $\gam_0$ and therefore invariant, and
for the unit vectors $\dis_{\pm\ecut}$ it follows from
\eref{rpp-pois-J}. The charge $J$ generates a rigid rotation of the
ADM surface with respect to the reference frame, without deforming it.
For such a transformation, the parameter $\rpar$ can be arbitrarily
large, and for $\rpar=2\pi$ we get a full rotation and thus the
identity. On the extended phase space $\esp$, it is therefore more
natural to call $J$ the total angular momentum, but for all physical
states its value is of course equal to $S$.

\subsubsection*{Global aspects}
There are some global features of the extended phase space which we
neglected so far. For example, until now we have fixed the graph
$\cuts$. But it is not possible to use the same graph for every
possible configuration of the particles. We therefore have to take
into account that the extended phase space consists of finitely many
disconnected components. The same, of course, applies to the
kinematical and the physical subspace,
\begin{equation}
  \label{rpp-esp-union}
  \esp = \bigcup_{\cuts} \esp_\cuts, \qquad 
   \ksp = \bigcup_{\cuts} \ksp_\cuts, \qquad 
   \psp = \bigcup_{\cuts} \psp_\cuts.
\end{equation}
On the other hand, the original phase space of the particles was a
connected manifold. We previously argued that this phase space is
equal to the quotient space $\ksp_\cuts/{\sim}$. But this identity
only holds locally. Each components of the quotient space only covers
a certain region of the original phase space, namely that region where
the graph $\cuts$ defines a proper triangulation of the ADM surface.

So, what we actually have is an atlas of finitely many charts covering
the original phase space of the particles. An alternative point of
view is to consider the transitions between different triangulations
as \emph{large gauge symmetries}. The quotient space $\ksp/{\sim}$,
where the large gauge symmetries are also divided out, is then
globally equal to the original phase space. We shall look at this
global structure of the phase space more closely in \sref{inter},
where it is schematically indicated in \fref{phsp}. 

It is also useful to keep in mind that following hierarchy of phase
spaces, which is very similar to the Hamiltonian formulation of
general relativity,
\begin{equation}
  \label{rpp-phase-array}
  \begin{array}{ccccc}
     \esp  & \supset & \ksp & \supset & \psp \cr
            & & \downarrow & & \downarrow \cr
        \phantom{\ksp/{\sim}}    &  & \ksp/{\sim} &  & \psp/{\sim}. 
  \end{array}
\end{equation}
The starting point is the extended phase space $\esp$. The kinematical
constraints (\ref{rpp-icon}--\ref{rpp-jcon}) define the kinematical
subspace $\ksp$, and the mass shell constraints \eref{rpp-mss-rel}
define the physical phase space $\psp$. At each level, the gauge
symmetries associated with the respective constraints can be divided
out. The quotient space $\ksp/{\sim}$ is the original phase space,
where the redundancies are removed. And the quotient space
$\psp/{\sim}$ is the reduced phase space, which is the set of all
classical trajectories, that is all physically inequivalent solutions
to the equations of motion.

Finally, there is another somewhat marginal point, which also has to
do with the global structure of the phase space. The kinematical
constraints alone are not sufficient to ensure that the geometry of
the ADM surface in \fref{fspc} is well defined. For the polygons in
\fref{fpol} to be well defined, the sequences of edges defined by the
vectors $\dis_\cut$ must be boundaries of spacelike surfaces. A
necessary condition is of course that all vectors $\dis_\cut$ are
spacelike. But there are more consistency conditions. For a compact
triangle, for example, the vector product of two successive edges must
be negative timelike, for the polygon surface to be spacelike with the
correct orientation. For higher polygons and also for non-compact
polygons there are more complicated conditions.

However, they can all be written as inequalities involving the vectors
$\dis_\cut$, and this is all we need to know. They can be treated at
the same level as the positive energy conditions, which we have to
impose together with the mass shell constraints. They pick out an open
subset of the actual constraint surface, which is the physically
accessible subset. The kinematical phase space $\ksp_\cuts$ is this
subset of the constraint surface, similar to the physical phase space
$\psp_\cuts$, which is that part of the constraint surface where the
positive energy conditions are satisfied. This further restriction has
no influence on the local structure of the phase space, for example
number of gauge degrees of freedom. But it imposes some restriction on
the parameters of the gauge transformations, as otherwise we drop out
of the kinematical phase space.

For example, we cannot rotate the external links arbitrarily far, thus
there is a restriction on the parameters $\jmul_\ecut$ in
\eref{rpp-tws-cdir}, as otherwise the external polygons are no longer
spacelike. This was also the reason why the transformation generated
by $S$ as a conserved charge is not well defined for all parameters.
And there is also a restriction on the multipliers $\mul_\prt$ for the
mass shell constraints. We cannot shift a particle along its world
line arbitrarily far, as otherwise the polygons attached to this
particle are no longer spacelike. Of course, this already follows from
the original restriction \eref{spacelike}, so this restriction has
already been present before we introduced the link variables and the
extended phase space.

\section{Spacetime geometry}
\label{inter}
The gravitational field of a massive point particle in three
dimensional Einstein gravity is a cone with a deficit angle of $8\pi
Gm$, where $0<m<\Mpl/4$ is the mass of the particle and $G$ is
Newton's constant, which is the inverse of the Planck mass $\Mpl=1/G$.
In the neighbourhood of the world line, a cylindrical coordinate
system $(t,r,\p)$ can be introduced, so that the metric becomes
\begin{equation}
  \label{prt-ds}
  \dd s^2 = - \dd t^2 + \dd r^2 + (1-4Gm)^2 \, r^2 \, \dd\p^2.
\end{equation}
This metric is locally flat for $r>0$, and there is a conical
singularity on the world line at $r=0$. A spacetime containing $\nprt$
particles can in principle be covered by an atlas of $\nprt$ such
conical coordinate systems, and the relative motion of the particles
can be read off from the respective transition functions.

And alternative and more appropriate way to introduce coordinates is
first to consider the spacetime between the particles. Let us remove
the world lines from the spacetime. It is then a locally flat
manifold, which is not simply connected. If a spacetime vector is
transported, say, along a path that winds around a world line in
clockwise direction, then we pick up a Lorentz rotation, which is
called the \emph{holonomy} of the particle. It does not depend on the
precise path, and is therefore a quantity that can be assigned to the
particle. If the deficit angle of the conical singularity is $8\pi
Gm$, then the holonomy is a rotation by $8\pi Gm$ about some timelike
axis, which is parallel to the world line.

From the holonomy we can read off both the mass and the direction of
the motion of the particle. It can be regarded as a generalized
momentum \cite{matwel}. So, we have in this picture a well defined
notion of a momentum. Massless particles can also be included. For a
massless particle, the world line is lightlike, and the holonomy is a
null rotation, whereas a conical coordinate system like \eref{prt-ds}
does not exist. It is also possible to introduce relative position
coordinates of the particles, very similar to the free particle
coordinates in the previous section. So let us make all this a little
bit more explicit.

\subsubsection*{Minkowski coordinates}
Suppose that the spacetime is foliated by a family of spacelike
slices, labeled by an ADM time coordinate $t$. Then we would like to
know, for example, what the relative position of two particles is at a
moment of time $t$. Consider the corresponding ADM surface, and some
neighbourhood of this surface in spacetime. If the world lines are
excluded, then this is a spacelike surface embedded into a locally
flat spacetime. In this locally flat spacetime, we can introduce an
atlas of \emph{Minkowski charts}. This can be done as follows. First,
we triangulate the ADM surface in the same way as before. We introduce
a graph $\cuts$, so that the links $\cut\in\cuts$ are spacetime
geodesics, and the ADM surface is divided into a collection $\pols$ of
simply connected polygons $\pol\in\pols$.

\begin{figure}[t]
  \begin{center}
    \epsfbox{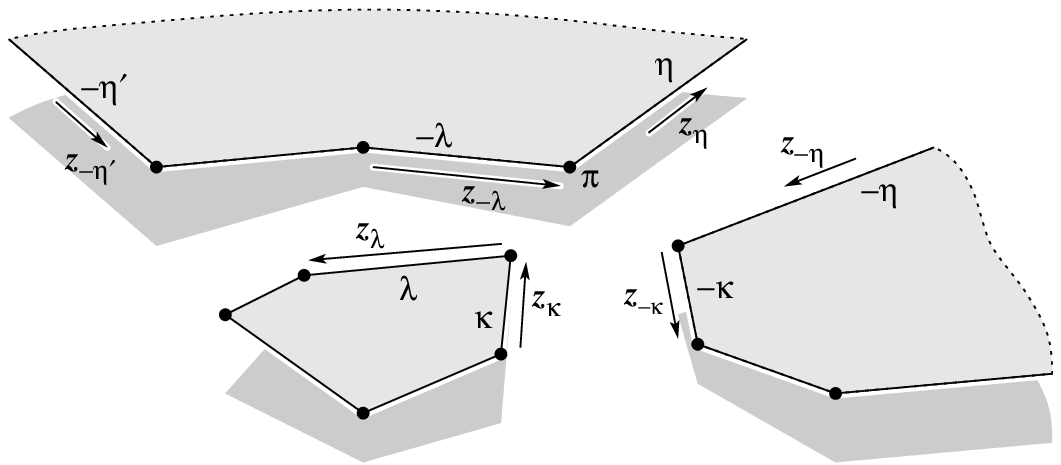}
    \xcaption{The ADM surface is a conical surface with $\nprt$ tips.
    It is divided into a collection of simply connected polygons. The
    polygons are embedded into an auxiliary Minkowski space, which
    provides an atlas of local coordinate charts. The edges $\pm\cut$
    of two adjacent polygons $\pol_{\pm\cut}$ are mapped onto each
    other by some Lorentz rotation. This ensures that the polygons can
    be glued together, forming a spacelike surface, which is embedded
    into a locally flat spacetime with $\nprt$ conical singularities.}
    \label{ipol}
  \end{center}
\end{figure}

For simplicity, let us assume that the topology of space is that of
$\RR^2$. Possible generalizations will be discussed later. Then, in
the neighbourhood of each polygon, we introduce a Minkowski coordinate
system. This is possible because the spacetime is locally flat, and
the neighbourhood of the polygon in spacetime is simply connected. The
polygons become spacelike surfaces, which are embedded into some
auxiliary Minkowski space, as indicated in \fref{ipol}. Locally, this
embedding Minkowski space can be identified with the spacetime. Each
edge $\cut\in\cuts_\pol$ of a polygon $\pol$ represents a geodesic
connecting two particles. It becomes a straight line in Minkowski
space. 

To this we can assign a vector $\dis_\cut\in\algsl(2)$, in the same
way as before. It represents the relative position of the particles
$\prt_\cut$ and $\prt_{-\cut}$ in spacetime. Vice versa, the vectors
$\dis_\cut$ determine the geometry of the individual polygons in
\fref{ipol}. There are only the following modifications, as compared
to polygons in \fref{fpol}. The vectors $\dis_\cut$ now refer to local
coordinates on the spacetime, and not to a fixed reference frame. We
are allowed to perform coordinate transformations. On the embedded
polygons in \fref{ipol}, a coordinate transformation acts as a Lorentz
rotation and a translation, and the vectors $\dis_\cut$ are also
Lorentz rotated. 

Moreover, for each pair of edges $\cut$ and $-\cut$, the vectors
$\dis_\cut$ and $\dis_{-\cut}$ still represent the relative position
of the same two particles. But now they refer to different local
coordinates systems. The vector $\dis_\cut$ is defined in the chart
that contains the polygon $\pol_\cut$, and the vector $\dis_{-\cut}$
refers to the chart containing the polygon $\pol_{-\cut}$. There is an
overlap region of the two charts, which contains the links $\cut$ and
$-\cut$. The transition function between the two charts must be an
isometry of Minkowski space, thus a Poincar\'e transformation. Let us
only consider the rotational component of this transition function. It
is represented by an element $\chol_\cut\in\grpSL(2)$ of the Lorentz
group.

In \fref{ipol}, this means that the edge $-\cut$ of the polygon
$\pol_\cut$ is mapped onto the edge $\cut$ of the polygon $\pol_\cut$
by a Lorentz rotation given by $\chol_\cut$. Of course,
$\chol_{-\cut}$ is then the inverse transformation. We have the
following relation between the relative position vectors and the
transition functions,
\begin{equation}
  \label{cut-inv-rel}
  \chol_{-\cut} = \chol\inv_\cut , \qquad
  \dis_{-\cut} = - \chol_\cut \, \dis_\cut \, \chol\inv_\cut, 
   \qquad \cut\in\cuts.
\end{equation}
This relation also applies to the external edges in \fref{ipol}. The
spacelike unit vectors $\dis_{\pm\ecut}$ define the directions of the
external edges in the embedding Minkowski space. And there is also a
transition function associated with these links, defining the relation
between two adjacent non-compact charts.

Now, suppose we are given the vectors $\dis_\cut$ and the transition
functions $\chol_\cut$ for all $\cut\in\cuts$. The geometry of the
polygons in \fref{ipol} is then determined up to smooth deformation of
the surfaces in the interior. The transition functions tell us how
these polygons are to be glued together. This determines the geometry
of the ADM surface, and the way it is embedded into the spacetime. So,
we conclude that the relative position vectors $\dis_\cut$ and the
transition functions $\dis_\cut$ specify the geometry of space at a
moment of time, up to some physically redundant gauge symmetries.  The
deformations are in fact gauge symmetries of general relativity,
namely spacetime diffeomorphisms acting on the foliation. 

It is useful to think about the polygons in \fref{ipol} as a
\emph{deformation} of the polygons in \fref{fpol}, with Newton's
constant being the deformation parameter. We can take a limit $G\to0$,
where the gravitational interaction is switched off, if we relate the
transition functions $\chol_\cut$ in a certain way to the relative
momentum vectors $\cmom_\cut$ of the free particles. Let us define
\begin{equation}
  \label{cmom-chol-exp}
  \chol_\cut = \expo{4\pi G \cmom_\cut} = 
              \one + 4\pi G \, \cmom_\cut + O(G)^2,
       \qquad \cut\in\cuts.  
\end{equation}
Newton's constant has to show up in this relation, because the vector
$\cmom_\cut$ has the physical dimension of a momentum, whereas
$\chol_\cut$ is dimensionless. The numerical factor will be explained
in a moment. If we insert this into \eref{cut-inv-rel}, and take the
limit $G\to0$, then in the leading order in $G$ we recover the
relations \eref{rpp-cut-inv}. Moreover, for all transition functions
we have $\chol_\cut\to\one$, so that the individual Minkowski charts
can be unified, providing one global chart. So, in this limit we
recover the triangulated ADM surface of the free particles, which is
globally embedded into a flat Minkowski space.

It is also useful to look at the way the link variables transform
under coordinate transformation in the Minkowski charts. In each
chart, we can act on the coordinates with some Poincar\'e
transformation. On the corresponding polygon in \fref{ipol}, the
transformation acts as a Lorentz rotation and a translation. Since
none of the link variables refers to the absolute position of the
polygons in the embedding Minkowski space, we can again on the Lorentz
rotations and ignore the translations. So, let
$\blpar_\pol\in\grpSL(2)$ be the Lorentz rotation acting on the chart
containing the polygon $\pol$. Then, one can easily verify that the
link variables transform as
\begin{equation} 
  \label{chol-dis-lor}
  \chol_\cut \mapsto 
       \blpar_{\pol_{-\cut}}\, \chol_\cut \, \blpar\inv_{\pol_\cut},
  \qquad 
  \dis_\cut \mapsto 
       \blpar_{\pol_\cut}\, \dis_\cut \, \blpar\inv_{\pol_\cut}.
\end{equation}
Note that the transition function $\chol_\cut$ sees the
transformations acting on both polygons $\pol_\cut$ and
$\pol_{-\cut}$, whereas the vector $\dis_\cut$ only refers to the
polygon $\pol_\cut$ and transforms accordingly.

To see what this coordinate transformation looks like when the
gravitational interaction is switched off, we replace the parameter
$\blpar_\pol$ by a vector $\blgen_\pol$, so that
\begin{equation}
  \label{lor-exp}
  \blpar_\pol = \expo{4\pi G \blgen_\pol} = 
              \one + 4\pi G \, \blgen_\cut + O(G)^2,
       \qquad \pol\in\pols .
\end{equation}
In the limit $G\to0$, the Lorentz rotation \eref{chol-dis-lor} then
reduces to the redundancy transformation \eref{rpp-cmom-lor} for the
relative momentum vectors of the free particles, again in the leading
order in $G$. The coordinate transformations in the Minkowski charts
are the redundancies of the interacting particle system, analogous to
the redundancies of the relative momenta for the free particles. 

\subsubsection*{Holonomies}
\begin{figure}[t]
  \begin{center}
    \epsfbox{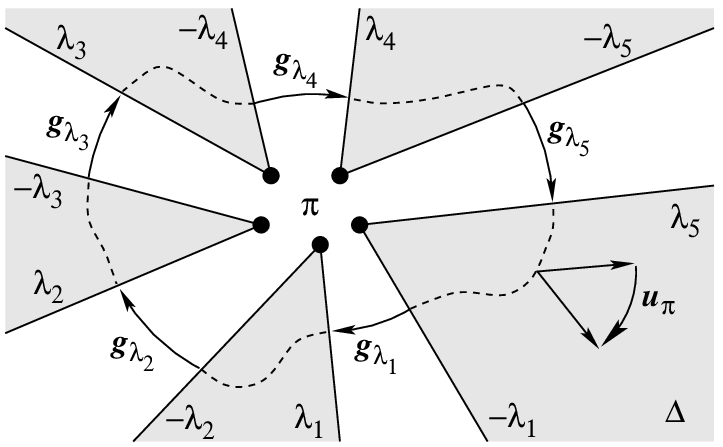}
    \xcaption{There is a conical singularity at each vertex of the
    triangulation. Its holonomy $\hol_\prt$, evaluated in the polygon
    $\pol$, is the product of the transitions functions $\chol_\cut$
    for $\cut\in\cuts_{\prt,\pol}$. In the given example, we have
    $\cuts_{\prt,\pol}=\{\cut_1,\cut_2,\cut_3,\cut_4,\cut_5\}$, because
    $-\cut_1$ and $\cut_5$ are the edges of the polygon $\pol$.}
    \label{holo}
  \end{center}
\end{figure}

When the polygons are glued together, then at each vertex of the
triangulation a conical singularity arises. Consider a spacetime
vector which is transported once around a particle $\prt$, in
clockwise direction. As indicated in \fref{holo}, the vector is
defined in some polygon $\pol$ adjacent to the particle $\prt$. The
path winds once around the particle. Whenever it hits a link
$-\cut\in\cuts_{-\prt}$ beginning at $\prt$, then it continues from
the link $\cut\in\cuts_\prt$ ending at $\prt$, in the next polygon.
The sequence of in which the links are crossed is given by the cyclic
ordering of the set $\cuts_\prt$. When the vector is transferred from
the link $-\cut$ to the link $\cut$, hence from one chart to another,
then we have to act on it with the transition function $\chol_\cut$.

The holonomy $\hol_\prt\in\grpSL(2)$ of the particle $\prt$ is the
product of the transition functions $\chol_\cut$ for all
$\cut\in\cuts_\prt$. The ordering of the factors is defined by the
cyclic ordering of the set $\cuts_\prt$. Which factor is the first
depends on the polygon $\pol$ in which the path begins and ends. Let
us introduce the following notation. If $\prt$ is a particle, and
$\pol$ is a polygon adjacent to it, then
$\cuts_{\prt,\pol}=\{\cut',\dots,\cut\}$ is the ordered set of all
links ending $\prt$, so that the first and last elements
$\cut,\cut'\in\cuts_\prt$ are those successive links ending at $\prt$,
which enclose the polygon $\pol=\pol_\cut=\pol_{-\cut'}$. We say that
the cyclic ordering of the set $\cuts_\prt$ is \emph{broken} at the
polygon $\pol$.

With this notation, the holonomy of the particle $\prt$, evaluated in
the polygon $\pol$, becomes 
\begin{equation}
  \label{prt-hol}
  \hol_{\prt,\pol} = \prodi{\cut\in\cuts_{\prt,\pol}} \chol_\cut.
\end{equation}
There are as many holonomies $\hol_{\prt,\pol}$ of the particle $\prt$
as there are polygons $\pol$ adjacent to the vertex $\prt$. They are,
however, just representatives of the same physical object in different
coordinate charts. To see this, consider the behaviour of the holonomy
under Lorentz rotations \eref{chol-dis-lor} of the Minkowski
charts, 
\begin{equation}
  \label{hol-lor}
  \hol_{\prt,\pol} \mapsto 
    \blpar_\pol \, \hol_{\prt,\pol} \, \blpar\inv_\pol. 
\end{equation}
Hence, $\hol_{\prt,\pol}$ transforms according to the Lorentz rotation
acting on the polygon $\pol$. Moreover, consider the representatives
of the holonomy $\hol_{\prt,\pol_{\cut}}$ and
$\hol_{\prt,\pol_{-\cut}}$ in the two polygons, sharing a common
boundary $\cut\in\cuts_\prt$ and $-\cut\in\cuts_{-\prt}$. For these
two representatives, the definition \eref{prt-hol} differs by the
position of the factor $\chol_\cut$. For $\pol=\pol_\cut$ it is the
last factor, whereas for $\pol=\pol_{-\cut}$ it is the first factor in
the product. This implies
\begin{equation}
  \label{hol-trans}
  \hol_{\prt,\pol_{\cut}} = 
    \chol\inv_\cut \, \hol_{\prt,\pol_{-\cut}} \, \chol_\cut. 
\end{equation}
But this is just the statement that $\hol_{\prt,\pol_{-\cut}}$ and
$\hol_{\prt,\pol_\cut}$ represent the same physical object in two
different coordinate charts, where $\chol_\cut$ is the transition
function between the two charts. Note that if there is only one link
$\cut$ attached to the particle $\prt$, then there is also only one
polygon $\pol$, and the relation \eref{hol-trans} is just a trivial
identity, because $\hol_{\prt,\pol}=\chol_\cut$ and
$\pol=\pol_\cut=\pol_{-\cut}$.

According to the arguments given in the beginning, we can now read off
the mass of the particle and the direction of motion from the
holonomy. The direction of motion is the axis of the Lorentz rotation
represented by the holonomy. This axis can be defined by
\emph{projecting} the group element $\hol_\prt$ onto a vector
$\mom_\prt$, which we call the \emph{momentum vector} of the particle.
According to \eref{project}, the projection $\grpSL(2)\to\algsl(2)$ is
defined by expanding a group element in terms of the unit and the
gamma matrix, and then dropping the term proportional to the unit
matrix. To give the momentum vector the correct physical dimension, we
introduce Newton's constant once again, and define
\begin{equation}
  \label{prt-mom}
  \hol_{\prt,\pol} = \hols_\prt \, \one + 
              4 \pi G \, \momv_{\prt,\pol}^a \, \gam_a , \qquad 
  \mom_{\prt,\pol} = \momv_{\prt,\pol}^a \, \gam_a .
\end{equation}
By definition, the momentum vector commutes, as a matrix, with the
holonomy, and therefore defines the axis of the Lorentz rotation in
Minkowski space. It transforms in the same way \eref{hol-lor} under
coordinate transformations, and between the representatives in the
different charts we also have the relation \eref{hol-trans}, with
$\hol_\prt$ replaced by $\mom_\prt$.

In the limit $G\to0$, and with \eref{cmom-chol-exp} inserted, the
product in \eref{prt-hol} can be replaced by a sum. In the leading
order in $G$, we recover the definition \eref{rpp-cmom-mom} of the
momentum vector of the free particle, which is the sum of the relative
momentum vectors. To explain the numerical factors appearing together
with Newton's constant, let us derive the relation between the
holonomy and the mass of the particle. For the mass to be $m_\prt$,
the holonomy must be a rotation by $8\pi Gm_\prt$ in clockwise
direction about some timelike axis. All such element of the group
$\grpSL(2)$ belong to the conjugacy class of $\expo{4\pi
Gm_\prt\gam_0}$, which represents a clockwise rotation about the
$\gam_0$-axis.

This conjugacy class of $\grpSL(2)$ is specified by the following
\emph{mass shell} and \emph{positive energy} condition. The scalar
$\hols_\prt$ in \eref{prt-mom}, which is half the trace of the
holonomy, must be equal to $\cos(4\pi Gm_\prt)$, and the
$\gam_0$-component of the momentum $\mom_\prt$ must be positive to fix
the direction of the rotation. Hence,
\begin{equation}
  \label{mss-hol}
  \hols_\prt = \cos(4\pi Gm_\prt), \qquad 
    \momv_\prt^0 = \ft12\Trr{\hol_\prt\gam^0} > 0.
\end{equation}
Note that the trace of the holonomy is a scalar. It is the same in
every polygon, because the trace is invariant under cyclic
permutations of the factors,
\begin{equation}
  \label{prt-hols}
  \hols_\prt = \ft12\Trr{\hol_\prt} = 
        \ft12\Trrr{ \prodi{\cut\in\cuts_\prt} \chol_\cut },
\end{equation}
and it transform trivially under coordinate transformations. The mass
shell condition is therefore a well defined, coordinate independent
equation. Actually, there are some subtleties regarding the global
structure of the group manifold $\grpSL(2)$. But these can be ignored
here, and we can take \eref{mss-hol} as a condition for the mass of
the particle to be $m_\prt$. A comprehensive discussion can be found
in \cite{matwel}.

It is now also possible to include massless particles. For $m_\prt=0$,
the holonomy becomes a null rotation and the momentum vector is a
lightlike vector. Thus a massless particle moves with the velocity of
light. On the other hand, there is an upper bound on the mass, because
the deficit angle of a conical singularity must be smaller than
$2\pi$. The upper bound is $\Mpl/4=1/4G$. This is also the value where
the cosine takes its minimum $-1$, and the holonomy becomes a full
rotation by $2\pi$. The allowed range of the mass parameters is
\begin{equation}
  \label{mss-range}
  0 \le m_\prt < \Mpl/4 = 1/4G.
\end{equation}
And finally, to explain the numerical factors, let us consider the
limit $G\to0$ once again. Between the trace of the holonomy and the
momentum vector we have the relation \eref{det-con}, 
\begin{equation}
  \label{hols-mom}
  \hols_\prt\^2 = 8 \pi^2 G^2 \, \Trr{\mom_\prt\^2} + 1.
\end{equation}
Therefore, the mass shell condition implies
\begin{equation}
   \ft12  \Trr{\mom_\prt\^2} = - \frac{\sin^2(4\pi Gm_k)}
                                      {16 \pi^2 G^2}
\end{equation}
In the limit $G\to0$, this obviously becomes the free particle mass
shell condition \eref{rpp-erg}. 

\subsubsection*{The conical infinity}
There is one feature of the free particle system which has so far no
counterpart for the interacting system. This is the \emph{reference
frame}, defined by the embedding Minkowski space. The Minkowski space
in \fref{ipol} is now only an auxiliary space, providing local
coordinates on the spacetime manifold. Since we are free to perform
coordinate transformations, rotating the individual polygons, this
Minkowski space does not provide a well defined reference frame. On
the other hand, a reference frame is needed if we want to set up a
proper Hamiltonian formulation in general relativity \cite{nico}.

To define a reference frame, we have to look at the asymptotic
structure of the spacetime at spatial infinity. Let us assume that,
far away from the particles, the spacetime looks like the
gravitational field of a single particle. We'll see in \sref{reduc},
that this can be formulated as a kind of asymptotical flatness
condition, hence a fall off condition imposed on the metric at
infinity. If this is the case, then an observer at infinity
effectively sees a \emph{fictitious} centre of mass particle, and the
rest frame of this particle defines the centre of mass frame of the
universe. We may therefore identify the reference frame with the
centre of mass frame, in the same way as we did this in the previous
section.

The gravitational field of the fictitious centre of mass particle is a
cone like \eref{prt-ds}. However, we have to allow this particle to
have a variable mass $M$, which represents the total energy of the
universe. And it also receives a spin $S$, which represents the total
angular momentum of the universe. It is then possible to introduce a
\emph{conical coordinate} system $(T,R,\cdir)$. It covers a certain
region of the spacetime, far away from the particle, so that the
metric becomes that of a \emph{spinning cone} \cite{djh},
\begin{equation}
  \label{spin-cone}
  \dd s^2 = - (\dd T + 4 G S \, \dd\cdir)^2 + \dd R^2 
            + (1 - 4 G M)^2 \, R^2 \, \dd\cdir^2 . 
\end{equation}
The spinning cone has a \emph{deficit angle} of $8\pi GM$, and a
\emph{time offset} of $8\pi GS$. There is an upper bound on $M$ which
is the same as \eref{mss-range}. For convenience, let us also assume
that $M$ is positive. This is actually not needed, but one can show
that the positive energy conditions for the particles imply that also
the total energy $M$ is positive. Hence
\begin{equation}
  \label{M-range}
  0 < M < \Mpl/4=1/4G. 
\end{equation}
The conical coordinate system $(T,R,\cdir)$ can then be regarded as a
fixed reference frame. It replaces the Minkowski frame of the free
particle system. The radial coordinate $R$ defines the distance from
the fictitious world line of the centre of mass, the time coordinate
$T$ defines the absolute time in the reference frame, and the angular
coordinate $\cdir$, which has a period of $2\pi$, defines the angular
orientation of the reference frame. If we take the limit $G\to0$, with
$M$ and $S$ fixed, the spinning cone becomes a flat Minkowski space,
and the fictitious world line of the centre of mass is the
$\gam_0$-axis. 

So, the reference frame is also a deformation of its free particle
counterpart. There is only the following technical difference. The
reference frame does not provide a \emph{global} coordinate system,
which covers the whole spacetime. It only provides another local
chart, which covers a \emph{neighbourhood of infinity}. This is the
region outside a cylinder surrounding all the world lines. To define
the absolute positions of the particles with respect to this reference
frame, we cannot just read off their coordinates. The particles are
not inside the chart where \eref{spin-cone} applies. We have to define
them indirectly. At each moment of time, we have to tell how the ADM
surface is embedded into the spinning cone at spatial infinity.

Given the geometry of the ADM surface, we have to fix two additional
parameters to define its embedding into the spinning cone, because
there are two Killing symmetries of the metric \eref{spin-cone}. These
are the time translations $T\mapsto T-\tpar$ and the spatial rotations
$\cdir\mapsto\cdir+\rpar$. Hence we have the same rigid symmetries of
the centre of mass frame as before. The symmetries are the possible
rotations and translations of the reference frame with respect to the
rest of the universe \cite{nico}. To fix this freedom, we have to
assign some additional variables to the external links, telling us how
they behave when the ADM surface is embedded into the reference frame.

Since we are still free to deform the ADM surface smoothly, without
affecting the locations of the particles, let us impose the following
restriction on the external links. As for the free particles, we
require them to be \emph{spatial half lines}. A spatial half line in
the spinning cone is a spacelike geodesic extending to infinity, which
is orthogonal to the fictitious axis at $R=0$. An alternative
definition is to say that a spatial half line is orthogonal to the
Killing field of time translations. This is obviously a
straightforward generalization of the definition in Minkowski space,
where a spatial half line is orthogonal to the $\gam_0$-axis.

One can then easily show that on every spatial half line the conical
coordinates $T$ and $\cdir$ then converge in the limit where $R$ goes
to infinity. So, on each external link we have
\begin{equation}
  \label{cone-T-cdir}
   T \to T_\ecut , \qquad \cdir\to\cdir_\ecut, 
   \qquad R \to \infty , \qquad \ecut\in\cuts_\infty. 
\end{equation}
The physical interpretation of the link variables $T_\ecut$ and
$\cdir_\ecut$ is almost the same as before. Let us consider an
observer sitting at the far end of the link $\ecut$, hence at spatial
infinity. The variable $\cdir_\ecut$ tells us where this observer is,
that is in which \emph{direction}, and the variable $T_\ecut$
represents the absolute time at the location of this observer, hence
it defines a \emph{clock}. 

On the other hand, given the geometry of the ADM surface, specified by
the link variables $\dis_\cut$ and $\chol_\cut$, and also the
variables $T_\ecut$ and $\cdir_\ecut$, we know how to embed the space
manifold into the spinning cone, and thus we have complete information
about the state of the interacting particle system at a given moment
of time. We know the relative positions of the particles with respect
to each other, and the absolute positions with respect to the
reference frame. 

\subsubsection*{Consistency conditions}
There are also some consistency conditions to be satisfied by the link
variables, in analogy to the kinematical constraints for the free
particles. First of all, for a polygon $\pol$ in \fref{ipol} to be
well defined, the sequence of edges given by the vectors $\dis_\cut$
for $\cut\in\cuts_\pol$ must be the boundary of spacelike surface. For
each compact polygon we must have
\begin{equation}
  \label{icond}
   \sumi{\cut\in\cuts_\pol} \dis_\cut = 0 , \qquad \pol\in\pols_0.
\end{equation}
Additionally, there are some inequalities to be satisfied, for example
all vectors $\dis_\cut$ must be spacelike, and the boundaries must be
correctly oriented. The precise form of these inequalities is still
not important, for the same reasons as before, which we discussed in
the very end of \sref{free}.
\begin{figure}[t]
  \begin{center}
    \epsfbox{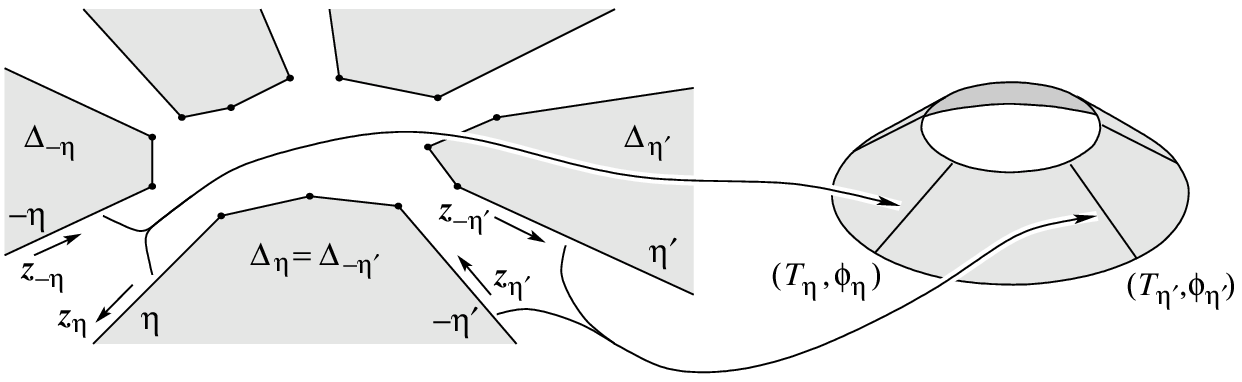}
    \xcaption{The non-compact polygons are embedded
    into the spinning cone. The spinning cone on the right defines the
    reference frame, whereas the Minkowski space on the left only
    provides local coordinates. The vectors $\dis_{\pm\ecut}$ are
    related to the conical directions $\cdir_\ecut$ of the links in
    the spinning cone, in such a way that the average orientation of
    the Minkowski frame is the same as that of the conical frame, so
    that effectively the embedding Minkowski space can also be
    regarded as a reference frame.}
    \label{cone}
  \end{center}
\end{figure}

Then there are also some consistency conditions for the ADM surface to
fit into the spinning cone at spatial infinity. It must be possible to
embed the non-compact polygons into the spinning cone, so that the
edges fit together correctly. We have indicated this, somewhat
schematically, in \fref{cone}. Each non-compact polygon
$\pol\in\pols_\infty$ covers a certain segment of the spinning cone,
which is bounded by the external links $\ecut$ and $\ecut'$. They are
spatial half lines with conical coordinates \eref{cone-T-cdir}. For
the polygons to fit in, there must be some relation between the
vectors $\dis_\cut$, defining the geometry of the polygons in the
embedding Minkowski space, and the external links variables $T_\ecut$
and $\cdir_\ecut$.

To find this relation, let us impose one further restriction. We are
still free to perform arbitrary Lorentz rotations in all Minkowski
charts, including the non-compact ones. On the other hand, each
non-compact chart has some overlap region with the conical chart at
infinity. Hence, there is a transition function between the Minkowski
coordinates and the conical coordinates. This transition function is a
local isometry between the spinning cone and Minkowski space. There is
a special class of such isometries. They have the property that the
axis of the spinning cone is mapped onto the $\gam_0$-axis of
Minkowski space. The most general such local isometry is given by
\begin{equation}
  \label{iso}
   (T,R,\cdir) \quad \mapsto \quad 
       ( T + \atau_\pol(\cdir) ) \, \gam_0  
            +  R  \, \gam( \cdir - \adir_\pol(\cdir)), 
\end{equation}
where $\adir$ and $\atau$ are two linear functions satisfying
\begin{equation}
  \label{atau-adir}
  \atau_\pol'(\cdir) = 4 G S  \qquad \adir_\pol'(\cdir) = 4 G M. 
\end{equation}
Note that this is only a local isometry, because it does not respect
the periodicity of $\cdir$. There is a two parameter family of such
local isometries, because we are free to add constants to $\atau_\pol$
and $\adir_\pol$. Clearly, this corresponds to a time translation and
a spatial rotation, either in the spinning cone or in Minkowski space.

Let us choose the Minkowski coordinates in the non-compact charts so
that the transition function to the conical coordinates is given by
such a local isometry. Hence, for each non-compact polygon
$\pol\in\pols_\infty$ there exists a pair of linear functions
$\atau_\pol$ and $\adir_\pol$, so that in the overlap region of the
respective Minkowski chart with the neighbourhood of infinity the
transition function \eref{iso} applies. What can we then say about the
unit vectors $\dis_{\pm\ecut}$ and the transition functions
$\chol_{\pm\ecut}$ associated with the external links?

Consider first the transition functions. Each pair of external links
$\pm\ecut$ represents a triple overlap region between two Minkowski
charts, containing the polygons $\pol_{\pm\ecut}$, and the conical
chart, which provides a globally defined chart in a neighbourhood of
infinity. There is thus a certain relation which involves the
transition functions $\chol_{\pm\ecut}$ between the two Minkowski
charts, and the transition functions \eref{iso} to the conical
coordinates. The Minkowski coordinates in the two adjacent charts
differ by a time translation, which is given by the difference
$\atau_{\pol_\cut}(\cdir)-\atau_{\pol_{-\cut}}(\cdir)$, and a spatial
rotation, which is given by the difference
$\atau_{\pol_\cut}(\cdir)-\atau_{\pol_{-\cut}}(\cdir)$.

Note that these differences are constant, because in both polygons the
functions $\atau_\pol(\cdir)$ and $\adir_\pol(\cdir)$ are linear
according to \eref{atau-adir}. The translation can again be ignored.
The rotation tells us that the transition function between the
Minkowski coordinates is
\begin{equation}
  \label{ecut-chol}
  \chol_\ecut = \expo{-4\pi GM_\ecut}, \qquad 
  \chol_{-\ecut} = \expo{4\pi GM_\ecut}, \qquad 
   \ecut\in\cuts_\infty,
\end{equation}
where the \emph{energy} $M_\ecut$ is given by
\begin{equation}
  \label{M-adir}
   8\pi G M_\ecut = 
       \adir_{\pol_{-\ecut}}(\cdir) - \adir_{\pol_\ecut}(\cdir).
\end{equation}
This can be evaluated at any point on the links $\pm\ecut$, again
because of \eref{atau-adir}. We can also take the limit
\eref{cone-T-cdir}, where the definition of $M_\ecut$ becomes  
\begin{equation}
  \label{bdir-cut}
  8 \pi G M_\ecut = \bdir^+_\ecut - \bdir^-_\ecut .
\end{equation}
The abbreviations $\bdir_\ecut^\pm$ are defined as
\begin{equation}
  \label{adir-ecut}
  \bdir^+_\ecut = - \adir_{\pol_\ecut}(\cdir_\ecut) , \qquad 
  \bdir^-_\ecut = - \adir_{\pol_{-\ecut}}(\cdir_\ecut) . 
\end{equation}
Of course, in the limit $G\to0$, the relation \eref{ecut-chol} reduces
to the free particle relation \eref{rpp-ecut-cmom}, stating that the
external relative position vector $\cmom_\cut$ is proportional to
$\gam_0$. To see that the variable $M_\ecut$ can still be interpreted
as an energy, consider the definition \eref{bdir-cut}. It is the usual
relation between the energy of a particle and the deficit angle, if it
is measured in a frame which is not the rest frame of the particle
\cite{matwel}. And in fact, the difference on the right hand side
defines a certain angle, which can be regarded as a deficit angle.

To see this, we have to consider the unit vectors $\dis_{\pm\ecut}$,
defining the directions of the external edges in \fref{cone}. First of
all, we observe that a local isometry \eref{iso} maps a spatial half
line in the spinning cone onto a spatial half line in Minkowski space.
So, the unit vectors $\dis_{\pm\ecut}$ are still orthogonal to the
$\gam_0$-axis. To find the directions of these vectors, we have to
evaluate the right hand side of \eref{iso} in the limit
\eref{cone-T-cdir}, on the external edges $\pm\ecut$. This gives 
\begin{equation}
  \label{ecut-dis}
  \dis_\ecut = \gam(\cdir_\ecut + \bdir^+_\ecut) , \qquad
  \dis_{-\ecut} = -\gam(\cdir_\ecut + \bdir^-_\ecut), \qquad
   \ecut\in\cuts_\infty.
\end{equation}
Obviously, the difference $\bdir^+_\ecut-\bdir^-_\ecut$ is the angle
between the edges $-\ecut$ and $\ecut$ in the embedding Minkowski
space, on the left hand side of \fref{cone}. We may call this the
deficit angle of the external link $\ecut$. It goes to zero in the
limit $G\to0$.

The variables $\bdir_\ecut^\pm$ are called the \emph{deviations}. They
tell us how much the directions of the vectors \eref{ecut-dis} in
Minkowski space, on the left hand side of \fref{cone}, deviate from
the directions of the external links in the spinning cone, on the
right hand side of the \fref{cone}. There is no counterpart of these
variables for the free particles, because the spinning cone as a
reference frame is then replaced by the embedding Minkowski space
itself, and therefore the deviations are all zero. However, we shall
now show that here the deviations are also not independent. They are
determined by the other external link variables.

We already have the relation \eref{bdir-cut}. Another relation follows
immediately from the definition \eref{adir-ecut}. If we replace
$\ecut$ by $\ecut'$ in the second equation, and use that
$\pol_\ecut=\pol_{-\ecut'}$ for two successive external links
$\ecut,\ecut'\in\cuts_\infty$, then it follows from \eref{atau-adir}
that
\begin{equation}
  \label{bdir-pol}
  \bdir^-_{\ecut'} - \bdir^+_\ecut = 
   - 4GM  \, (\cdir_{\ecut'} - \cdir_\ecut).
\end{equation}
This relation has a simple geometric interpretation in \fref{cone}.
For the polygon $\pol$ to fit into the spinning cone between the
spatial half lines with conical coordinates $\cdir_\ecut$ and
$\cdir_{\ecut'}$, the \emph{opening angle}, hence the angle between
the external edges $\ecut$ and $-\ecut'$ in Minkowski space, must be
equal to $1-4GM$ times the opening angle in conical coordinates, thus
the difference $\cdir_{\ecut}-\cdir_{\ecut}$. This is the factor that
shows up in \eref{spin-cone} in front of the angular coordinate, thus
it relates a conical angle to a metric angle.

Now, suppose we are given the energies $M_\ecut$ and the directions
$\cdir_\ecut$ of the external links in the spinning cone. Then we can
solve the system of linear equations \eref{bdir-cut} and
\eref{bdir-pol} for the deviations. We have $2\ncut_\infty$ equations
and $2\ncut_\infty$ unknown variables. However, the equations are not
all independent. If we add them all, then the deviations drop out. The
differences $\cdir_{\ecut'}-\cdir_\ecut$ add up to $2\pi$, and what we
get is 
\begin{equation}
  \label{cone-M}
  M = \sumi{\ecut\in\cuts_\infty} M_\ecut.
\end{equation}
Hence the total energy is again the sum of the energies of the
external links. Or, in a geometric language, the deficit angle of the
spinning cone on the right hand side of \fref{cone} is the sum of the
deficit angles of the external links on the left hand side. 

So, given the energies $M_\ecut$ and the conical directions
$\cdir_\ecut$, the deviations are determined only up to an overall
constant $\bdir_\ecut^\pm\mapsto\bdir_\ecut^\pm+\bdir$, where
$\bdir\in\RR$ is an unknown angle. In \fref{cone}, this means that we
are still free to perform an overall rotation of all non-compact
polygons in the embedding Minkowski space on the left hand side. We
need one extra condition to fix this freedom, so that the deviations
are given uniquely as functions of the other link variables. And of
course, we want that in the limit $G\to0$ we get
$\bdir_\ecut^\pm\to0$.

The extra condition that we impose is the following \emph{average
condition}. We want that, in a certain sense, the overall orientation
of the Minkowski frame on the left hand side in \fref{cone} is the
same as the overall orientation of the spinning cone on the right hand
side. This condition can be formulated as follows. Consider a certain
angular direction $\cdir$ in the spinning cone, not necessary the
direction of one of the external links. A radial line pointing into
this direction is mapped onto a radial line in Minkowski space, which
points into the direction $\cdir-\adir_\pol(\cdir)$, where $\pol$ is
the polygon that contains this radial line. This can be read off from
\eref{iso}. 

Hence, $\adir_\pol(\cdir)$ tells us how much a particular angular
direction in Minkowski space deviates from the corresponding angular
direction in the spinning cone. The average condition now requires
that on average this deviation is zero. More precisely, the integral
$\adir_\pol(\cdir)$ over all angular directions $\cdir$ vanishes.
Hence,
\begin{equation}
  \label{deviation-1}
  \sum_{\pol\in\pols_\infty} 
       \intl{\cdir_\ecut}{\cdir_{\ecut'}}
           \, \dd\cdir \, \adir_\pol(\cdir) = 0 .
\end{equation}
Here we took into account that each non-compact polygon
$\pol\in\pols_\infty$ covers a certain range of the conical
coordinates, namely the one between the external edges
$\ecut,\ecut'\in\cuts_\infty$, where $\pol=\pol_\ecut=\pol_{-\ecut'}$.

This average condition imposes one extra restriction on the Minkowski
coordinates in the non-compact charts. To see that it fixes the values
of the deviations, let us simplify it a little bit. Since $\adir_\pol$
is a linear function, we can evaluate the integral in
\eref{deviation-1}. The value of the integral of a linear function is
half the sum of the values of function at the end points, multiplied
by the length of the integration interval. With the irrelevant factor
of one half dropped, and the definitions \eref{adir-ecut} inserted, we
get
\begin{equation}
  \label{bdir-ave-cdir}
  \sumi{\pol\in\pols_\infty}  
      ( \bdir_{\ecut'}^- + \bdir_{\ecut}^+ )
     \, ( \cdir_{\ecut'} - \cdir_\ecut ) = 0. 
\end{equation}
We can then also make use of the relations \eref{bdir-pol} and
\eref{bdir-cut}, and rearrange the summation slightly, which gives the
equivalent condition
\begin{equation}
  \label{bdir-ave-M}
  \sumi{\ecut\in\cuts_\infty}
     M_\ecut \, (\bdir^+_\ecut + \bdir^-_\ecut) = 0 . 
\end{equation}
We see that this provides an additional linear equation, which fixed
the absolute values of the deviations. In the limit $G\to0$, it is
easy to verify that the unique solution to the system of linear
equations \eref{bdir-cut}, \eref{bdir-pol}, and \eref{bdir-ave-M} is
$\bdir_\ecut^\pm=0$.

So, after this somewhat technical derivation, what is the conclusion?
The deviations $\bdir_\ecut^\pm$, and thus the unit vectors
$\dis_{\pm\ecut}$ are uniquely specified by the external link
variables $\cdir_\ecut$ and $M_\ecut$. The geometry of the ADM surface
and its embedding into the reference frame is finally specified by
almost same independent link variables as previously for the free
particles. We have the relative position vectors $\dis_\cut$, and the
transition functions $\chol_\cut$ for $\cut\in\cuts_+$, and the
energies $M_\ecut$, the clocks $T_\ecut$, and the directions
$\cdir_\ecut$ for $\cut\in\cuts_\infty$. We shall later also introduce
an auxiliary phase space variable $L_\ecut$, representing the angular
momentum conjugate to $\cdir_\ecut$, but for the time being we do not
need this.

There is then, finally, another consistency condition which involves
the clocks. It ensures that the ADM surface can be embedded into the
spinning cone, so that the external links are in fact spatial half
lines with the given conical time coordinates. Consider again a
non-compact polygon $\pol\in\pols_\infty$, and the external links
$\ecut,\ecut'\in\cuts_\infty$ with $\pol=\pol_\ecut=\pol_{-\ecut'}$.
Both edges are spatial half lines in Minkowski space. The constant
$\gam_0$-coordinates of these edges are given by
\begin{equation}
  T_\ecut + \atau_\pol(\cdir_\ecut), \qquad
  T_{\ecut'} + \atau_\pol(\cdir_{\ecut'}),
\end{equation}
respectively. This follows again from \eref{iso}, evaluated in the
limit \eref{cone-T-cdir}. The difference between these coordinates is
called the \emph{time offset} of the polygon $\pol$. It is given by
\begin{equation}
  \label{pol-offset}
  T_{\ecut'} - T_\ecut + 
     \atau_\pol(\cdir_{\ecut'}) - \atau_\pol(\cdir_\ecut) 
  = T_{\ecut'} - T_\ecut + 4 G S ( \cdir_{\ecut'} - \cdir_\ecut ). 
\end{equation}
Now, the same difference is given by the sum of the
$\gam_0$-components of the relative position vectors $\dis_\cut$
defining the internal edges $\cut\in\cuts_\pol\cap\cuts_0$ of the
polygon $\pol$. Hence, we have the relation 
\begin{equation}
  \label{econd}
  T_{\ecut'} - T_{\ecut} + 4 G S ( \cdir_{\ecut'} - \cdir_{\ecut} ) 
  +  \sumi{\cut\in\cuts_\pol\cap\cuts_0} \disv_\cut^0 = 0 .  
\end{equation}
This is obviously a generalization of \eref{rpp-econd}, to which it
reduces in the limit $G\to0$. We can use it to compute the parameter
$S$ of the spinning cone, thus the total angular momentum of the
universe. We have to sum over all non-compact polygons in
\eref{econd}. Then the differences between the clocks
$T_{\ecut'}-T_\ecut$ drop out, and the conical angles
$\cdir_{\ecut'}-\cdir_{\ecut}$ add up to $2\pi$. The result is
\begin{equation}
  \label{S-epol}
  8 \pi G S = -  \sum_{\pol\in\pols_\infty} 
      \sum_{\cut\in\cuts_\pol\cap\cuts_0} \!\!\! \disv_\cut^0  .
\end{equation}
On the right hand side we have to sum over all internal edges of all
non-compact polygons. To simplify this, we may equally well sum over
all compact polygons as well. For each compact polygon, the sum over
its edges is zero, according to \eref{icond}. But then the sum just
goes over all internal links, thus
\begin{equation}
  \label{S-cone}
  S = - \frac1{8\pi G} \, \sumi{\cut\in\cuts_0} \disv^0_\cut 
    = \sumi{\cut\in\cuts_+} L_\cut, \txt{where}
   L_\cut = -\frac{1}{8\pi G} \, ( \disv^0_\cut + \disv^0_{-\cut}).
\end{equation}
This looks very similar to \eref{rpp-M-S}. The total angular momentum
is distributed over the internal links. To recover the free particle
expression for $L_\cut=L_{-\cut}$, we use the relation
\eref{cut-inv-rel}, which tells us that
\begin{equation}
  L_\cut = \frac1{16\pi G} \, 
   \Trr{ (  \chol_\cut \, \dis_\cut \, \chol\inv_\cut - \dis_\cut )
         \, \gam^0 }.
\end{equation}
We can then write the transition function $\chol_\cut$ as an
exponential of the relative momentum vector $\cmom_\cut$, and expand
this up to the first order in $G$ as in \eref{cmom-chol-exp}. In the
leading order in $G$, we then recover \eref{rpp-M-S}. 

So, we finally see that not only the geometry of the ADM surface and
its embedding into the reference frame is determined by the link
variables, but also the parameters $M$ and $S$ of the spinning cone.
And we have the same kind of consistency conditions, a vector equation
\eref{icond} for every compact polygon, and a scalar equation
\eref{econd} for every non-compact polygon.

\subsubsection*{Redundancies}
In the beginning we already saw that the redundancy transformations of
the free particles system are replaced by the coordinate
transformations, hence the Lorentz rotations in the Minkowski charts
associated with the polygons.  We found that the link variables
transform as
\begin{equation}
  \label{chol-dis-lor-x}
  \chol_\cut \mapsto 
       \blpar_{\pol_{-\cut}}\, \chol_\cut \, \blpar\inv_{\pol_\cut},
  \qquad
  \dis_\cut \mapsto 
       \blpar_{\pol_\cut}\, \dis_\cut \, \blpar\inv_{\pol_\cut},
\end{equation}
where $\blpar_\pol\in\grpSL(2)$ represents the Lorentz rotation acting
on the polygon $\pol$. There is now a restriction on these parameters,
because the Minkowski coordinates in the non-compact charts are
related in a certain way to the conical coordinates. We are only
allowed to perform spatial rotations, thus for non-compact polygons
$\pol\in\pols_\infty$ we must have
\begin{equation}
  \label{ecut-blpar}
  \blpar_\pol = \expo{-4\pi G \emul_\pol\gam_0} , \qquad 
    \pol\in\pols_\infty, 
\end{equation}
where $\emul_\pol\in\RR$ specifies the angle rotation for the polygon
$\pol$. This is also the restriction that we had for the free
particles. The definition \eref{ecut-blpar} is chosen so that the
energies transform in the same way as before, thus
\begin{equation}
  \label{M-lor}
  M_\ecut \mapsto M_\ecut 
           + \emul_{\pol_{-\ecut}} - \emul_{\pol_{\ecut}}.
\end{equation}
Clearly, the link variables $T_\ecut$ and $\cdir_\ecut$ are invariant
under such a transformation, because we only change the local
Minkowski coordinates, but not the embedding of the ADM surface into
the spinning cone. There is then, however, a transformation of the
deviations. For the transformation \eref{chol-dis-lor-x} to apply also
to the unit vector $\dis_{\pm\ecut}$ in \eref{ecut-dis}, we must have 
\begin{equation}
  \label{bdir-lor}
  \bdir_\ecut^+ \mapsto \bdir_\ecut^+ - 
                    8 \pi G \, \emul_{\pol_\ecut}, \qquad
  \bdir_\ecut^- \mapsto \bdir_\ecut^- - 
                    8 \pi G \, \emul_{\pol_{-\ecut}}. 
\end{equation}
Now, one can easily check that this is consistent with \eref{bdir-cut}
and \eref{bdir-pol}, hence with the definition of the deviations as
functions of the other link variables. But it is not consistent with
\eref{bdir-ave-M}. So, there is something wrong with this
transformation.

Before we look at this more closely, let us consider a second class of
redundancies. As for the free particles, we are free to rotate the
external links within a certain range, which includes a smooth
deformation of the ADM surface. However, due to the screw like
geometry of the spinning cone, a rotation of the external link $\ecut$
not only affects the conical angular coordinate $\cdir_\ecut$, but
also the conical time coordinate $T_\ecut$. It is not difficult to see
that a spatial half line in the spinning cone is always a line of
constant $T+4GS\cdir$. Thus if we fix the start point of a spatial
half line and rotate it about this point, then its coordinates, in the
limit $R\to\infty$, transform as
\begin{equation}
  \label{tws-T-cdir}
  \cdir_\ecut \mapsto \cdir_\ecut + \jmul_\ecut, \qquad
  T_\ecut \mapsto T_\ecut - 4 G S \, \jmul_\ecut, \qquad
   \ecut\in\cuts_\infty, 
\end{equation}
where $\jmul_\ecut\in\RR$ defines the angle of rotation in conical
coordinates. To see how such a transformation acts on the embedded
polygons in Minkowski space, we use the definition \eref{adir-ecut} of
the deviations, and the property \eref{atau-adir} of the functions
$\adir_\pol$. Note that these functions are now fixed, because we only
rotate the external link, but we do not change the transition function
between the conical coordinates and the Minkowski coordinates. It
follows that
\begin{equation}
  \label{bdir-tws}
  \bdir_\ecut^+ \mapsto \bdir_\ecut^+ -4GM \, \jmul_\ecut,
  \qquad
  \bdir_\ecut^- \mapsto \bdir_\ecut^- -4GM \, \jmul_\ecut.  
\end{equation}
Together with \eref{tws-T-cdir}, this implies that the unit vectors
$\dis_{\pm\ecut}$ are rotated by $(1-4GM)\jmul_\ecut$ about the
$\gam_0$-axis in the embedding Minkowski space. Again, we recover the
factor $1-4GM$, which relates a conical angle to the corresponding
angle in Minkowski space. And we also see that there is a certain
restriction on the parameters $\jmul_\ecut$, because the non-compact
polygons are twisted when the external edges are rotated, and for too
large parameters they are no longer spacelike. 

One can again verify that the transformations \eref{bdir-tws} are
compatible with \eref{bdir-cut} and \eref{bdir-pol}, but not with
\eref{bdir-ave-M}. To understand this, let consider the infinitesimal
generator of the most general redundancy transformation. To define the
generator of a Lorentz rotation acting on a compact polygon
$\pol\in\pols_0$, we introduce a vector $\blgen_\pol\in\algsl(2)$. It
is related to the actual parameter of the Lorentz rotation by
$\blpar_\pol=\expo{4\pi G \blgen_\pol}$. For a non-compact polygon
$\pol\in\pols_\infty$, we set $\blgen_\pol=-\emul_\pol\gam_0$ for some
$\emul_\pol\in\RR$, which is then also compatible with
\eref{ecut-blpar}. And finally, the generator of a rotation of the
external links is defined by $\jmul_\ecut$, which is also the
parameter in \eref{tws-T-cdir}.

For the transition functions and the relative position vectors we find
the following variations,
\begin{equation}
  \label{dis-chol-gen}
  \delta \chol_\cut = 4\pi G \,
      (\blgen_{\pol_{-\cut}}\, \chol_\cut 
        - \chol_\cut \, \blgen_{\pol_\cut}) , 
  \qquad
  \delta \dis_\cut = 4\pi G \,
      \comm{\blgen_{\pol_\cut}}{\dis_\cut}, \qquad \cut\in\cuts_0. 
\end{equation}
This is the infinitesimal generator of \eref{chol-dis-lor-x}. For the
external link variables we find
\begin{equation}
  \label{M-T-gen}
  \delta M_\ecut =
        \emul_{\pol_{-\ecut}} - \emul_{\pol_{\ecut}}, \qquad
  \delta T_\ecut = - 4 G S \, \jmul_\ecut, \qquad
  \delta \cdir_\ecut = \jmul_\ecut, \qquad \ecut\in\cuts_\infty,
\end{equation}
which is the generator of \eref{M-lor} and \eref{tws-T-cdir},
respectively. And finally, for the deviations this implies 
\begin{equation}
  \label{bdir-gen}
  \delta\bdir_\ecut^+ = -4GM \, \jmul_\ecut   
                   -  8 \pi G \, \emul_{\pol_\ecut}, \qquad
  \delta\bdir_\ecut^- = -4GM \, \jmul_\ecut 
                   -  8 \pi G \, \emul_{\pol_{-\ecut}}. 
\end{equation}
In the limit $G\to0$, these are precisely the transformations
\eref{rpp-kin-pois}. However, there is one crucial difference. For the
free particles, we found that a certain combination of redundancy
transformation was trivial. Setting $\jmul_\ecut=0$ and
$\blgen_\pol=-\emul\gam_0$ for all polygons, with $\emul\in\RR$, we
found that the transformation was void. But here we still get a
non-trivial transformation, namely
\begin{equation}
  \label{sim-rot}
  \delta\chol_\cut = 4\pi G \, \emul \, \comm{\dis_\cut}{\gam_0}
     , \qquad
  \delta \dis_\cut = 4\pi G \, \emul \, \comm{\dis_\cut}{\gam_0}
     , \qquad
  \delta\bdir_\ecut^\pm = - 8\pi G\, \emul.  
\end{equation}
On the left hand side in \fref{cone}, this is a \emph{simultaneous}
rotation of all polygons in the embedding Minkowski space, by $8\pi
G\emul$ in clockwise direction about the $\gam_0$-axis. It is this
transformation that is not compatible with the average condition
\eref{bdir-ave-M}.

It is now obvious what this means. Since we wanted the overall
orientation of the Minkowski frame to be the same as that of the
conical frame, we are not allowed to perform such an overall rotation.
There is a restriction on the parameters $\emul_\pol$ and
$\jmul_\ecut$. The condition is that the average condition
\eref{bdir-ave-M} must be preserved. Inserting the variation of the
energies $M_\ecut$ and the deviations $\bdir_\ecut^\pm$ from above,
one can easily derive the following restriction to be imposed on the
parameters $\emul_\pol$ and $\jmul_\ecut$,
\begin{equation}
  \label{tws-lor-restrict}
  \sumi{\pol\in\pols_\infty} 
   \emul_\pol \, (\cdir_{\ecut'}-\cdir_\ecut)
     + \sumi{\ecut\in\cuts_\infty} \jmul_\ecut \, M_\ecut = 0 .
\end{equation}
As for the free particles, we have one redundancy less than there are
parameters. It is only a technical difference that this is now due to
a restriction on the parameters, whereas for the free particles a
certain combination of the parameters drops out. We shall look at this
again at the phase space level in \sref{phase}, and there we shall see
that the analogy to the free particle system is indeed very close.

\subsubsection*{Inserting and removing links}
There is another class of redundancy transformations, which we more or
less ignored so far. Of course, we can also change the triangulation
of the ADM surface, and this should not affect the state of the
particles. If we consider the time evolution, for example, then we
have to do this from time to time, since not every configuration of
the particles admits same triangulation. For example, if we have a
graph $\cuts$ where the particles $\prt_\cut$ and $\prt_{-\cut}$ are
connected by an internal link $\cut$, then it is in general not
possible to use this triangulation for a state where these particles
are far apart, with many particles in between.

What typically happens is that a spacelike geodesic connecting these
particles intersects with the light cones emerging from the other
particles in between. This is actually not a feature introduced by the
gravitational interaction. The problem also arises for the free
particle system, where we just ignored it. So, let us show how to
transform from one triangulation to another. We shall later consider
this as a \emph{large} gauge transformation, in contrast to the
redundancy transformations above, which are \emph{smoothly generated}
gauge symmetries. A convenient way to describe a general transition
between two different triangulations is to decompose it into
elementary steps.

\begin{figure}[t]
  \begin{center}
    \epsfbox{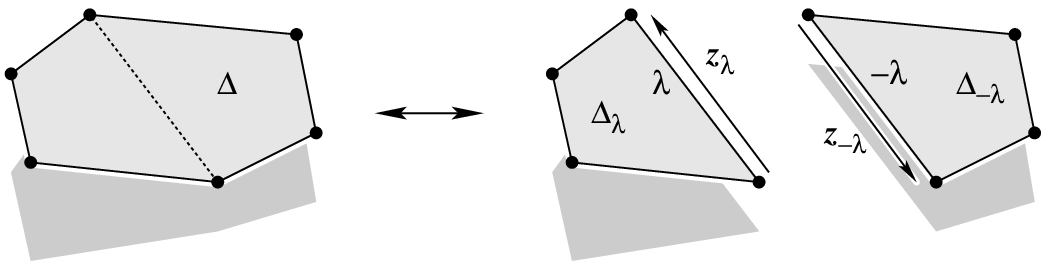}
    \xcaption{A compact polygon can be split into two compact polygons
    by inserting a pair of links $\pm\cut$. The new link variables
    $\chol_{\pm\cut}$ and $\dis_{\pm\cut}$ are determined by
    \eref{new-chol} and \eref{new-dis}. For the reverse
    transformation, one first has to perform a redundancy
    transformation, so that $\chol_{\pm\cut}=\one$. The two Minkowski
    charts can then be unified, the polygons can be glued together,
    and the links $\pm\cut$ can be removed.}
    \label{link}
  \end{center}
\end{figure}

An elementary step is the insertion or the removal of a single link.
Inserting a new internal link is very simple. Suppose we have a graph
$\cuts$ with an associated collection $\pols$ of polygons, and let
$\pol\in\pols$ by a polygon with more than three edges. It can be
either a compact or a non-compact polygon. For a compact one, the
typical situation in sketched in \fref{link}. Consider one of the
diagonals of this polygon, and suppose that the polygon surface can be
deformed, so that the diagonal is contained in the surface. Then it is
possible to cut the polygon into two new polygons.

The insertion is defined as follows. The new graph
$\newcuts=\cuts\cup\{\cut,-\cut\}$ is obtained by adding a pair of
internal links to the original graph $\cuts$. The new collection
$\newpols$ of polygons is obtained by removing $\pol$ and adding
$\pol_\cut$ and $\pol_{-\cut}$. The edges $\newcuts_{\pol_{\cut}}$ and
$\newcuts_{\pol_{\cut}}$ of the new polygons are defined in the
obvious way. The links $\cut$ is also added to the set
$\newcuts_{\prt_{\cut}}$, and $-\cut$ is inserted into
$\newcuts_{\prt_{-\cut}}$ at the appropriate place in the cyclic
ordering. All other subsets of links $\newcuts_\pol=\cuts_\pol$ and
$\newcuts_\prt=\cuts_\prt$ are unchanged.

Regarding the new link variables $\chol_{\pm\cut}$ and
$\dis_{\pm\cut}$, we have to define them in such a way that the
geometry of the ADM surface is the same as before. Since both polygons
are originally embedded into the same Minkowski chart, it is clear
that the newly arising transition function is trivial, hence
\begin{equation}
  \label{new-chol}
  \chol_\cut = \chol_{-\cut} = \one.
\end{equation}
The relative position vector $\dis_\cut$ is determined as follows. At
least one of the newly arising polygons is compact, even if the
original polygon $\pol$ was non-compact. Without loss of generality,
let us assume that $\pol_\cut$ is compact. Hence,
$\newcuts_{\pol_{\cut}}$ contains the edge $\cut$, and a sequence of
at least two other internal edges. The new vectors $\dis_{\pm\cut}$
are then given by
\begin{equation}
  \label{new-dis}
  \dis_{\cut} = - \dis_{-\cut} = 
    - \sumi{\bcut\in\newcuts_{\pol_{\cut}}-\{\cut\}} \dis_\bcut.
\end{equation}
This immediately implies that the consistency condition \eref{icond}
for the new polygon $\pol_\cut$ is satisfied. Moreover, if the
original polygon $\pol$ was compact, the the same holds for the new
compact polygon $\pol_{-\cut}$. This follows from \eref{new-dis} and
the consistency condition \eref{icond} for the original polygon.
Similarly, if the original polygon $\pol$ was non-compact, the the new
polygon $\pol_{-\cut}$ is also non-compact, and the consistency
condition \eref{econd} for the new polygon follows from the same
consistency condition for the original polygon.

So, the conclusion is that, whenever it is possible to deform the ADM
surface so that it contains the diagonal of a polygon, then it is
possible to add this as a new pair of internal links to the graph. If
we define the new link variables according to \eref{new-chol} and
\eref{new-dis}, then the new triangulation defines the same local
geometry of the ADM surface, up to smooth deformations, and thus the
same physical state. For the free particle system, the insertion of a
new internal link works in the very same way, we just have to replace
\eref{new-chol} by $\cmom_\cut=\cmom_{-\cut}=0$.

Inserting a new external link into a non-compact polygon is equally
straightforward. Suppose that we have a graph $\cuts$ and
$\cuts_\infty=\{\dots,\ecut,\ecut',\dots\}$ is the set of external
links. Suppose further that the non-compact polygon
$\pol_{\ecut}=\pol_{-\ecut'}$ has more than one internal edge.
Provided that the polygon surface can be deformed appropriately, we
can then introduce a new external link $\fcut$ between $\ecut$ and
$\ecut'$, so that
$\newcuts_\infty=\{\dots,\ecut,\fcut,\ecut',\dots\}$. The original
polygon $\pol_\ecut=\pol_{-\ecut'}$ splits into two new non-compact
polygons $\pol_\ecut=\pol_{-\fcut}$ and $\pol_\fcut=\pol_{-\ecut'}$.

The new link variables associated with $\fcut$ are determined as
follows. In contrast to the insertion of a new internal link, there is
now some freedom regarding the direction of the new link. The new
conical direction $\cdir_\fcut$ can be any direction in the range
\begin{equation}
  \label{new-cdir}
  \cdir_\ecut \le \cdir_\fcut \le \cdir_{\ecut'}. 
\end{equation}
It has to lie between the original links $\ecut$ and $\ecut'$, but
we are free to rotate it, because there is no fixed end point. But
once the conical direction $\cdir_\fcut$ is chosen, all other link
variables are fixed. For the same reason as before, we have to set
\begin{equation}
  \label{new-M}
  M_\fcut = 0 \follows
   \chol_\fcut = \chol_{-\fcut} = \one,
\end{equation}
The transition function between the two new Minkowski charts is
trivial. The clock $T_\fcut$ is determined by the consistency
condition \eref{econd}, which has to be satisfied for both new
polygons $\pol_\fcut$ and $\pol_{-\fcut}$, thus
\begin{equation}
  \label{new-T}
    T_{\fcut}  
    = T_{\ecut} - 4 G S ( \cdir_{\fcut} - \cdir_{\ecut} ) 
    - \sumi{\cut\in\newcuts_{\pol_{-\fcut}}\cap\newcuts_0} 
                                                 \disv_\cut^0 
    = T_{\ecut'} + 4 G S ( \cdir_{\ecut'} - \cdir_{\fcut} ) 
    + \sumi{\cut\in\newcuts_{\pol_\fcut}\cap\newcuts_0} 
                                                 \disv_\cut^0 .  
\end{equation}
The last equality is the consistency condition \eref{econd} for the
original polygon $\pol_\ecut=\pol_{-\ecut'}$, so we may use either
expression to define $T_\fcut$. 

Finally, consider the deviations $\bdir^\pm_\fcut$. Since they are
determined by system of linear equations, involving all external
links, it is not immediately obvious whether inserting a new external
link affects the deviations of the other external links or not. So,
consider the system of linear equations \eref{bdir-cut},
\eref{bdir-pol}, and \eref{bdir-ave-M}, after the insertion, and let
us write down only those equation that involve the new links
$\pm\fcut$. First of all, we see that the relevant term in the average
condition \eref{bdir-ave-M} drops out, because $M_\fcut=0$. So, at
least this equation is unaffected.

There are then two equations of the form \eref{bdir-pol}, namely those
that define the opening angles of the polygons $\pol_{\pm\fcut}$,
\begin{equation}
  \bdir_{\ecut'}^- - \bdir_{\fcut}^+ = 
   -4GM \, (\cdir_{\ecut'} - \cdir_\fcut ) , \qquad
  \bdir_{\fcut}^- - \bdir_{\ecut}^+ = 
   -4GM \, (\cdir_{\fcut} - \cdir_\ecut ) , 
\end{equation}
and one equation of the form \eref{bdir-cut}, namely that for the
deficit angle of the new link,  
\begin{equation}
  \bdir_\fcut^+ - \bdir_\fcut^- = 8\pi G M_\fcut = 0 .
\end{equation}
First of all, we see that once we chose the conical direction
$\cdir_\fcut$ of the new link, these equations can be used to
determine the deviations $\bdir_\ecut^\pm$. Moreover, if we then
eliminate the variables $\bdir_\ecut^\pm$ from three given equations
above, we get
\begin{equation}
  \bdir_{\ecut'}^- - \bdir_{\ecut}^+ = 
   -4GM \, (\cdir_{\ecut'} - \cdir_\ecut ) .
\end{equation}
But this is exactly the equation \eref{bdir-pol} for the original
polygon $\pol_\ecut=\pol_{-\ecut}$, before the insertion. Hence, it
follows that the other external links are not affected by the
insertion.

So, we can insert internal and external links, whenever it is possible
to deform the polygons appropriately so that the new link becomes a
spacetime geodesic contained in the ADM surface. And what about
removing links? Suppose that $\pm\cut$ is a pair of internal or
external links, so that $\pol_\cut\neq\pol_{-\cut}$. Hence the links
do not represent a self overlap region of a single coordinate chart.
Then it is possible to perform a coordinate transformation in one
chart, so that the transition function
$\chol_\cut=\chol_{-\cut}=\one$ becomes trivial. Note that this is not
possible if $\pol_\cut=\pol_{-\cut}$, because then the transformation
\eref{chol-dis-lor-x} is a conjugation, which cannot be used to
transform a group element into the unit element. 

So, in order to remove a pair of links, we first have to perform a
smoothly generated redundancy transformation, or \emph{fix a gauge},
so that the associated transition function becomes trivial. Then the
link can be removed, and the two Minkowski charts can be unified into
a single one. We can remove as many links as we like, until there is
only a single polygon left, and the graph is a tree. Vice versa, the
maximal number of links is reached when all polygons are triangles.
And finally, we can transform between any two triangulations using the
given elementary steps, and the previously considered smoothly
generated redundancy transformations, thus the Lorentz rotations of the
polygons and rotations of the external links. 

\section{Phase space geometry}
\label{phase}
After this somewhat technical derivation, let us now come to the
actual subject of this article, the phase space structure of the multi
particle model. In the previous section we have seen how the geometry
of the ADM surface, and its embedding into the reference frame, is
specified by a triangulation and a set of link variables. It is
possible to set up the same hierarchy of phase spaces, which we also
encountered in \sref{free} for the free particles,
\begin{equation}
  \label{phase-array}
  \begin{array}{ccccc}
     \esp  & \supset & \ksp & \supset & \psp \cr
            & & \downarrow & & \downarrow \cr
        \phantom{\ksp/{\sim}}    &  & \ksp/{\sim} &  & \psp/{\sim}. 
  \end{array}
\end{equation}
The \emph{extended} phase space $\esp$ is spanned by the link
variables. The \emph{kinematical} subspace $\ksp\subset\esp$ is
defined by a set of kinematical constraints. They are the consistency
conditions derived in the previous section, and the associated gauge
symmetries are the redundancy transformation. The quotient space
$\ksp/{\sim}$ is basically the set of all possible geometries of the
ADM surface, and all possible embeddings into the reference frame. 

In analogy to general relativity, we can think of the extended phase
space $\esp$ as spanned by the usual ADM variables, the spatial metric
and its conjugate momentum, the external curvature. The kinematical
constraints are then the diffeomorphism constraints, the associated
gauge symmetries are the coordinates transformation on the space
manifold, and consequently the quotient space is, roughly speaking,
the set of all spatial geometries. Following this analogy, we can then
regard the mass shell constraints as a the Hamiltonian constraints of
general relativity. They define the \emph{physical} subspace
$\psp\subset\ksp$, and the quotient space $\psp/{\sim}$ is the set of
all spacetimes, or the set of all physically inequivalent solutions to
the equations of motion.

As for the free particles, the phase space consists of a finite number
of disconnected components, one for each possible graph $\cuts$ with
$\nprt$ particles,
\begin{equation}
  \label{esp-union}
  \esp = \bigcup_{\cuts} \esp_\cuts, \qquad 
   \ksp = \bigcup_{\cuts} \ksp_\cuts, \qquad 
   \psp = \bigcup_{\cuts} \psp_\cuts.
\end{equation}
To make the quotient spaces $\ksp/{\sim}$ and $\psp/{\sim}$ connected
manifold, we have to include large gauge transformations into the
definition of the equivalence relation, and these are of course the
transitions between the triangulations, as defined in the very end of
the previous section. In the context of general relativity they are
analogous to the large diffeomorphism.

The whole phase space structure can be regarded as a \emph{deformed}
version of the free particle phase space, again with Newton's constant
as a deformation parameter. At any stage of the derivation we can get
back to the free particles by taking the limit $G\to0$. Using this we
can see how the phase space structures are actually affected when the
gravitational interaction is switched on. The most interesting
deformation is that of the symplectic structure, which has been
mentioned in the introduction. The Poisson brackets of the components
of the relative position vectors are no longer zero, which means that
when the system is quantized these become non-commuting objects.

\subsubsection*{The extended phase space}
The definition of the extended phase space $\esp$ is almost the same
as before. There are finitely many disconnected components
$\esp_\cuts$. Each component is specified by a graph $\cuts$ with
$\nprt$ particles, and it is spanned by the following link variables,
\begin{itemize}
\item[-] a \emph{transition function} $\cmom_\cut\in\grpSL(2)$
  and a \emph{relative position vector} $\dis_\cut\in\algsl(2)$ for
  every internal link $\cut\in\cuts_0$, and
\item[-] an \emph{energy} $M_\ecut$, a \emph{clock} $T_\ecut$, an
  \emph{angular momentum} $L_\ecut$, and a \emph{direction}
  $\cdir_\ecut$ for every external link.  
\end{itemize}
The only modification, as compared to the free particles, is that the
relative momentum vector $\cmom_\cut$ is replaced by the transition
function $\chol_\cut$. There is still a relation between the variables
associated with an internal link $\cut$ and the reversed link $-\cut$.
According to \eref{cut-inv-rel} we have 
\begin{equation}
  \label{cut-inv}
  \chol_{-\cut} = \chol\inv_\cut , \qquad
  \dis_{-\cut} = - \chol_\cut \, \dis_\cut \, \chol\inv_\cut, \qquad 
   \cut\in\cuts_0.
\end{equation}
Thus actually only half of the internal link variables are
independent. There is also a transition function $\chol_\ecut$ for
each external link, which is given by
\begin{equation}
  \label{ecut-rot}
  \chol_\ecut = \expo{-4\pi G M_\ecut\gam_0}, \qquad
  \chol_{-\ecut} = \expo{4\pi G M_\ecut\gam_0}, \qquad
         \ecut\in\cuts_\infty.    
\end{equation}
And finally we have also defined a vector $\dis_\ecut$ for each
external link, which is a spatial unit vector given by 
\begin{equation}
  \label{ecut-vec}
  \dis_\ecut = \gam(\cdir_\ecut + \bdir^+_\ecut) , \qquad
  \dis_{-\ecut} = -\gam(\cdir_\ecut + \bdir^-_\ecut), \qquad
   \ecut\in\cuts_\infty.
\end{equation}
The quantities $\bdir_\ecut^\pm$, which we called the
\emph{deviations}, are implicitly given by the following system of
linear equations
\begin{equation}
  \label{bdir-eqs}
  \bdir^-_{\ecut'} - \bdir^+_\ecut = 
   - 4GM  \, (\cdir_{\ecut'} - \cdir_\ecut),  \qquad 
  \bdir^+_\ecut - \bdir^-_\ecut =  8\pi G M_\ecut, 
\end{equation}
and the \emph{average condition}
\begin{equation}
  \label{bdir-ave}
  \sumi{\ecut\in\cuts_\infty}
     M_\ecut \, (\bdir^+_\ecut + \bdir^-_\ecut) = 0 .
\end{equation}
This implies that the relation \eref{cut-inv} also holds for external
links. Provided that some consistency conditions are satisfied by the
link variables, which we shall impose below as kinematical
constraints, it is then possible to construct the polygons in
\fref{ipol}, glue them together, and embed the resulting ADM surface
into the spinning cone, as indicated in \fref{cone}. This spinning
cone defines the centre of mass frame, and also the reference frame.
The geometry of the spinning cone depends on two parameters $M$ and
$S$. They represent the total energy and the total angular momentum of
the universe, and they are given by 
\begin{equation}
  \label{M-S}
  M = \sumi{\ecut\in\cuts_\infty} M_\ecut, \qquad
  S = \sumi{\cut\in\cuts_+} L_\cut, \txt{where}
   L_\cut = -\frac1{8\pi G}\,(\disv^0_\cut + \disv^0_{-\cut}).
\end{equation}
This is the definition of the extended phase space $\esp_\cuts$. It is
a $6\ncut_0+4\ncut_\infty$ dimensional manifold. It is a
\emph{deformed} version of the free particle phase space in the sense
that the vectors $\cmom_\cut\in\algsl(2)$ are replaced by the group
elements $\chol_\cut\in\grpSL(2)$. Using the relation
\eref{cmom-chol-exp} in the limit $G\to0$, we see how this deformation
disappears when the gravitational interaction is switched off.

\subsubsection*{Non-commutative coordinates}
Now need to know what the symplectic structure is, and to say
something about the dynamics of the particles we also need to know the
Hamiltonian. Both will be derived in \sref{reduc} from the Einstein
Hilbert action. So, let us here only give a definition without any
further motivation. The symplectic structure is in fact the most
natural one that can be defined on the given phase space. It also
agrees very nicely with the symplectic structure derived for the
single particle model in \cite{matwel}, of which this multi particle
model is a straightforward generalization.

The external link variables still provide two canonical pairs
$(M_\ecut,T_\ecut)$ and $(L_\ecut,\cdir_\ecut)$ for of each external
link $\ecut\in\cuts_\infty$. For each internal link $\cut\in\cuts_0$,
the pair $(\chol_\cut,\dis_\cut)$ can be regarded as an element of the
cotangent bundle of the group manifold
$\tng_*\grpSL(2)=\grpSL(2)\times\algsl(2)$. The most natural
symplectic structure is therefore the canonical symplectic structure
on this cotangent bundle, which is fixed up to a constant. This
constant has to be proportional to the inverse of Newton's constant,
in order to provide the correct physical dimension of the symplectic
potential, which is that of an action. The numerical factor can then
be derived from the condition that in the limit $G\to0$ we have to
recover the free particle expression \eref{rpp-pot-ext}.

All together, this implies that the most natural symplectic potential
on the extended phase space $\esp_\cuts$ is
\begin{equation}
  \label{pot-tot}
  \pot = \sumi{\ecut\in\cuts_\infty}  
          ( T_\ecut \, \dd M_\ecut 
                  +  L_\ecut \, \dd \cdir_\ecut  ) 
       - \frac1{8\pi G} \, \sumi{\cut\in\cuts_+} 
             \Trr{\chol\inv_\cut \, \dd \chol_\cut \, \dis_\cut}.
\end{equation}
To see that it reduces to \eref{rpp-pot-ext} in the limit $G\to0$, we
have write the transition function $\chol_\cut$ as an exponential
\eref{cmom-chol-exp} of the relative momentum vector $\cmom_\cut$, and
expand the result up to the first order in $G$. The given expression
is a well defined one-form on $\esp_\cuts$, because it is independent
of the chosen decomposition $\cuts_0=\cuts_+\cup\cuts_-$. Using the
relations \eref{cut-inv}, we easily see that
\begin{equation}
  \label{pot-inv}
  \Trr{\chol\inv_{-\cut} \, \dd \chol_{-\cut} \, \dis_{-\cut}}
  =  \Trr{\chol\inv_\cut \, \dd \chol_\cut \, \dis_\cut}.
\end{equation}
So, we find that the free particle symplectic potential has a natural
generalization. But we should emphasize that this is just an \emph{ad
hoc} definition. The only way to derive this expression is from the
Einstein Hilbert action, and this is what we are going to do in
\sref{reduc}.

Let us then derive the Poisson brackets. These are of course the most
interesting objects when we want to learn something about the
quantized model, without actually performing the quantization. Nothing
particular happens for the external link variables, 
\begin{equation}
  \label{pois-ecut}
  \pois{T_\ecut}{M_\ecut} = 1, \qquad 
  \pois{L_\ecut}{\cdir_\ecut} = 1,   \qquad \ecut\in\cuts_\infty.
\end{equation}
More interesting are the internal links. Here we can directly
generalize the results from the single particle system. For each
internal link, we have a pair $(\chol_\cut,\dis_\cut)$, with the
symplectic potential being the same as (4.3) in \cite{matwel}. We just
have to generalize the Poisson brackets accordingly, which are given
by (4.11) in \cite{matwel}, and insert Newton's constant. The vector
$\dis_\cut$ generates on the group element $\chol_\cut$ a
multiplication from the right, and $\dis_{-\cut}$ generates a
multiplication from the left,
\begin{equation}
  \label{pois-chol-dis}
  \pois{\chol_\cut}{\disv_\cut^a} = 
                4\pi G \, \chol_\acut  \gam^a , \qquad
  \pois{\chol_\cut}{\disv_{-\cut}^a} =  
              - 4 \pi G \, \gam^a  \chol_\cut , \qquad
      \cut\in\cuts_0.   
\end{equation} 
To avoid confusion, we shall here and in the following use the vector
component notation inside the Poisson bracket, and never write
brackets with more than one matrix entry. Since the brackets between
different components of the group elements $\chol_\cut$ are all zero,
it is sufficient to expand the vectors
$\dis_\cut=\disv_\cut^a\,\gam_a$. The vector components themselves
provide a representation of the Lorentz algebra,
\begin{equation}
  \label{pois-dis-dis}
  \pois{\disv_\cut^a}{\disv_\cut^b} 
          = 8\pi G \, \eps^{ab}\_c \, \disv_\cut^c, 
   \qquad \cut\in\cuts_0.  
\end{equation}
The brackets between the components of $\dis_\cut$ and $\dis_{-\cut}$
are zero, because the generators of left and right multiplication in
\eref{pois-chol-dis} commute. The brackets
(\ref{pois-ecut}--\ref{pois-dis-dis}) are therefore the only
non-vanishing brackets involving the basic phase space variables. As a
cross check, one can verify that the brackets are compatible with
\eref{cut-inv}. We can either consider $\chol_\cut$ and $\dis_\cut$ as
a pair of independent phase space variables, or $\chol_{-\cut}$ and
$\dis_{-\cut}$.

The crucial difference to the free particle system is obviously that
the components of the relative position vectors $\dis_\cut$ have
non-vanishing Poisson brackets with each other. At the classical
level, this is just a special feature of the symplectic structure. But
at the quantum level it implies that the components of the relative
position vectors of the particles do not commute. The particles are
effectively moving in a kind of \emph{non-commutative} spacetime. We
can also say that the geometry of the space manifold defined by the
polygons in \fref{ipol} becomes a \emph{non-commutative} geometry.

What this means to a quantized point particle, and the quantum
spacetime that this particle effectively sees, has been studied in
some detail for the single particle model in \cite{matwel}. Since we
are here not going to quantize the model, we can only say that some
qualitatively similar effects are expected for a multi particle
system. The technical details are however more involved, because for a
proper quantization we also have to solve the various constraints
defined below. For a two particle system, this can still be done
explicitly. And in fact, one finds some interesting features of the
quantized model. For example, it is impossible to localize the
particles at a point in space, and it is also impossible to bring them
closer together than a certain minimal distance, which is of the order
of the Planck length \cite{matlou}.

\subsubsection*{Kinematical constraints}
The kinematical subspace $\ksp_\cuts\subset\esp_\cuts$ is defined in
the same way as for the free particles. We have to impose the
consistency conditions derived in the previous section as kinematical
constraints. For each compact polygon $\pol\in\pols_0$, we found the
relation \eref{icond}, stating that the edges form a piecewise
straight, closed curve in Minkowski space, 
\begin{equation}
  \label{icon}
  \pcon_\pol = \sumi{\cut\in\cuts_\pol} \dis_\cut \approx 0 , 
   \qquad \pol\in\pols_0.
\end{equation}
For each non-compact polygon $\pol\in\pols_\infty$, there was a
consistency condition \eref{econd}. It relates the $\gam_0$-components
of the vectors representing the internal edges
$\cut\in\cuts_\pol\cap\cuts_0$ to the conical coordinates of the two
external edges $\ecut\in\cuts_\pol\cap\cuts_\infty$ and
$-\ecut'\in\cuts_\pol\cap\cuts_{-\infty}$,
\begin{equation}
  \label{econ}
  \econ_{\pol} =   T_{\ecut'} - T_{\ecut} 
  + 4 G S ( \cdir_{\ecut'} - \cdir_{\ecut} ) 
  + \sumi{\cut\in\cuts_\pol\cap\cuts_0} \disv_\cut^0 \approx 0 , \qquad 
         \pol\in\pols_\infty.
\end{equation}
Finally, there is also a constraint which defines the value of the
angular momentum $L_\ecut$, which is introduced as an auxiliary
variable to obtain a well defined phase space. The constraint is a
deformed version of \eref{rpp-jcon},
\begin{equation}
  \label{jcon}
  \jcon_\ecut = L_\ecut + 4 G S M_\ecut \approx 0, \qquad 
      \ecut\in\cuts_\infty. 
\end{equation}
At this point, there is no particular motivation for this relation
between the angular momentum $L_\ecut$ and the energies $M_\ecut$.
This is what comes out in \sref{reduc}, from the phase space reduction
applied to the Einstein Hilbert action. The only possible motivation
for this particular constraint is that the associated gauge symmetry
is a rotation \eref{tws-T-cdir} of the external link $\ecut$ in the
spinning cone, and that in the limit $G\to0$ it reduces to the free
particle constraint \eref{rpp-jcon}.

This applies to all the kinematical constraints. There is also a
relation between them, which is the same as \eref{rpp-kin-sum-zero}.
If we add the $\gam_0$-components of the vector constraints
$\pcon_\pol$ for all $\pol\in\pols_0$, and the scalar constraints
$\econ_\pol$ for all $\pol\in\pols_\infty$, then we get
\begin{equation}
  \label{kin-sum-zero}
     \sumi{\pol\in\pols_0} \pconv_\pol^0 
   + \sumi{\pol\in\pols_\infty} \econ_\pol 
   = 8 \pi G S  + \sumi{\cut\in\cuts_0} \disv^0_\cut = 0.
\end{equation} 
The first equality holds because the conical time differences
$T_{\ecut'}-T_\ecut$ add up to zero, and the conical angles
$\cdir_{\ecut'}-\cdir_{\ecut}$ add up to $2\pi$. What remains is a sum
of the $\gam_0$-components of the vectors $\dis_\cut$ for all internal
links. But this is just the definition of the angular momentum $S$ in
\eref{M-S}. So, there is one independent kinematical constraint less
then the number of equations given by (\ref{icon}-\ref{jcon}). We'll
do a precise counting later on.

All kinematical constraint are first class constraints. This is now
less trivial because they involve non-commuting phase space variables.
But a straightforward calculation shows that they form a closed
algebra. Every bracket between two constraints is again a linear
combination of constraints. The only interesting bracket is
\begin{equation}
  \label{pois-icon-icon}
  \pois{\pconv^a_\pol}{\pconv^b_\pol} 
          = 8\pi G \, \eps^{ab}\_c \, \pconv^c_\pol, 
   \qquad \pol\in\pols_0. 
\end{equation}
For each compact polygon $\pol$, the components of the vector
constraint $\pcon_\pol$ provide a representation of the Lorentz
algebra. This is quite reasonable, since we expect the associated
gauge symmetries to be the Lorentz rotations of the polygons in the
embedding Minkowski space, thus the coordinate transformations in the
Minkowski charts.

To derive the associated gauge symmetries, it is again useful to
define a general linear combination of the kinematical constraints. We
introduce a vector valued multiplier $\blgen_\pol\in\algsl(2)$ for
each compact polygon $\pol\in\pols_0$, a scalar multiplier
$\emul_\pol\in\RR$ for each non-compact polygon $\pol\in\pols_\infty$,
and another scalar multiplier $\jmul_\ecut\in\RR$ for each external
link $\ecut\in\cuts_\infty$. Then we define a linear combination like
\eref{rpp-kin},
\begin{equation}
  \label{kin}
  \kin = \ft12 \sumi{\pol\in\pols_0} \Trr{\blgen_\pol\pcon_\pol} 
        + \sumi{\pol\in\pols_\infty} \emul_\pol \, \econ_\pol
        + \sumi{\ecut\in\cuts_\infty} \jmul_\ecut \, \jcon_\ecut.
\end{equation}
Inserting the constraints, and rearranging the summation, this can be
written as
\begin{eqnarray}
  \label{kin-sum-S}
   \kin &=& \ft12 \sumi{\cut\in\cuts_0} 
         \Trrr{\blgen_{\pol_\cut} \dis_\cut} 
       - \sumi{\ecut\in\cuts_\infty} 
         (\emul_{\pol_\ecut} - \emul_{\pol_{-\ecut}} ) \, T_\ecut
       + \sumi{\ecut\in\cuts_\infty} 
         \jmul_\ecut \, L_\ecut + \nwl && {} \quad
       {} + 4 G S \, \Big( \,\, \sumi{\,\,\pol\in\pols_\infty} 
         \emul_{\pol} (\cdir_{\ecut'} - \cdir_{\ecut} ) 
    + \sumi{\ecut\in\cuts_\infty} \jmul_\ecut \, M_\ecut \, \Big). 
\end{eqnarray}
Again, we defined $\blgen_\pol=-\emul_\pol\gam_0$ for non-compact
polygons $\pol\in\pols_\infty$, because the first sum also involves
contributions from the non-compact polygons.  The first line is the
free particle expression \eref{rpp-kin-sum}. The second line is the
total contribution that is proportional to $S$, which comes from both
\eref{econ} and \eref{jcon}. It disappears in the limit $G\to0$. 

Since $S$ is also a function of the link variables, the constraints
are no longer linear, and this makes things a little bit more
complicated. But we can use the following trick. According to the
identity \eref{kin-sum-zero}, there is some redundancy in the
definition of the multipliers. If we make the replacement
$\emul_\pol\mapsto\emul_\pol+\emul$ for $\pol\in\pols_\infty$, and
$\blgen_\pol\mapsto\blgen_\pol-\emul\gam_0$ for $\pol\in\pols_0$,
where $\emul\in\RR$ is some fixed real number, then the linear
combination \eref{kin} is unchanged. This can be verified explicitly
in \eref{kin-sum-S}, where we once again recover the definition of $S$
when we make this replacement. We can therefore, without loss of
generality, choose the multipliers so that
\begin{equation}
  \label{kin-mul}
     \sumi{\pol\in\pols_\infty} 
         \emul_{\pol} (\cdir_{\ecut'} - \cdir_{\ecut} ) 
    +  \sumi{\ecut\in\cuts_\infty} \jmul_\ecut \, M_\ecut = 0. 
\end{equation}
Doing so, the expression for $\kin$ simplifies to 
\begin{equation}
  \label{kin-sum}
   \kin = \ft12 \sumi{\cut\in\cuts_0} 
         \Trrr{\blgen_{\pol_\cut} \dis_\cut} 
       - \sumi{\ecut\in\cuts_\infty} 
         (\emul_{\pol_\ecut} - \emul_{\pol_{-\ecut}} ) \, T_\ecut
       + \sumi{\ecut\in\cuts_\infty} 
         \jmul_\ecut \, L_\ecut ,
\end{equation}
which is just the free particle expression. Note that this is still
the most general linear combination of the kinematical constraints,
although we are no longer free to choose all multipliers
independently. 

It is then straightforward to derive the following Poisson brackets of
the phase space variables with $\kin$. For the internal link
variables, we have to apply the brackets \eref{pois-chol-dis} and
\eref{pois-dis-dis}, and what we get is
\begin{equation}
  \label{pois-kin-cut}
  \pois{\kin}{\chol_\cut} = 
        4\pi G \, ( \blgen_{\pol_{-\cut}} \, \chol_\cut 
                   - \chol_\cut \, \blgen_{\pol_{\cut}} ), \qquad
  \pois{\kin}{\dis_\cut} = 
        4\pi G \, \comm{\blgen_{\pol_\cut}}{\dis_\cut} ,
                 \qquad  \cut\in\cuts_0.
\end{equation}
This is obviously the infinitesimal generator \eref{dis-chol-gen} of a
Lorentz rotation of the polygons in the embedding Minkowski space. For
the external link variables, we use the brackets \eref{pois-ecut} and
find
\begin{eqnarray}
  \label{pois-kin-ecut}
  \pois{\kin}{M_\ecut} &=& 
      \emul_{\pol_{-\ecut}} - \emul_{\pol_{\ecut}}  , \hspace*{5.5em}
  \pois{\kin}{T_\ecut} = - 4 G S \, \jmul_\ecut , \nwl
  \pois{\kin}{L_\ecut} &=&  
      - 4 G S  ( \emul_{\pol_{-\ecut}} - \emul_{\pol_{\ecut}}),
   \qquad
  \pois{\kin}{\cdir_\ecut} = \jmul_\ecut ,  
                    \qquad  \ecut\in\cuts_\infty.
\end{eqnarray}
These are the infinitesimal generators \eref{M-T-gen}, with the
appropriate transformation of the auxiliary variables $L_\ecut$ added,
so that the constraints \eref{jcon} are preserved.

We conclude that the vector constraint $\pcon_\pol$ generates a
Lorentz rotations of the compact polygon $\pol\in\pols_0$, the scalar
constraint $\econ_\pol$ generates a rotation about the $\gam_0$-axis
of the non-compact polygon $\pol\in\pols_\infty$, and the auxiliary
constraint $\jcon_\ecut$ generates a rotations of the external link
$\ecut\in\cuts_\infty$ in the spinning cone. And we also recover the
restriction \eref{tws-lor-restrict} on the parameters of these gauge
symmetries, which is the same as \eref{kin-mul}. This ensures that the
average condition \eref{bdir-ave} is preserved, hence the overall
orientation of the Minkowski frame and the conical frame in
\fref{cone}.

\subsubsection*{Mass shell constraints}
The definition of the physical phase space $\psp_\cuts$ is also
analogous to the free particle system, but technically again a little
bit more involved. We have to replace the mass shell constraints by
the appropriate deformed versions. First we recall the definition
\eref{prt-hol} of the \emph{holonomy} of the particle $\prt$,
evaluated in an adjacent polygon $\pol$,
\begin{equation}
  \label{hol-chol}
  \hol_{\prt,\pol} = \prodi{\cut\in\cuts_{\prt,\pol}} \chol_\cut, 
\end{equation}
and also the definition \eref{prt-mom} of the momentum vector, which is
the projection of the holonomy,
\begin{equation}
  \label{prt-mom-proj}
  \hol_{\prt,\pol} = \hols_\prt \, \one + 
              4 \pi G \, \momv_{\prt,\pol}^a \, \gam_a , \qquad 
  \mom_{\prt,\pol} = \momv_{\prt,\pol}^a \, \gam_a .
\end{equation}
They transform under coordinate transformations in the Minkowski
charts according to \eref{hol-lor}. This is now expressed in the
following brackets with $\kin$, 
\begin{equation}
  \label{pois-kin-hol}
  \pois{\kin}{\hol_{\prt,\pol}} = 
      4 \pi G \, \comm{\blgen_\pol}{\hol_{\prt,\pol}}, \qquad
  \pois{\kin}{\mom_{\prt,\pol}} = 
      4 \pi G \, \comm{\blgen_\pol}{\mom_{\prt,\pol}}.
\end{equation}
The physical phase space $\psp_\cuts\subset\ksp_\cuts$ is defined as a
subset of the kinematical phase space, by imposing the mass shell
constraints and positive energy conditions \eref{mss-hol},
\begin{equation}
  \label{mss-con}
  \hols_\prt = \cos(4\pi Gm_\prt), \qquad 
    \momv_\prt^0 = \ft12\Trr{\hol_\prt\gam^0} > 0.
\end{equation}
As for the free particles, the Hamiltonian becomes a linear
combination of the mass shell constraints. To obtain the correct limit
$G\to0$, we have to rescale the constraints by a certain power of $G$,
\begin{equation}
  \label{ham}
  \ham = \sumi\prt \mul_\prt \, \con_\prt , \qquad
   \con_\prt = \frac{\hols_\prt - \cos(4\pi G m_\prt)}
                    {16\pi^2G^2}  \approx 0 .
\end{equation}
To see that this provides the correct limit $G\to0$, one has to use
the relation \eref{hols-mom} between the scalar $\hols_\prt$ and the
momentum vector $\mom_\prt$, and expand the cosine up to second order
in $G$. 

It is also not difficult to see that the mass shell constraints are
still first class constraints. The trace of the holonomy $\hols_\prt$
commutes with $\kin$, and thus with all kinematical constraints. And
the scalars $\hols_\prt$ also commute with each other, because they
only depend on the transition functions $\chol_\cut$ and the energies
$M_\ecut$. To derive the time evolution equations, we have to find the
brackets of $\hols_\prt$ with the other phase space variables. The
only variables that do not commute with the mass shell constraints are
the relative position vectors $\dis_\cut$ for $\cut\in\cuts_0$, and
the clocks $T_\ecut$ for $\ecut\in\cuts_\infty$.

Consider first the relative position vectors. We have to derive the
brackets
\begin{equation}
  \pois{\hols_\prt}{\disv_\cut^a} = 
    \ft12 \pois{\Trrr{\prodi{\bcut\in\cuts_\prt} \chol_\bcut }}
               {\disv_\cut^a}.
\end{equation}
This bracket is zero unless $\cut\in\cuts_\prt$ or
$-\cut\in\cuts_\prt$. Consider the case $\cut\in\cuts_\prt$ first. The
transition function $\chol_\cut$ is then one of the factors in the
product, and according to \eref{pois-chol-dis} the action of
$\disv^a_\cut$ is to insert a factor of $4\pi G\gam^a$ into the
product, behind the factor $\chol_\cut$. So, what we have to do is to
break up the cyclic product behind the factor $\chol_\cut$, and insert
a gamma matrix. The result is
\begin{equation}
  \pois{\hols_\prt}{\disv_\cut^a} =  2\pi G \, 
    \Trrr{\gam^a \prodi{\bcut\in\cuts_{\prt,\pol_\cut}} \chol_\bcut }
    =  2 \pi G \, \Trrr{\gam^a \, \hol_{\prt,\pol_\cut} }
    = 16 \pi^2 G^2 \, \momv_{\prt,\pol_\cut}^a .
\end{equation}
If we have $-\cut\in\cuts_\prt$ instead, then we have to use the
second bracket in \eref{pois-chol-dis}. It is then a factor of $-4\pi
G\gam^a$, which is inserted into the cyclic product in front of the
factor $\chol_{-\cut}$. In this case, the result is
\begin{equation}
  \pois{\hols_\prt}{\disv_\cut^a} = - 2\pi G \, 
    \Trrr{\gam^a  \prodi{\bcut\in\cuts_{\prt,\pol_\cut}} \chol_\bcut}
    = -  2 \pi G \, \Trrr{\gam^a \, \hol_{\prt,\pol_\cut} }
    = - 16 \pi^2 G^2 \, \momv_{\prt,\pol_\cut}^a.
\end{equation}
All together, and again in matrix notation, we get
\begin{equation}
  \label{pois-mss-dis}
  \pois{\con_\prt}{\dis_\cut} 
     = \cases{    
   \phantom{-} \mom_{\prt,\pol_\cut} & if $\prt=\prt_\cut$, \cr 
            -  \mom_{\prt,\pol_\cut} & if $\prt=\prt_{-\cut}$, \cr
   \phantom{-}  0       & otherwise. }
\end{equation}
This is formally the same as \eref{rpp-pois-mss}, and it has the same
interpretation. The mass shell constraint $\con_\prt$ generates the
motion of the particle $\prt$ along its world lines. The direction of
this world line is specified by the momentum vector $\mom_{\prt,\pol}$
in the chart containing the polygon $\pol$. For $\pol=\pol_\cut$, this
is also the chart in which the vector $\dis_\cut$ is defined. The
relative position vector $\dis_\cut$ sees the motion of the particles
$\prt_\cut$ and $\prt_{-\cut}$.

There is also a non-trivial action of the mass shell constraints on
the clocks $T_\ecut$. They have non-vanishing brackets with the
energies $M_\ecut$, and thus with the holonomies $\chol_{\pm\ecut}$ of
the external links. It follows from \eref{ecut-rot} that for
$\ecut\in\cuts_\infty$ and $-\ecut\in\cuts_{-\infty}$ we have
\begin{equation}
  \pois{\chol_\ecut}{T_\ecut} = 
        - 4\pi G \, \chol_\ecut \, \gam^0, \qquad
  \pois{\chol_{-\ecut}}{T_\ecut} = 
          4\pi G \, \gam^0 \, \chol_{-\ecut} .
\end{equation}
Formally, $T_\ecut$ acts on the transition functions
$\chol_{\pm\ecut}$ as if it was the $\gam_0$-component of a fictitious
relative position vector assigned to $-\ecut$. Performing the same
calculation again, it follows that
\begin{equation}
  \label{pois-mss-T}
  \pois{\con_\prt}{T_\ecut} = 
    \cases{ \momv_{\prt,\pol_\ecut}^0 = \momv_{\prt,\pol_{-\ecut}}^0 
               & if $\prt=\prt_{-\ecut}$, \cr
            0 & otherwise.}
\end{equation}
In contrast to \eref{pois-mss-dis}, the case $\prt=\prt_\ecut$ does
not occur, because the external link $\ecut$ has no end point. And
furthermore, we can here evaluate the momentum vector $\mom_\prt$ in
either of the two adjacent polygons $\pol_\ecut$ or $\pol_{-\ecut}$.
The transition function between the two associated charts is
$\chol_\ecut$, and this is a rotation about the $\gam_0$-axis.
Consequently, the $\gam_0$-component of the vector $\mom_\prt$ is the
same in both adjacent polygons $\pol_\ecut$ and $\pol_{-\ecut}$.

Finally, if we ignore the dependence of the various vectors on the
polygons, hence on the coordinate charts with respect to which they
are defined, then the time evolution equations are the same as
\eref{rpp-evolve-rel} for the free particles,
\begin{eqnarray}
  \label{evolve-rel}
  \dot\dis_\cut = \pois{\ham}{\dis_\cut} &=& 
                   \mul_{\prt_\cut}    \, \mom_{\prt_\cut} 
                 - \mul_{\prt_{-\cut}} \, \mom_{\prt_{-\cut}}, \qquad 
  \dot T_\ecut = \pois{\ham}{T_\ecut} = 
      \mul_{\prt_{-\ecut}} \, \momv^0_{\prt_{-\ecut}}.  
\end{eqnarray}
And clearly, the physical interpretation of these equations is also
the same as before. The particle $\prt$ moves along its world line, by
an amount that is specified by the multiplier $\mul_\prt$. Thereby,
the geometry of the ADM surface changes, and also its embedding into
the spinning cone. If all multipliers are positive, this finally
provides a proper foliation of the spacetime manifold. 

\subsubsection*{Symmetries}
In the Hamiltonian formulation of general relativity, every isometry
of the asymptotic metric at spatial infinity should be realized as a
rigid symmetry at the phase space level \cite{nico}. Let us check this
for the time translations $T\mapsto T-\tpar$ and the spatial rotations
$\cdir\mapsto\cdir+\rpar$ of the spinning cone \eref{spin-cone}. And
let us also derive the associated conserved charges. By definition,
these are the total energy and the total angular momentum. First we
have to find out how the symmetries act on the phase space variables.
Then we can check whether these are symmetries of the symplectic
structure, derive the charges, and finally we have to show that the
charges commute with all constraints.

A time translation $T\mapsto T-\tpar$ of the spinning cone does not
affect the geometry of space. The whole ADM surface is just shifted
backwards in time, with respect to the reference frame. Only the
clocks $T_\ecut$ refer to the absolute time coordinate $T$ of the
spinning cone, so only they transform,
\begin{equation}
  \label{T-trans}
  T_\ecut \mapsto T_\ecut - \tpar.
\end{equation}
Since $\tpar$ is a constant, this is obviously a symmetry of the
symplectic structure defined in \eref{pot-tot}. The symplectic
potential changes by a total derivative $\pot\mapsto\pot-\tpar\,\dd
M$, but the two-form $\sym=\dd\pot$ is invariant. The associated
charge is also easy to find. It is the previously defined total energy
$M$, which has vanishing brackets with all phase space variables,
except for the clocks,
\begin{equation}
  \label{pois-M}
  M = \sumi{\ecut\in\cuts_\infty} M_\ecut \follows
  \pois{M}{T_\ecut} = - 1.
\end{equation}
It is also immediately obvious that $M$ commutes with all kinematical
and dynamical constraints, thus the total energy is in fact a conserved
charge.

The same applies to a spatial rotation $\cdir\mapsto\cdir+\rpar$ of
the spinning cone. The local geometry of space is unchanged, while the
whole ADM surface is rotated with respect to the spinning cone by an
angle $\rpar$ in conical coordinates. As only the angular directions
$\cdir_\ecut$ of the external links refer to the conical coordinate
$\cdir$, let us make the following ansatz,
\begin{equation}
  \label{rot-cdir}
  \cdir_\ecut \mapsto \cdir_\ecut + \rpar.
\end{equation}
Now, suppose we apply this transformation. Then the deviations
$\bdir_\ecut^\pm$ are invariant. They are implicitly defined in
\eref{bdir-eqs} and \eref{bdir-ave}, and these equations are invariant
under the given transformation. But if we then look at the definition
\eref{ecut-vec} of the unit vectors $\dis_{\pm\cut}$, we find that
these vectors are also rotated by an angle $\rpar$, but now this
rotation takes place in the auxiliary Minkowski space, where the
polygons are embedded.

The reason can be seen in \fref{cone}. If we rotate all external links
simultaneously with respect to the spinning cone on the right, then we
also have to rotate the external edges in the embedding Minkowski
space on the left. This is because we required the overall, or average
orientation of the two frames to coincide. But then, if we want the
local geometry of the ADM surface to be unchanged, we also have to
rotate the internal edges of all non-compact polygons. Otherwise the
polygons are twisted, and the ADM surface is deformed. So, to define a
proper spatial rotation of the particles with respect to the reference
frame, we not only have the rotate the external links with respect to
the spinning cone. We also have to rotate all non-compact polygons in
the embedding Minkowski space by the same angle $\rpar$.

But then it is actually simpler to rotate all polygons, including the
compact ones. This just involves another kinematical gauge
transformation. In other words, we have to apply the same rotation
about the $\gam_0$-axis to all Minkowski charts, thus rotate all
polygons in \fref{ipol} by the same angle. And finally, we also have
to adjust the transition functions. If we rotate the Minkowski
coordinates in all charts, then we also have to apply this rotation to
the transition functions. All together, we get the following
transformation of the link variables, in addition to \eref{rot-cdir},
\begin{equation}
  \label{rot-J}
  \chol_\cut \mapsto 
   \expo{\rpar\gam_0/2} \chol_\cut \, 
       \expo{-\rpar\gam_0/2}, \qquad
  \dis_\cut \mapsto 
   \expo{\rpar\gam_0/2} \dis_\cut \,
       \expo{-\rpar\gam_0/2}, \qquad
   \cut\in\cuts.
\end{equation}
For internal links, this defines the transformation of the phase space
variables $\dis_\cut$ and $\chol_\cut$. For external links, it is
consistent with the definitions \eref{ecut-rot} and \eref{ecut-vec},
and the transformation \eref{rot-cdir}.

Given these transformations, it is straightforward to check that the
symplectic structure \eref{pot-tot} is invariant. Thus we also have a
symmetry of the phase space. To find the total angular momentum, let
us first consider the internal links, and the definition \eref{M-S} of
the angular momenta $L_\cut$. It is not difficult to derive the
brackets
\begin{equation}
  \label{pois-L-cut}
  \pois{L_\cut}{\dis_{\pm\cut}} 
     = \ft12\comm{\gam_0}{\dis_{\pm\cut}}, \qquad 
   \pois{L_\cut}{\chol_{\pm\cut}} 
     = \ft12\comm{\gam_0}{\chol_{\pm\cut}}, \qquad
       \cut\in\cuts_0.
\end{equation}
The phase space function $L_\cut$ generates a counter clockwise
rotation of the internal edges $\pm\cut$ in the embedding Minkowski
space. This was also the defining property of the angular momenta
$L_\cut$ in \eref{rpp-pois-L-cut}, for the free particles. But now we
also have to rotate the external links. The generator of the rotation
\eref{rot-cdir} is the angular momentum $L_\ecut$,
\begin{equation}
  \label{pois-L-ecut}
  \pois{L_\ecut}{\cdir_\ecut} = 1 \qquad \ecut\in\cuts_\infty.
\end{equation}
To act on all links at the same time, we just have to sum over all
internal and external links. Hence, the total angular momentum becomes
\begin{equation}
  \label{J}
  J = \sumi{\cut\in\cuts_+} L_\cut + 
       \sumi{\ecut\in\cuts_\infty} L_\ecut.
\end{equation}
This is formally the same as \eref{rpp-J} for the free particles. To
see that it is a conserved charge, we have to compute the brackets of
$J$ with the constraints. It turns out that the only non-vanishing
brackets are those with the vector constraints for the compact
polygons,
\begin{equation}
  \label{pois-S-icon}
  \pois{J}{\pcon_{\pol}} 
     = \ft12\comm{\gam_0}{\pcon_{\pol}} \approx 0 ,
       \qquad \pol\in\pols_0.
\end{equation}
This is again proportional to the same constraint, so that $J$ is at
least weakly conserved. This is of course sufficient to define a
conserved charge. We conclude that all symmetries of the spacetime
are in fact realized as symmetries of the phase space.

But how is $J$ related to the parameter $S$ of the spinning cone? It
is no longer so that $J$ is weakly equal to $S$. Instead,
\begin{equation}
  \label{J-S-M-rel}
  J  = \sumi{\cut\in\cuts_+} L_\cut 
      + \sumi{\ecut\in\cuts_\infty} L_\ecut
     \approx S - 4 G S \sumi{\ecut\in\cuts_\infty} M_\ecut
     = (1-4GM) \, S.
\end{equation}
So, the total angular momentum is not the geometric parameter $S$,
which defines the time offset of the spinning cone. It is rescaled by
a factor $1-4GM$.

The definition of the total angular momentum is actually somewhat
ambiguous. Of course, $S$ is also a conserved charge. It is a function
of $J$ and $M$, at least up to terms proportional to the constraints.
So, there is also a symmetry associated with $S$. It must be some
combination of a time translation and a spatial rotation, thus a kind
of \emph{screw rotation}. And in fact, the definition of the rotational
Killing vector of the spinning cone is also somewhat ambiguous. The
Killing vector associated with time translations is uniquely defined
as that Killing vector which is parallel to the fictitious world line
of the centre of mass of the universe. Hence the definition of the
total energy $M$ is unique. 

However, there are two alternative definitions of the rotational
Killing vector, and both are quite natural. They coincide only if the
spinning cone is actually static, thus if the angular momentum is
zero. One possible definition is to choose the unique Killing vector
which has closed orbits of affine length $2\pi$. The associated
angular momentum is then $J$, which has the property that a symmetry
transformation with parameter $\rpar=2\pi$ is the identity. But we may
also define the rotational Killing vector to be the unique Killing
vector which is orthogonal to the Killing vector of time translations.
On a spinning cone, this Killing vector generates a screw rotation,
and it is this symmetry which is associated with the conserved charge
$S$.

It is not clear which of the two is the \emph{correct} angular
momentum, $S$ or $J$. Which one is the more appropriate one depends on
the context. For example, if we quantize a particle model like this,
then it is $J$ whose eigenvalues are quantized in steps of $\hbar$,
because the essential property is then the existence of closed orbits
of affine length $2\pi$. On the other hand, if we want to describe the
conical geometry of the spacetime at infinity, then $S$ is more
natural, because it is this which appears in the metric
\eref{spin-cone} of the spinning cone, and which defines the time
offset $8\pi GS$.

An interesting consequence of this interplay between quantization and
geometry and the charges $J$ and $S$ is discussed in the very end of
\cite{matlou}. It turns out that, independent of the matter content of
the universe, only those conical geometries can be realized in a
quantized three dimensional universe, which fulfill the following
Dirac like quantization condition,
\begin{equation}
  \label{q-cone}
  \frac{(1-4GM) \, S}\hbar \in \ZZ .
\end{equation}
Since here were are not going to quantize anything, we shall not go
into any details and derive this quantization condition. But it is
more or less obvious that this follows from the fact that $J$ is
quantized in integer multiples of $\hbar$. This result is independent
of the non-commutative structure of spacetime implied by the brackets
\eref{pois-dis-dis}, but it also provides a kind of quantized
geometry.

\subsubsection*{Parity transformations}
Let us also briefly consider a discrete symmetry. Apparently, the
Poisson algebra \eref{pois-dis-dis} is not invariant under parity
transformations, thus under reflections of space, because the Levi
Civita tensor shows up. But on the other hand we know that the
Einstein equations are invariant, and thus for every spacetime there
exists a reflected spacetime. It is therefore not immediately obvious
that this symmetry is also realized at the phase space level. And in
fact, for the previously considered single particle model this was not
the case, due to a somewhat ambiguous definition of the reference
frame \cite{matwel}. 

For a single particle, we had the same Poisson algebra
\eref{pois-dis-dis}, but there was only one vector $\dis$,
representing the absolute position of the particle with respect to the
reference frame. Now each vectors $\dis_\cut$ is the relative position
vector of two particles, and there are several such vectors. But
still, the Levi Civita tensor shows up, and we should therefore check
whether the problem is still present or not. First of all, we have to
find out how the various phase space variables actually transform
under a reflection of space. We have to start at the very beginning,
the definition of a triangulation. When we perform a reflection of
space, we have to replace the graph $\cuts$ by its mirror image.

This is a non-trivial operation, because the cyclic orderings of the
subsets $\cuts_\prt$, $\cuts_\infty$ and $\cuts_\pol$ already refer to
the orientation of space. For each particle $\prt$, we have to reverse
the cyclic ordering of the set $\cuts_\prt$ of links ending at $\prt$,
and the same applies to the set $\cuts_\infty$ of links pointing
towards infinity. For each polygon $\pol$, we not only have to reverse
the cyclic ordering of its edges in $\cuts_\pol$, but also reverse the
orientation of the edges. For example, if a polygon $\pol$ is bounded
by the edges $\cuts_\pol=\{\cut_1,\cut_2,\dots,\cut_n\}$ before the
parity transformation, then after the parity transformation we have
$\cuts_\pol=\{-\cut_n,\dots,-\cut_2,-\cut_1\}$.
\begin{figure}[t]
  \begin{center}
    \epsfbox{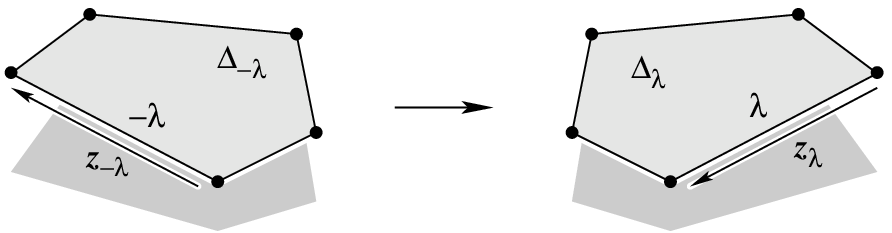}
    \xcaption{A parity transformation acts on a polygon in two
    ways. The polygon is first reflected, and then the orientation of
    the boundary is reversed. If the link $-\cut$ is an edge of the
    original polygon $\pol_{-\cut}$, which is represented by the
    vector $\dis_{-\cut}$, then the reversed link $\cut$ becomes an
    edge of the reflected polygon $\pol_\cut$, which is represented by
    the transformed vector $\dis_\cut$.}
    \label{flip}
  \end{center}
\end{figure}

Now, consider the reflection of a polygon in the embedding Minkowski
space, which is indicated in \fref{flip}. The edge $-\cut$ of the
polygon $\pol_{-\cut}$ becomes, after the reflection, the edge $\cut$
of the polygon $\pol_{\cut}$. The edge was originally represented by
the vector $\dis_{-\cut}$. The polygon is then reflected, and
additionally the orientation of the boundary is reversed. Hence, the
new vector $\dis_\cut$ is obtained from the original vector
$\dis_{-\cut}$ by a reflection and an overall change of sign. Let us
choose the reflection in Minkowski space so that
$\gam_2\mapsto-\gam_2$, hence a reflection at the
$\gam_0$-$\gam_1$-plane, which is given by \eref{reflection}

Combining this with an overall change of sign, and deriving the
appropriate transformation of the transition functions from
\eref{cut-inv} gives 
\begin{equation}
  \label{par-dis-chol}
  \dis_\cut \mapsto \gam_2 \, \dis_{-\cut} \, \gam_2 , \qquad
  \chol_\cut \mapsto  \gam_2 \, \chol_{-\cut} \, \gam_2 . 
\end{equation}
The external link variables also transform in a non-trivial way.  It
follows from the definition \eref{M-S} of the angular momentum that
$S\to-S$. Thus the reflected spinning cone is spinning into the
opposite direction. The conical coordinates are reflected by the
transformation $\cdir\mapsto-\cdir$. From this we infer that
\begin{equation}
  \label{par-ecut}
  T_\ecut \mapsto T_\ecut , \qquad
  \cdir_\ecut \mapsto -\cdir_\ecut, \qquad
  M_\ecut \mapsto M_\ecut , \qquad
  L_\ecut \mapsto - L_\ecut.  
\end{equation}
The last transformation follows from $S\mapsto-S$, and the constraint
equation $\jcon_\ecut\approx0$ defined in \eref{jcon}, which must be
satisfied before and after the reflection. The vector constraint
$\pcon_\pol\approx0$ in \eref{icon} is obviously also preserved. And
for the constraint $\econ_\pol\approx0$ defined in \eref{econ}, we
have to take into account that the external links $\ecut$ and $\ecut'$
are interchanged, because the ordering of $\cuts_\infty$ is reversed,
so that this is also preserved.

The symplectic potential \eref{pot-tot} can then be shown to be
invariant. The external terms $T_\ecut\,\dd M_\ecut$ and
$L_\ecut\,\dd\cdir_\ecut$ are obviously invariant under the
transformation \eref{par-ecut}. For the internal links, this is not
immediately obvious. However, we have $\gam_2\gam_2=\one$, and
therefore
\begin{equation}
  \label{par-cut}
  \Trr{\chol\inv_\cut \dd\chol_\cut \, \dis_\cut}
   \mapsto \Trr{ \chol\inv_{-\cut} \, \dd\chol_{-\cut} \, \dis_{-\cut} }.
\end{equation}
But this just means that we have to replace $\cuts_+$ by $\cuts_-$ in
\eref{pot-tot}. And since we know that the symplectic potential is
independent of the decomposition $\cuts_0=\cuts_+\cup\cuts_-$, it
follows that it is also invariant under parity transformations.

Finally, we also have to check the mass shell constraints, hence the
traces of the holonomies $\hols_\prt$ of the particles. In the
definition \eref{prt-hols}, each factor $\chol_\cut$ in the product is
replaced by $\gam_2\chol_{-\cut}\gam_2$, and the ordering of the
factors is reversed. Again, the factors $\gam_2$ drop out because
$\gam_2\gam_2=\one$. So, the group element under the trace is
effectively replaced by its inverse. But the trace of the inverse of
an element of $\grpSL(2)$ is equal to the trace of the element itself.

So, we find that all the phase space structures are invariant under
parity transformations. But what is then the difference to the
previously considered single particle system, where this was not the
case? The difference is that the definition of the relative position
vector $\dis_\cut$ already refers to a fixed orientation of the
spacetime. It defines the edge of a polygon, where the boundary is
traversed in counter clockwise direction. This is why the Levi Civita
tensor can show up in the Poisson algebra, without breaking the parity
invariance. 

\subsubsection*{Open and closed universes}
So far we assumed that the topology of space is $\RR^2$. But of
course, general relativity allows a more general topology. The space
manifold can be any two dimensional, orientable surface, and it can be
either compact or non-compact. Let us for example consider a
\emph{closed} universe, where the space manifold is a Riemann surface
of genus $\genus$. In this case, the graph $\cuts=\cuts_0$ consists of
internal links only, and all polygons are compact, thus
$\pols=\pols_0$. When the Riemann surface is triangulated, then we
have the following relation between the number of vertices, links,
polygons, and the genus,
\begin{equation}
  \label{genus-cls}
   \ncut_0 - \npol_0 = \nprt + 2 \genus - 2 . 
\end{equation}
We can use this to calculate the number of physical degree of freedom
of a closed universe. The extended phase space $\esp_\cuts$ is spanned
by the independent link variables $\chol_\cut\in\grpSL(2)$ and
$\dis_\cut\in\algsl(2)$ for $\cut\in\cuts_+$, hence
\begin{equation}
  \label{dim-esp-cls}
  \dim(\esp_\cuts) = 6 \ncut_0 .
\end{equation}
Then we have to impose the kinematical constraints. For a closed
universe, we only have the vector constraints $\pcon_\pol$ for
$\pol\in\pols_0$, and the associated gauge symmetries are the Lorentz
rotation of the polygons. There is no relation like
\eref{kin-sum-zero}, and there is no restriction on the parameters of
the Lorentz rotations. Thus we have $3\npol_0$ independent
kinematical constraints, and also $3\npol_0$ independent gauge
symmetries,
\begin{equation}
  \label{dim-ksp-cls}
  \dim(\ksp_\cuts) = 6 \ncut_0 - 3 \npol_0 , \qquad
  \dim(\ksp_\cuts/{\sim}) = 
      6 \ncut_0 - 6 \npol_0 = 6 \nprt + 12 \genus - 12 .
\end{equation}
This is the number of independent kinematical degrees of freedom, thus
the phase space dimension before the mass shell constraints are
imposed. To obtain the physical phase space, we have to impose $\nprt$
additional mass shell constraints, and subtract $\nprt$ additional
dynamical gauge symmetries. Hence, the number of physical degrees of
freedom is
\begin{equation}
  \label{dim-psp-cls}
  \dim(\psp_\cuts) = 6 \ncut_0 - 3 \npol_0 - \nprt , \qquad
  \dim(\psp_\cuts/{\sim}) = 
      6 \ncut_0 - 6 \npol_0 - 2 \nprt = 4 \nprt + 12 \genus - 12 .
\end{equation}
For each particle we have $4$ physical degrees of freedom. The
particle is moving in a two dimensional space, thus we have two
position and two momentum coordinates.

Additionally, the closed universe also has some topological degrees of
freedom. Each \emph{handle} of the Riemann surface contributes with
$12$ degrees of freedom. The topological degrees of freedom are the
holonomies of the fundamental, non-contractible loops in space. There
are two such loops for each handle, and regarding the holonomy we have
to take into account both the rotational and the translational
components \cite{witten,matrev}. Hence, for each non-contractible loop
we have one element of the six dimensional Poincar\'e group, and all
together this makes $12\genus$ topological degrees of freedom, since
the genus $\genus$ counts the number of handles.

Finally, there is a consistency condition, which must be satisfied for
the universe to be closed. This explains the minus twelve in the end.
As an example, consider a sphere with $\genus=0$. Then we have
$4(\nprt-3)$ independent physical degrees of freedom. We need at least
four particles to get a non-trivial dynamical system. In this case,
the space at a moment of time looks like a tetrahedron, and we have
four independent degrees of freedom. The relative motion of two
particles already determines the motion of the other two particles.
For higher genus surfaces, we need less particles or even no particles
at all to get a non-trivial dynamical system.

For an \emph{open} universe, we may also consider more general
topologies. The space manifold can be represented as a
\emph{punctured} Riemann surface, thus a surface of genus $\genus$
with $\ninf\ge1$ points taken away. Each puncture represents a region
that extends to spatial infinity. For each such infinity, there is a
conical frame, and a set of external links and non-compact polygons
covering a neighbourhood of this particular infinity. Moreover, each
conical frame has its own total energy and total angular momentum, and
from each infinity an external observer can see the particles and also
the other observers. The case that we discussed so far, and which
reduces to the free particle system in the limit $G\to0$, is the
special open universe with $\genus=0$ and $\ninf=1$. The closed
universes are also special cases, with $\ninf=0$.

Let us count the number of physical degrees of freedom for an open
universe. First we have to derive the relation between the number of
vertices, links, and polygons. It is the same as \eref{genus-cls}, but
now we have a vertex not only for every particle but also for every
infinity. Moreover, we have to distinguish between internal and
external links. The external links are now those that extend to any of
the $\ninf$ infinities. But there are still as many external links are
there are non-compact polygons. Hence, 
\begin{equation}
  \label{genus-opn}
  \ncut_0 - \npol_0 = \nprt + \ninf + 2 \genus - 2 ,
  \qquad \ncut_\infty = \npol_\infty. 
\end{equation}
The phase space dimension is counted in the same way as before. The
extended phase space $\esp_\cuts$ is spanned by $6\ncut_0$ independent
components of the group elements $\chol_\cut$ and the vectors
$\dis_\cut$ for $\cut\in\cuts_0$, and $4\ncut_\infty$ real variables
$M_\ecut$, $L_\ecut$, $T_\ecut$, $\cdir_\ecut$ for
$\ecut\in\cuts_\infty$, thus 
\begin{equation}
  \label{dim-esp-opn}
  \dim(\esp_\cuts) = 6 \ncut_0 + 4 \ncut_\infty.
\end{equation}
Then we have to count the number of kinematical constraints. There are
$3\npol_0$ components of the vector constraints $\pcon_\pol$ for
$\pol\in\pols_0$, $\npol_\infty$ scalar constraints $\econ_\pol$ for
$\pol\in\pols_\infty$, and finally $\ncut_\infty$ scalar constraints
$\jcon_\ecut$ for $\ecut\in\cuts_\infty$. But now they are not all
independent. For each infinity, we have a relation like
\eref{kin-sum-zero}. This implies that the total number of independent
kinematical constraints is $3\npol_0+\npol_\infty+\ncut_\infty-\ninf$,
and this is also the number of independent redundancy transformations.
It follows that
\begin{eqnarray}
  \label{dim-ksp-opn}
  \dim(\ksp_\cuts) &=& 6 \ncut_0 - 3 \npol_0 + 2 \ncut_\infty + \ninf,
  \nwl 
  \dim(\ksp_\cuts/{\sim}) &=& 
      6 \ncut_0 - 6 \npol_0 + 2\ninf =
      6 \nprt + 12 \genus + 8 \ninf - 12. 
\end{eqnarray}
This is the number of independent kinematical degrees of freedom. To
get the physical degrees of freedom, we have to impose $\nprt$
additional mass shell constraints, and subtract another $\nprt$ gauge
symmetries. This gives
\begin{eqnarray}
  \label{dim-psp-opn}
  \dim(\psp_\cuts) &=& 
  6 \ncut_0 - 3 \npol_0 + 2 \ncut_\infty - \nprt + \ninf,
  \nwl
  \dim(\psp_\cuts/{\sim}) &=& 
      6 \ncut_0 - 6 \npol_0  - 2 \nprt + 2 \ninf =
      4 \nprt + 12 \genus + 8 \ninf - 12. 
\end{eqnarray}
Once again, we have four degrees of freedom for every particle, and
twelve degrees of freedom for every handle. And obviously there are
eight degrees of freedom associated with every infinity. This can be
understood as follows. There are two charges $M$ and $S$, defining the
conical geometry, which can be different at each infinity. The other
six degrees of freedom represent the relative orientation of the
conical frames with respect to each other.

Consider for example two infinities and the associated conical frames.
Think of them as the rest frames of two different observers. When the
observers talk to each other, they find out that they do not agree
about the fictitious position of the centre of mass of the universe,
and in general they are also in relative motion with respect to each
other. Hence, every infinity has its own centre of mass frame. The
relation between two such frames is, at least locally in phase space,
parameterized by a Poincar\'e transformation. If we fix one reference
frame, then for every other there are six degrees of freedom,
representing the relative position and orientation with respect to the
fixed one.

The minus twelve can now be explained by looking at the special case
$\ninf=1$ and $\genus=0$, which reduces to the free particle system in
the limit $G\to0$. In this case, the dimension of the reduced phase
space is $4(\nprt-1)$. This is the number of physical degrees of
freedom of the relative motion of $\nprt$ particles in a two
dimensional space. To get a non-trivial dynamical system, we need at
least two particles. In general, each additional infinity, and also
each handle already provides sufficiently many topological degrees of
freedom, so that we do not need any particles at all to obtain a
non-trivial dynamical system. These are the usual, say, cosmological
toy models, where only the topological degrees of freedom are present,
and they were the first quantized toy models in three dimensional
gravity \cite{witten}.

\subsubsection*{Large gauge symmetries}
Finally, let us also have a look at the global structure of the phase
space. We have seen that the extended phase space $\esp$ consists of
finitely many disconnected components $\esp_\cuts$, one for each
possible graph $\cuts$ with $\nprt$ particles. According to
\eref{dim-esp-opn}, they even have different dimensions. It is
therefore not possible to consider the components $\esp_\cuts$ of the
extended phase space $\esp$ as overlapping coordinate charts of a
single manifold. However, we can still glue them together in a certain
way, so that the quotient spaces $\ksp/{\sim}$ and $\psp/{\sim}$
become connected manifold. So far, we only considered the gauge
symmetries generated by the constraints.

Two states $\state_1,\state_2\in\ksp_\cuts$, belonging to the same
component of the kinematical subspace, are defined to be physically
equivalent, $\state_1\sim\state_2$, if they can be smoothly deformed
into each other by a transformation which is generated by the
kinematical constraints. This is equivalent to the condition that they
lie on the same \emph{gauge orbit}, that is the same null orbit of the
symplectic structure on $\ksp_\cuts$. However, there are also states
that belong to different components of $\ksp$, but which are
nevertheless equivalent because they define the same geometry of
space. They are then related by a \emph{large} gauge transformation.
In the end of \sref{inter}, we have seen how these transformations
look like. They can be decomposed into elementary steps, namely
inserting and removing links.
\begin{figure}[t]
  \begin{center}
    \epsfbox{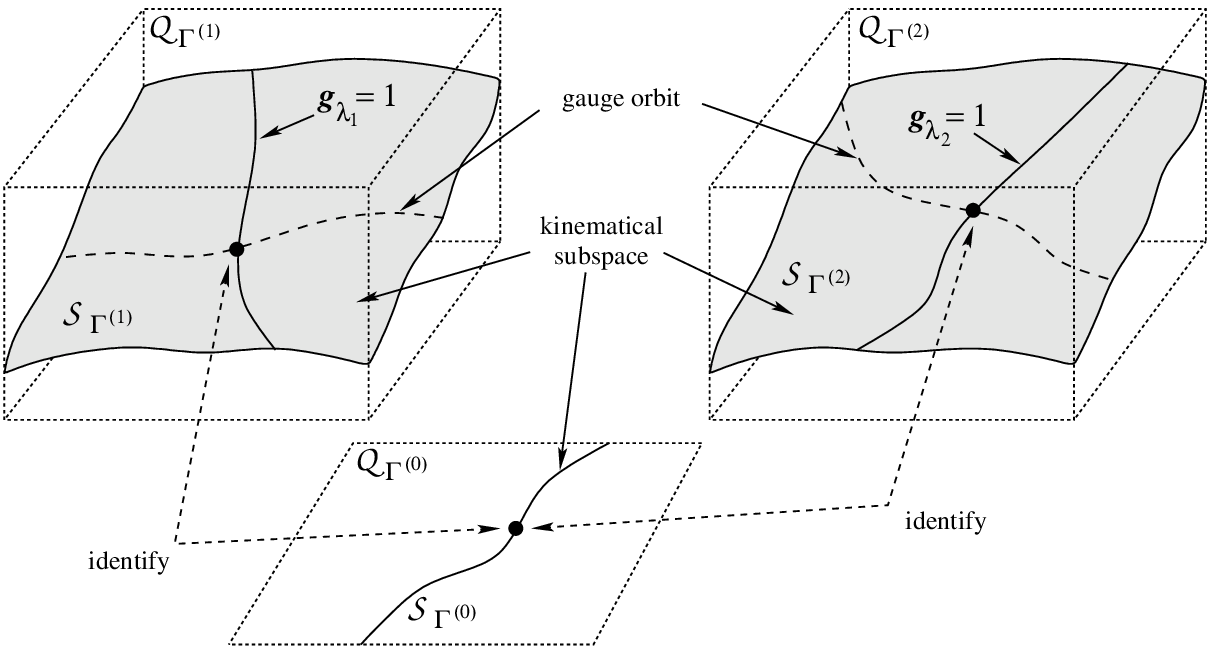}
    \xcaption{The extended phase space $\esp$ consists of finitely many 
    disconnected components $\esp_\cuts$ of different dimension. A
    large gauge symmetry embeds a lower dimensional component
    $\ksp_\cuts$ of the kinematical phase space into the respective
    higher dimensional component. The embedding is given by the
    insertion of a new link $\cut$, as shown in \fref{link}. The image
    of the lower dimensional component in the higher dimensional one
    is the subspace defined by the gauge condition $\chol_\cut=\one$.}
    \label{phsp}
  \end{center}
\end{figure}

From the phase space point of view, the typical situation is shown in
\fref{phsp}. There we see three components of the extended phase space
$\esp$, corresponding to the graphs $\cuts^{(0)}$, $\cuts^{(1)}$, and
$\cuts^{(2)}$. Let us assume that both $\cuts^{(1)}$ and $\cuts^{(2)}$
are obtained from $\cuts^{(0)}$ by inserting a new pair of links,
$\pm\cut_1$ and $\pm\cut_2$, respectively. For example, the new links
may be two different diagonals of the polygon in \fref{link}.  The
component $\esp_{\cuts^{(0)}}$ is the lower dimensional one, and the
components $\esp_{\cuts^{(1)}}$ and $\esp_{\cuts^{(2)}}$ have the same
higher dimension. The same applies to the kinematical subspaces
$\ksp_{\cuts^{(0)}}$, $\ksp_{\cuts^{(1)}}$, and $\ksp_{\cuts^{(2)}}$,
which are submanifolds of the respective components of $\esp$.

The transition from the graph $\cuts^{(0)}$ to $\cuts^{(1)}$ is a
large gauge symmetry, which maps a state
$\state^{(0)}\in\ksp_{\cuts^{(0)}}$ onto an equivalent state
$\state^{(1)}\in\ksp_{\cuts^{(1)}}$. And a transition from
$\cuts^{(0)}$ to $\cuts^{(2)}$ maps the state
$\state^{(0)}\in\ksp_{\cuts^{(0)}}$ onto an equivalent state
$\state^{(2)}\in\ksp_{\cuts^{(2)}}$. We can think of an
\emph{embedding} of the lower dimensional component
$\ksp_{\cuts^{(0)}}$ into the higher dimensional components
$\ksp_{\cuts^{(1)}}$ and $\ksp_{\cuts^{(2)}}$. The image of this
embedding is a submanifold of the higher dimensional component of the
kinematical phase space. It is the submanifold which is defined by the
\emph{gauge condition} $\chol_{\cut_1}=\one$ in $\ksp_{\cuts^{(1)}}$,
and $\chol_{\cut_2}=\one$ in $\ksp_{\cuts^{(2)}}$, respectively.

A general kinematical gauge transformation can be described as
follows. As long as we stick to a given triangulation, say
$\cuts^{(1)}$, the gauge symmetries are generated by the kinematical
constraints. We follow a gauge orbit, hence a null direction of the
symplectic structure on the kinematical subspace $\ksp_{\cuts^{(1)}}$.
Now, suppose that we want to remove the pair of links $\pm\cut_1$.
Then, we first have to fix a gauge where $\chol_{\cut_1}=\one$. Thus
we have to follow the gauge orbit to some state where this gauge
condition is satisfied. This state $\state^{(1)}\in\ksp_{\cuts^{(1)}}$
is the image of some state $\state^{(0)}\in\ksp_{\cuts^{(0)}}$ in the
lower dimensional component, and the large gauge transformation takes
us there.

If we want, we can then perform another large gauge transformation,
inserting a new link $\cut_2$. Then we have a state $\state^{(2)}$ in
the component $\ksp_{\cuts^{(2)}}$ of the kinematical phase space.
There we can again perform a smoothly generated gauge transformation,
following a gauge orbit provided by the kinematical constraints, and
so on. All components of the phase space are finally connected in this
way, and consequently the quotient space $\ksp/{\sim}$ becomes a
connected manifold. That it is a proper manifold follows from the fact
that all components $\ksp_\cuts/{\sim}$ of this quotient space have
the same dimension. According to \eref{dim-ksp-opn}, this dimension
only depends on the number of particles and the topology, but not on
the triangulation.

However, what we still have to show is that $\ksp/{\sim}$ is in fact a
proper symplectic manifold, and that the Hamiltonian, or the mass
shell constraints are well defined functions thereon. In other words,
we have to show that the symplectic structure is invariant under large
gauge symmetries. This is the case if the symplectic potential on the
lower dimensional component $\ksp_{\cuts^{(0)}}$ is equal to the one
that is induced by the embedding into the higher dimensional
components $\ksp_{\cuts^{(1)}}$ and $\ksp_{\cuts^{(2)}}$. And the same
has to apply to the holonomies of the particles. It is not difficult
to see that this is indeed the case. If the inserted link is an
internal link $\cut$, then the image of the lower dimensional
component in the higher dimensional one is the subspace defined by the
gauge condition $\chol_\cut=\one$. If we look at the symplectic
potential \eref{pot-tot}, we find that on this subspace the terms
involving the link $\cut$ are just absent.

Hence, the symplectic potential is the same as if the link was not
there. The same holds if the inserted link is an external link
$\ecut$. In this case, the image of the lower dimensional component in
the higher dimensional one is the subset defined by $M_\ecut=0$, and
consequently also $L_\ecut=0$, which follows from the constraint
\eref{jcon}. Again, this implies that the terms involving the link
$\ecut$ are not present in the expression \eref{pot-tot} for the
symplectic potential. Since we also know that no other link variables
are affected when we insert a new link, we conclude that the
symplectic structure on the lower dimensional component of the
kinematical phase space is equal to that on its image in the higher
dimensional component.

The same holds for the mass shell constraints and the Hamiltonian,
which is given by \eref{ham}. If any of the transition functions is
trivial, thus $\chol_\cut=\one$, then it does not contribute to the
holonomies in \eref{hol-chol}. Once again, it is therefore irrelevant
for the definition of the holonomy whether the new link has been
inserted or not. And the value of the Hamiltonian is consequently also
invariant under large gauge symmetries. All together, this implies
that the quotient space $\ksp/{\sim}$ is a well defined symplectic
manifold, and the Hamiltonian, which is actually a linear combination
of the mass shell constraints, is a well defined function thereon.

For the free particle system, we had an explicit definition of this
manifold. It was the original, $6\nprt-4$ dimensional phase space
spanned by the positions $\pos_\prt$ and the momenta $\mom_\prt$, with
the restriction to the centre of mass frame imposed. Here we do not
have such an explicit definition. All we have is a finite atlas, where
each chart is a labeled by a graph $\cuts$, and where the coordinates
are equivalence classes of kinematical states. Hence, both at the
spacetime level and at the phase space level we only have local
coordinates. A global chart on the phase space can only be defined if
we also introduce a global coordinate chart on the spacetime manifold
\cite{bcv,ms}. But then we also have to give up the simple physical
interpretation of the phase space coordinates. It is not possible to
use the relative position vectors and the conjugate transition
functions as global coordinates on the phase space \cite{matcs}.

\section{Einstein gravity}
\label{reduc}
In this section we are going to derive the phase space structures
defined in \sref{phase} from the Einstein Hilbert action. The starting
point is the definition of the particle model at the level of general
relativity as a field theory on a fixed spacetime manifold $\M$. It is
convenient to stick to the ADM formulation from the very beginning, so
that $\M=\RR\times\N$ is foliated by a space manifold $\N$ evolving
with respect to some unphysical ADM time coordinate $t$. Generalizing
the definition of the previously considered single particle model, we
define $\N$ to be a two dimensional surface, which is either a plane
or a sphere, and we cut out with $\nprt$ open discs, so that $\nprt$
circular boundaries arise, representing the locations of the particles
in space.

This provides a proper regularization of the conical singularities, and
it formally converts the matter degrees of freedom associated with the
particles into topological degrees of freedom of the gravitational
field. A set of boundary conditions is imposed to make the boundaries
look like points in space, and boundary terms are added to the action,
to define the coupling of the particles to the gravitational field. A
comprehensive description can be found in section~2 of \cite{matwel}.
For an open universe, we finally impose a fall off condition at
spatial infinity. This is a kind of asymptotical flatness condition.
It defines the centre of mass frame of the universe as a reference
frame in the sense of \cite{nico}. 

To the so defined model we shall then apply the phase space reduction.
We set up the Hamiltonian framework, identify the phase space, derive
and solve the constraints, and finally the gauge degrees of freedom
can be divided out. We'll find a finite number of \emph{observables},
representing the physical degrees of freedom of the model, and these
are the link variables used in \fref{phase} to define the phase space.
We shall also recover the symplectic structure and the various
kinematical constraints defined in the beginning of \sref{phase},
which are then shown to be derived from the Einstein Hilbert action.

\subsubsection*{The space manifold and the particles}
\begin{figure}[t]
  \begin{center}
    \epsfbox{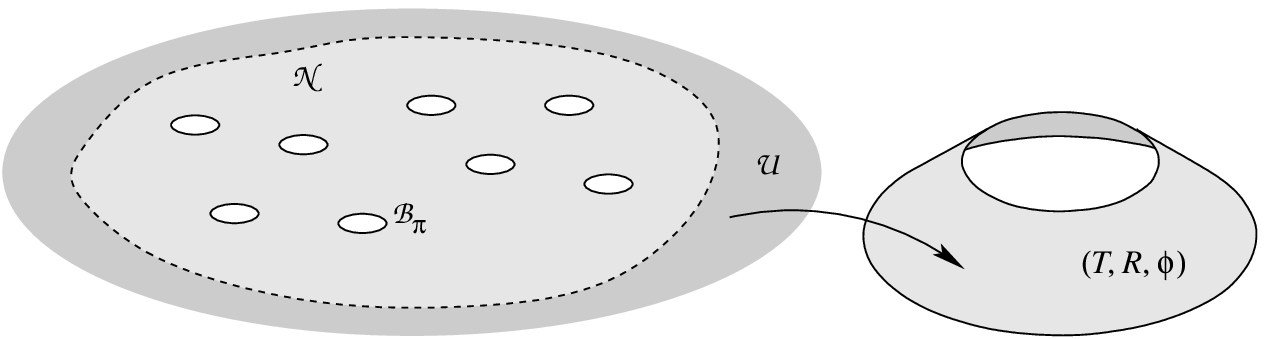}
    \xcaption{The space manifold $\N$ for an open universe is a plane
    $\RR^2$ with $\nprt$ open discs cut out. For a closed universe,
    the same open discs are cut out from a sphere $\SP^2$. The
    circular boundary $\bnd_\prt$ represents the location of the
    particle $\prt$ in space. The metric is degenerate on the
    boundaries, so that they look like points in space. The fall off
    condition at infinity requires the existence of a neighbourhood of
    infinity $\U\subset\N$, which is embedded into a spinning cone.
    This defines the centre of mass frame and serves as a reference
    frame.}
    \label{mfld}
  \end{center}
\end{figure}

The ADM space manifold $\N$ is a two dimensional, orientable surface
with $\nprt$ circular boundaries, representing the locations of the
particles in space. The global topology of $\N$ is in principle
arbitrary. But for simplicity, and also for the reasons given in
\sref{phase}, we shall here restrict to two special cases. For an
\emph{open} universe, the space manifold $\N$ is a plane $\RR^2$ with
$\nprt$ open discs cut out, as shown in \fref{mfld}. For a
\emph{closed} universe, the same $\nprt$ open discs are cut out from a
sphere $\SP^2$.

In the first order dreibein formulation of general relativity, the
basic field variables are the dreibein $\ee_\mu$ and the spin
connection $\om_\mu$, where $\mu,\nu,\dots$ are formal tangent indices
on $\M=\RR\times\N$. For both open and closed universes, the tangent
bundle of the spacetime manifold is trivial. Both fields are therefore
$\algsl(2)$-valued one-forms. We don't have to define them as sections
in a bundle. The metric and the dreibein determinant are given by
\begin{equation}
  \label{met-det}
  g_{\mu\nu} = \ft12\Trr{\ee_\mu\ee_\nu}, \qquad
  e = \ft1{12} \eps^{\mu\nu\rho} \, \Trr{\ee_\mu\ee_\nu\ee_\rho} > 0 ,
\end{equation}
where $\eps^{\mu\nu\rho}$ is the Levi Civita tensor on $\M$. We
require the metric to be invertible, and the determinant to be
positive everywhere in the \emph{interior} of $\M$. What happens at
the boundaries will be explained in a moment.

In the ADM framework, the fields are decomposed into one-forms $\ee_i$
and $\om_i$ on $\N$, where $i,j,\dots$ are formal tangent indices, and
scalars $\ee_t$ and $\om_t$. The induced metric on $\N$ and the normal
vector are given by
\begin{equation}
  \label{norm}
  g_{ij} = \ft12\Trr{\ee_i\ee_j}, \qquad
  \norm = \ft12 \eps^{ij} \, \comm{\ee_i}{\ee_j} .  
\end{equation}
For the Levi Civita tensor on $\N$ we have $\eps^{ij}=\eps^{tij}$. For
the foliation to by spacelike, and the time orientation of the
dreibein to coincide with that of the ADM time coordinate $t$, the
normal vector $\norm$ is required to be \emph{negative timelike}. A
future pointing timelike vector $\xi^\mu$ is then mapped onto a
positive timelike Minkowski vector $\xi^\mu\ee_\mu$ by the dreibein.

The \emph{particle boundaries} of $\N$ are denoted by $\bnd_\prt$. For
a circular boundary to look like a point in space, its circumference
has to vanish. This is the case if and only if the tangent component
of the dreibein vanishes on the boundary. Let $(r,\p)$ by a polar
coordinate system in the neighbourhood of the boundary, so that the
boundary is at $r=0$, and $\p$ increases in counter clockwise
direction with a period of $2\pi$. The \emph{point particle condition}
then requires
\begin{equation}
  \label{prt-con}
  \ee_\p \big|_{\bnd_\prt} = 0.
\end{equation}
This implies that both the normal vector $\norm$ and the dreibein
determinant $e$ are zero on the boundary. Therefore, the restriction
to negative timelike normal vectors and invertible metrics applies in
the interior of $\N$ only. For a one-dimensional boundary of the space
manifold to look like a point, the metric on the boundary has to be
degenerate.

From the spacetime point of view, the two-dimensional cylindrical
boundary $\RR\times\bnd_\prt$ represents a one-dimensional object, the
world line of the particle $\prt$, and therefore the spacetime metric
on the boundary has to be degenerate as well. But this is not going to
be a problem, because in the first order formulation of Einstein
gravity in three dimensions, the inverse dreibein never shows up.
Blowing up the world line to a cylindrical boundary in fact removes
all the singularities, which are otherwise present in the dreibein and
the spin connection describing the gravitational field of a point
particle \cite{matwel}.

\subsubsection*{Action and boundary terms}
The next step is to define the action functional for the particle
model. In the first order dreibein formulation, the ADM Lagrangian
depends on the fields $\ee_\mu$ and $\om_\mu$ and their time
derivatives. It is the sum of a bulk term, a boundary term for each
particle boundary, and for an open universe there is also a boundary
term at spatial infinity,
\begin{equation}
  \label{lag-tot}
  \lag = \lag_0 + \lag_\infty - \sumi{\prt} \mul_\prt \, \con_\prt.
\end{equation}
The bulk term $\lag_0$ is the usual first order Einstein Hilbert
Lagrangian, which is given by
\begin{equation}
  \label{lag-0}
  \lag_0 = \frac1{8\pi G} \inti{\N} \dd^2x \, \eps^{ij} \, 
       \Trr{\dot \om_i \, \ee_j 
            + \ft12 \om_t \, \TT_{ij} + \ft12 \ee_t \, \FF_{ij} }.
\end{equation}
The dot denotes the derivative with respect to $t$, and $\FF_{ij}$ and
$\TT_{ij}$ are the spatial components of the curvature and the
torsion, respectively,
\begin{eqnarray}
  \label{FF-TT}
   \FF_{\mu\nu} &=&
     \del_\mu\om_\nu - \del_\nu\om_\mu + \comm{\om_\mu}{\om_\nu}, \nwl
   \TT_{\mu\nu} &=&
     \del_\mu\ee_\nu - \del_\nu\ee_\mu + \comm{\om_\mu}{\ee_\nu}
                                       - \comm{\om_\nu}{\ee_\mu} .
\end{eqnarray}
This is the same as (3.1) in \cite{matwel}, with Newton's constant
restored. It is the standard form of the Einstein Hilbert Lagrangian
in three spacetime dimensions. The variation of the bulk term with
respect to the dreibein and the spin connection, and with all boundary
terms neglected, implies that 
\begin{equation}
  \label{vac}
  \FF_{\mu\nu}=0 , \qquad \TT_{\mu\nu}=0.
\end{equation}
These are the vacuum Einstein equations, stating that the spacetime is
locally flat outside the matter sources. 

The coupling of the particles to the gravitational field is introduced
as follows. At each particle boundary $\bnd_\prt$, we add a \emph{mass
shell constraint} $\con_\prt$ to the Lagrangian, with a multiplier
$\mul_\prt$ as a coefficient. Formally, the mass shell constraints are
given by \eref{ham},
\begin{equation}
  \label{hol-con}
  \con_\prt = \frac{\hols_\prt - \cos(4\pi Gm_\prt)}{16\pi^2G^2}
\end{equation}
But the holonomy of the particle is now defined as a function of the
spin connection on the boundary,
\begin{equation}
  \label{bnd-hol}
  \hols_\prt = \ft12\Trrr{\Pexp \inti{\bnd_\prt} \dd \p \, \om_\p}.
\end{equation}
The path ordered exponential under the trace represents the Lorentz
rotation acting on a vector which is transported once around the
particle.

So far, this is a straightforward generalization of the previously
defined single particle model. The mass shell constraint serves in at
this level a two-fold purpose. It depends on the spin connection and
therefore provides a boundary term for the Lagrangian. One can show
that this, together with the point particle condition and the vacuum
Einstein equations, implies that the particle is moving on a geodesic,
and that there is a conical singularity on this geodesic. On the other
hand, the actually mass shell constraint, which is the equation of
motion for the multiplier $\mul_\prt$, implies that this geodesic is
timelike if the particle is massive and lightlike if it is massless,
and it fixed the deficit angle of the conical singularity. All this is
shown in \cite{matwel}, and it now applies to each particle
independently.

\subsubsection*{Asymptotical flatness}
For a closed universe, we now already have a well defined dynamical
system. The configuration space is spanned by the fields $\ee_\mu$ and
$\om_\mu$, and the multipliers $\mul_\prt$. The fields are defined on
a fixed space manifold $\N$, and they are subject to various boundary
conditions. We have a well defined Lagrangian, which depends on the
configuration variables and their time derivatives. The boundary term
$\lag_\infty$ is absent for a closed universe, and the space manifold
$\N$ is compact, so that the integral defining the bulk term
\eref{lag-0} is finite.

For an open universe, this is not the case. We have to impose some
fall off conditions at infinity, and we also have to add an
appropriate boundary term $\lag_\infty$. Let us require that, far away
from the particles, the spacetime looks like the gravitational field
of a single particle, whose mass $M$ is the total energy, and whose
spin $S$ is the total angular momentum of the universe. The metric at
infinity is then that of a \emph{spinning cone}, which we already
encountered in \sref{inter},
\begin{equation}
  \label{asm-ds}
  \dd s^2 = - (\dd T - 4 G S \, \dd \cdir)^2 + \dd R^2 
            + (1- 4 G M)^2 \, R^2 \, \dd \cdir^2.
\end{equation}
Due to the absence of physical degrees of freedom, or local
excitations of the gravitational field, we do not have to think about
the actual fall off behaviour of the metric at infinity.

The spinning cone is, up to coordinate transformations, the most
general solution to the vacuum Einstein equations in the considered
region of spacetime, which admits the definition of a centre of mass
frame. The condition for a centre of mass frame to exist is the same
as for the free particle system. The total momentum must be timelike.
For the gravitating particles, this total momentum is actually the
total holonomy. It is defined like the holonomy of the particles, but
with the spin connection integrated along the circle at infinity. If
the total holonomy is not timelike, then the universe does not even
even admit a proper causal structure \cite{gott,steif}. On the other
hand, if it is timelike, then it is always possible to choose
coordinates so that the metric becomes \eref{asm-ds}
\cite{matrev,djh}.

The region where the spinning cone metric applies is a
\emph{neighbourhood of infinity}. A neighbourhood of infinity in space
is a subset $\U\subset\N$, which has the topology of
$\RR_+\times\SP^1$, and whose complement is compact. A typical
neighbourhood of infinity is shown in \fref{mfld}. A neighbourhood of
infinity in spacetime is a subset of $\M=\RR\times\N$, so that at
each moment of time the intersection with the space manifold $\N$ at
that moment of time is a neighbourhood of infinity in space. So, the
fall off condition to be imposed on the metric is that, at each moment
of ADM time $t$, there exists a neighbourhood of infinity $\U$, and in
$\U$ it is possible to introduce a conical coordinate system
$(T,R,\cdir)$ so that \eref{asm-ds} holds.

Let us translate this into a fall off condition for the dreibein and
the spin connection. A possible representation of the spinning cone
metric is given by
\begin{eqnarray}
  \label{cone-ee-om}
  \ee_\mu &=& ( \del_\mu T + 4 G S \, \del_\mu \cdir ) \, \gam_0 
          + \del_\mu R \,\gam(\cdir) 
          + (1- 4GM) \, R \, \del_\mu \cdir \, \gam'(\cdir), \nwl  
  \om_\mu &=& - 2 G M \,  \del_\mu \cdir \, \gam_0. 
\end{eqnarray}
One can easily check that this implies $\FF_{\mu\nu}=0$ and
$\TT_{\mu\nu}=0$, so that the spinning cone is indeed a solution to
the vacuum Einstein equations. However, it is not possible to impose
this directly as a boundary condition on the basic field variables,
because it still involves time derivatives. But it turns out to be
sufficient to impose only the spatial components of \eref{cone-ee-om}
as boundary conditions at infinity.

The precise fall off condition is given as follows. At each moment of
ADM time $t$, there exists a neighbourhood of infinity $\U\subset\N$,
a pair of real numbers $M$ and $S$, and a set of conical coordinates
$(T,R,\cdir)$ on $\U$, so that
\begin{eqnarray}
   \label{asm-ee-om}
    \ee_i &=& ( \del_i T + 4 G S \, \del_i \cdir ) \, \gam_0 
          + \del_i R \,\gam(\cdir) 
          + (1- 4GM) \, R \, \del_i \cdir \, \gam'(\cdir), \nwl  
    \om_i &=& - 2 G M \,  \del_i \cdir \, \gam_0.
\end{eqnarray}
Note that we do not fix the conical coordinates in any way, or require
the conical time $T$ to be related in any way to the ADM time $T$. The
conical coordinates are instead considered as additional field
variables supported on $\U$, which are related to the dreibein and the
spin connection by this fall off condition. And it is also useful to
note that $M$ and $S$ are implicitly defined as functions of the
spatial components of the dreibein and the spin connection. 

Now, suppose that the dreibein and the spin connection satisfy this
condition at each moment of time. Is the bulk term \eref{lag-0} of the
Lagrangian then well defined? The last two terms are finite. The
torsion $\TT_{ij}$ and the curvature $\FF_{ij}$ both vanish on $\U$,
and therefore these terms have a compact support. What remains is the
first term, involving the time derivative. In $\U$, it is given by
\begin{equation}
  \label{asm-lag}
  \frac1{8\pi G} \, \eps^{ij} \, \Trr{\dot \om_i \, \ee_j } 
  = \frac{M}{2\pi} \, \eps^{ij} \, \del_i [
                ( \dot T  + 4 G S \, \dot \cdir) 
                  \, \del_j \cdir ].
\end{equation}
Here we added a total time derivative, which simplifies the result
slightly. This is of course allowed, as we may at any time add a total
time derivative to the Lagrangian. So, within the neighbourhood of
infinity $\U$, the integrand in \eref{lag-0} can obviously be written
as total spatial derivative. To obtain a well defined finite
Lagrangian, we just have to subtract the resulting boundary term at
infinity. Hence, the appropriate term $\lag_\infty$ to be added in
\eref{lag-tot} is
\begin{equation}
  \label{lag-inf}
  \lag_\infty = - \frac M{2\pi} \inti\infty \dd s \, \del_s\cdir \, 
            ( \dot T + 4 G S \, \dot \cdir ) .
\end{equation}
The notation means that the integral is evaluated along a circle at
infinity, in counter clockwise direction. More precisely, the total
Lagrangian is defined as follows. We first integrate \eref{lag-0} over
a connected compact subset $\N_0\subset\N$, which is sufficiently
large, so that $\N=\N_0\cup\U$, and so that the boundary
$\del\N_0$ is a loop in $\U$. Then we add the term \eref{lag-inf},
integrated along the boundary $\del\N_0$. According to \eref{asm-lag},
the result is invariant under deformations of the boundary $\del\N_0$.
Thus we may finally also take the limit $\N_0\to\N$, where the
boundary $\del\N_0$ becomes a circle at infinity.

With the fall off condition imposed and the boundary term added, the
Lagrangian becomes finite. However, it is no longer only a function of
the dreibein and the spin connection. The conical coordinates and
their time derivatives show up explicitly in the boundary term. We have
to include them into the definition of the configuration space. For a
given geometry of the ADM surface, the conical coordinates are only
fixed up to a time translation $T\mapsto T-\tpar$, and a spatial
rotation $\cdir\mapsto\cdir+\rpar$. These are the Killing symmetries
of the spinning cone. So, we have to add two extra degrees of freedom
to the configuration space, in order to obtain a well defined
Lagrangian. 

This is a typical feature of general relativity in the Lagrangian, or
Hamiltonian framework \cite{nico}. The additional degrees of freedom
represent a \emph{reference frame}, and the Killing symmetries are the
possible translations and rotation of the rest of the universe, thus
in this case the particles, with respect to this reference frame. By
definition, this reference frame coincides with the centre of mass
frame of the universe. It is this feature which is responsible for the
various ambiguities associated with the reference frame to disappear,
when we go over from a single to a multi particle model. In
\cite{matwel}, we found that the boundary term at infinity is not
uniquely determined by the condition that the Lagrangian should be
finite. 

To be precise, this is also not the case here. We are still free to
add any function of $M$ and $S$ to the Lagrangian. However, one can
show that the given boundary term is uniquely fixed by the addition
condition that the reference frame must not be translated or rotated
when time evolves. If we derive the equations of motion and take into
account that the conical coordinates are now also configuration
variables, we find in addition to the vacuum Einstein equations the
following time evolution equations for the conical coordinates, 
\begin{eqnarray}
  \label{asm-dot}
  \ee_t &=& ( \dot T + 4 G S \, \dot \cdir ) \, \gam_0 
          + \dot R \,\gam(\cdir) 
          + (1- 4GM) \, R \, \dot \cdir \, \gam'(\cdir), \nwl  
  \om_t &=& - 2 G M \,  \dot \cdir \, \gam_0. 
\end{eqnarray}
Clearly, these are just the time components of \eref{cone-ee-om}. If
we would add a function of $M$ or $S$ to the Lagrangian, then there
would be additional time translation or rotations of the conical
coordinates modifying the equations \eref{asm-dot}. Physically, this
would mean that the reference frame is translated or rotated while the
universe evolves, and this is of course not what we want. We therefore
conclude that the Lagrangian is uniquely fixed by the condition that
it must be finite, and that it has to provide the correct equations of
motion for the reference frame. 

\subsubsection*{Hamiltonian formulation}
Let us now go over from the Lagrangian to the Hamiltonian formulation.
This is actually only a formal step, because the Lagrangian is already
of first order in time derivatives. The phase space is spanned by the
following field variables on $\N$,
\begin{itemize}
\item[-] the spatial components $\ee_i$ and $\om_i$ of the dreibein
  and the spin connection, 
\item[-] the conical coordinates $(T,R,\cdir)$ defined in some
  neighbourhood of infinity $\U$.
\end{itemize}
They are subject to the following boundary conditions,
\begin{itemize}
\item[-] in the interior of $\N$, the normal vector
  $\norm=\ft12\eps^{ij}\comm{\ee_i}{\ee_j}$ is negative timelike,
\item[-] on each particle boundary $\bnd_\prt$, the tangent component
  of the dreibein $\ee_\p$ vanishes,
\item[-] in the neighbourhood of infinity $\U$, the relation
  \eref{asm-ee-om} is satisfied for some $M$ and $S$.
\end{itemize}
From the total Lagrangian, we can read off the symplectic potential
and the Hamiltonian. The symplectic potential is the one-form $\pot$,
which is given by those terms of $\lag$ which are linear in time
derivatives. There is a contribution from the bulk term \eref{lag-0},
and from the boundary term \eref{lag-inf} at infinity,
\begin{equation}
  \label{pot-ee-om}
  \pot = \frac1{8\pi G} \inti{\N} \dd^2x \, \eps^{ij} \, 
                  \Trr{\dd\om_i \, \ee_j } 
      - \frac M{2\pi} \inti\infty \dd s \, \del_s \cdir
            \, ( \dd T  +  4 G S \, \dd \cdir  ).
\end{equation}
Here and in the following, the upright letter $\dd$ always denotes the
exterior derivative on the phase space, except when it appears
immediately behind an integral, where it denotes the measure on $\N$
or along a curve on $\N$.  The Hamiltonian $\ham$ is defined by those
terms on $\lag$ which do not contain time derivatives. There is a
contribution from the bulk term, and one contribution from each
particle boundary,
\begin{equation}
  \label{ham-ee-om}
  \ham = \sumi\prt \mul_\prt \, \con_\prt
         - \frac1{16\pi G} \inti{\N} \dd^2x \, \eps^{ij} \, 
                \Trr{\om_t \, \TT_{ij} + \ee_t \, \FF_{ij} }.
\end{equation}
This is obviously a linear combination of primary constraints, with
the multipliers $\ee_t$, $\om_t$, and $\mul_\prt$ as coefficients.
These are not considered as phase space variables, but rather as free
parameters entering the Hamiltonian. The \emph{kinematical}
constraints are those associated with the multipliers $\ee_t$ and
$\om_t$, 
\begin{equation}
  \label{kin-con}
  \FF_{ij} \approx 0 , \qquad 
  \TT_{ij} \approx 0 .
\end{equation}
The \emph{dynamical} constraints are the mass shell constraints,
associated with the multipliers $\mul_\prt$,
\begin{equation}
  \label{dyn-con}
  \con_\prt = \frac{\hols_\prt - \cos(4\pi Gm_\prt)}{16\pi^2G^2}
              \approx 0. 
\end{equation}
The phase space reduction will now be applied to the kinematical
constraints. We are going to solve the constraints \eref{kin-con}, and
divide out the associated gauge symmetries. The mass shell constraints
will not be affected by this reduction. It is then already obvious
that the reduced Hamiltonian is a linear combination of the mass
shell constraints, as all other terms in \eref{ham-ee-om} are zero
when the kinematical constraints are solved. We may therefore in
following only focus on the symplectic structure.

\subsubsection*{Minkowski coordinates}
The general solution to the kinematical constraints \eref{kin-con} is
well known. Within any simply connected subset of the space manifold,
it can be parameterized by a group valued field $\gx:\N\to\grpSL(2)$
and a vector field $\fx:\N\to\algsl(2)$, so that
\cite{witten,matrev}
\begin{equation}
  \label{loc-sol}
  \om_i = \gx^{-1} \del_i \gx , \qquad
  \ee_i = \gx^{-1} \del_i \fx \, \gx.
\end{equation}
It follows that the metric is given by
\begin{equation}
  \label{emb-metric}
  g_{ij}= \ft12\Trr{\ee_i\ee_j} 
      = \ft12\Trr{\del_i\fx\del_j\fx},
\end{equation}
and the normal vector becomes 
\begin{equation}
  \label{norm-emb}
  \norm  = \ft12 \eps^{ij} \, \comm{\ee_i}{\ee_j}
   = \ft12 \eps^{ij} \, \gx^{-1}  
        \comm{\del_i \fx}{\del_j \fx}  \, \gx . 
\end{equation}
The field $\fx$ defines an \emph{isometric embedding} of the space
manifold into Minkowski space. This is just a different way to define
the Minkowski charts that we already introduced in \sref{inter}. The
components of $\fx$ are the Minkowski coordinates. The field $\gx$
represents a Lorentz rotation at each point in space. It tells us how
the local frame defined by the dreibein at a point in space is related
to the Minkowski frame defined in the respective chart.

\begin{figure}[t]
  \begin{center}
    \epsfbox{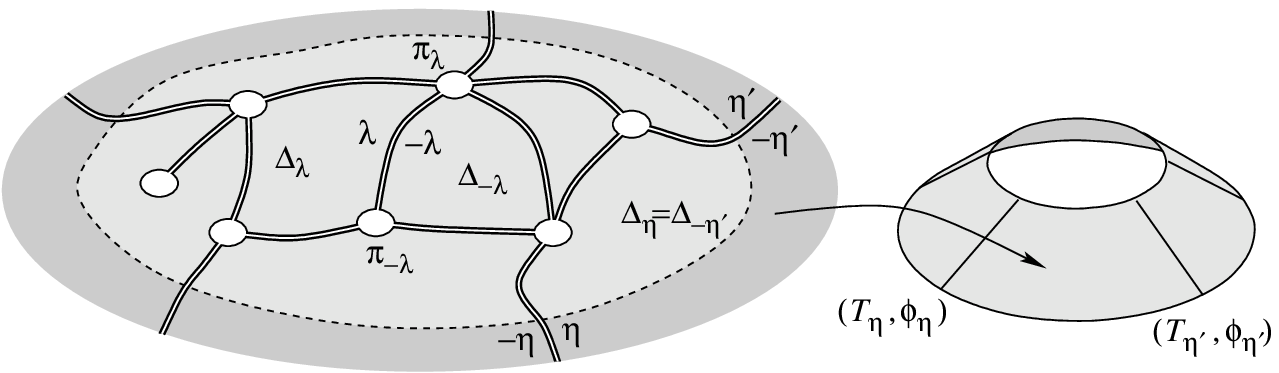}
    \xcaption{The triangulated space manifold $\N$ is covered by a
    collection of simply connected polygons $\pol\in\pols$. A
    neighbourhood of infinity $\U$ is divided into simply connected
    segments $\U\cap\pol$, one for each non-compact polygon
    $\pol\in\pols_\infty$. The links are assumed to be spacetime
    geodesics, and in $\U$ the external links $\ecut\in\cuts_\infty$
    are spatial half lines in the spinning cone, which conical
    coordinates $T\to T_\ecut$ and $\cdir\to\cdir_\ecut$.}
    \label{ispc}
  \end{center}
\end{figure}

If we want to parameterize the solutions to the kinematical
constraints globally on $\N$, then we have to cover the space manifold
by a collection of simply connected subsets. We therefore introduce a
\emph{triangulation}, in the very same way as in \sref{free}. The only
difference is that the links $\cut\in\cuts$ are now oriented curves on
$\N$, and the vertices are the particle boundaries $\bnd_\prt$. The
internal links $\cut\in\cuts_0$ are curves beginning at a particle
boundary $\bnd_{\prt_{-\cut}}$, and ending at a particle boundary
$\bnd_{\prt_\cut}$. The external links $\ecut\in\cuts_\infty$ extend
from a particle boundary $\bnd_{\prt_{-\ecut}}$ to infinity. The
polygons $\pol\in\pols$ are simply connected, closed subsets of $\N$,
overlapping along the links. A typical such triangulation is shown in
\fref{ispc}.

Now, suppose we are given a field configuration $\ee_i$ and $\om_i$ on
$\N$, satisfying the kinematical constraints $\FF_{ij}=0$ and
$\TT_{ij}=0$, and a graph $\cuts$ defining a triangulation of $\N$. We
can then introduce a pair of fields $\gx_\pol:\pol\to\grpSL(2)$ and
$\fx_\pol:\pol\to\algsl(2)$ in each polygon, so that
\begin{equation}
  \label{emb-sol}
  \om_i \big|_\pol = \gx\inv_\pol \del_i \gx_\pol, \qquad
  \ee_i \big|_\pol = \gx\inv_\pol \del_i \fx_\pol \, \gx_\pol.
\end{equation}
The image $\fx_\pol(\pol)$ becomes a spacelike surface embedded into
Minkowski space. We call this the \emph{embedded polygon}. The
embedded polygons are those in \fref{ipol}, provided that we impose
some further boundary condition on the fields $\gx_\pol$ and
$\fx_\pol$.

The first observation is that the fields $\gx_\pol$ and $\fx_\pol$ are
not uniquely determined by the dreibein and the spin connection. The
right hand sides of \eref{emb-sol} are invariant under the
transformations
\begin{equation}
  \label{pol-pnc}
  \gx_\pol \mapsto \blpar_\pol \, \gx_\pol , \qquad 
  \fx_\pol \mapsto \blpar_\pol \,
                    (\fx_\pol-\btpar_\pol) \, \blpar\inv_\pol, 
\end{equation}
where $\blpar_\pol\in\grpSL(2)$ and $\btpar_\pol\in\algsl(2)$ are two
independent \emph{constants} for each polygon $\pol\in\pols$. This is
a Poincar\'e transformation, thus a coordinate transformation in the
respective coordinate charts. On the embedded polygon
$\fx_\pol(\pol)$, it acts as a Lorentz rotation by $\blpar_\pol$, and
a translation by $\btpar_\pol$. Clearly, these are the redundancy
transformations already considered in \sref{inter}.

\subsubsection*{Link variables}
Referring to the embedding fields $\gx_\pol$ and $\fx_\pol$, we can now
also define the link variables. When considered as a subset of $\N$, a
polygon $\pol$ is not only bounded by the edges $\cut\in\cuts_\pol$,
but also by some segments of the particles boundaries
$\pol\cap\bnd_\prt$. The point particle condition \eref{prt-con}
implies that the field $\fx_\pol$ is constant on these segments,
\begin{equation}
  \label{prt-pos}
  \ee_\p \big|_{\pol\cap\bnd_\prt} = 
  \gx\inv_\pol \del_\p \fx_\pol \, \gx_\pol  \big|_{\bnd_\prt} 
        = 0 \follows
   \fx_\pol  \big|_{\bnd_\prt} = \pos_{\prt,\pol} . 
\end{equation}
The constant $\pos_{\prt,\pol}$ is called the \emph{position} of the
particle $\prt$ in the polygon $\pol$. There are as many positions
$\pos_{\prt,\pol}$ of the particle $\prt$ as there are polygons $\pol$
adjacent to $\prt$, and these position are the corners of the embedded
polygons in \fref{ipol}. We also see that the point particle condition
indeed implies that the particles are pointlike.

Now, let us assume that all the links are geodesics in spacetime. This
can be regarded as a gauge condition in general relativity, restricting
the possible foliations and also the embeddings of the triangulated
space manifold $\N$ into the spacetime. If this is the case, then the
image $\fx_{\pol_\ecut}(\ecut)$ of each link is a straight line in
Minkowski space. Let us introduce on $\cut$ an affine parameter $s$.
For internal links $\cut\in\cuts_0$, it is chosen so that
$s\in[0,1]$, and $\cut(0)=\prt_{-\cut}$ and $\cut(1)=\prt_\cut$. For
external links, we choose $s\in[0,infty)$, and for reversed external
links $s\in(-\infty,0]$, and in these cases $s$ measures the proper
length. Then we define
\begin{equation}
  \label{dis-emb-ecut}
  \dis_\ecut = \del_s \fx_{\pol_\ecut}, 
   \qquad \ecut\in\cuts. 
\end{equation}
The vector $\dis_\cut$ is a constant, because the link $\cut$ is a
geodesic. For external links, it is a spacelike unit vector. For
internal links, it is the relative position vector,
\begin{equation}
  \label{dis-emb-cut}
  \dis_\cut = \pos_{\prt_\cut,\pol_\cut} 
            - \pos_{\prt_{-\cut},\pol_\cut}, 
   \qquad \cut\in\cuts_0.  
\end{equation}
The transition functions are defined as follows. Consider a pair of
links $\pm\ecut$, and the adjacent polygons $\pol_{\pm\ecut}$. The
both links represent the same curve on $\N$, and therefore the
dreibein and the spin connection are equal on $\cut$ and $-\cut$. But
the fields $\gx_{\pol_\cut}$ and $\fx_{\pol_\cut}$ on $\cut$, and
$\gx_{\pol_{-\cut}}$ and $\fx_{\pol_{-\cut}}$ on $-\cut$ are in
general different. On the other hand, we have seen above that if two
embeddings define the same dreibein and the same spin connection, then
they are related by a Poincar\'e transformation of the form
\eref{pol-pnc}. Hence, for each pair of links $\pm\ecut$ there exists
a constant $\chol_\cut\in\grpSL(2)$ and a constant
$\cang_\cut\in\algsl(2)$ so that
\begin{equation}
  \label{cut-con}
  \gx_{\pol_{-\cut}} = \chol_\cut \, \gx_{\pol_\cut}, \qquad 
  \fx_{\pol_{-\cut}} = \chol_\cut \,
        ( \fx_{\pol_\cut} - \cang_\cut) \, \chol\inv_\cut ,
                        \qquad \cut\in\cuts.                    
\end{equation}
Exchanging $\pol_\cut$ and $\pol_{-\cut}$ in the first equation, and
differentiating the second equation with respect to the curve
parameter $s$, where we have to take into account that the reversed
link is oriented into the opposite direction, we derive the following
relations,
\begin{equation}
  \label{cut-inv-con}
  \chol_{-\cut} = \chol\inv_\cut , \qquad
  \dis_{-\cut} = - \chol_\cut \, \dis_\cut \, \chol\inv_\cut, \qquad 
   \cut\in\cuts.
\end{equation}
Hence $\chol_\cut$ is the transition function, and the relation
\eref{cut-inv} is recovered for both internal and external links. The
translational component $\cang_\cut$ of the transition function has
never shown up in \sref{inter}, because there are link variables that
refer to the absolute positions of the polygon in the embedding
Minkowski space.

\subsubsection*{The conical frame}
There is one more gauge condition that we have to impose. The external
links must be spatial half lines in the spinning cone. Hence, if we
follow a curve $\ecut\in\cuts_\infty$ on $\N$ to infinity, then at
some point the curves enters the neighbourhood of infinity $\U$, and
in the limit the conical coordinates are given by 
\begin{equation}
  \label{T-cdir-cone}
  T \to T_\ecut , \qquad \cdir \to \cdir_\ecut,
   \qquad  R\to\infty, \qquad \ecut\in\cuts_\infty. 
\end{equation}
Moreover, there is also a further condition to be imposed on the
embedding of the non-compact polygons. Let us choose the neighbourhood
of infinity $\U$ so that, for each non-compact polygon
$\pol\in\pols_\infty$, the intersection $\U\cap\pol$ is a simply
connected segment, as indicated in \fref{ispc}. This is always
possible, because given any neighbourhood of infinity we can always go
over to a smaller one which has this property.

The segment $\U\cap\pol$ is then not only embedded into the spinning
cone, but also into Minkowski space. The first embedding is provided
by the conical coordinates $(T,R,\cdir)$, the second by the field
$\fx_\pol$. And we still have the freedom to apply a Poincar\'e
transformation to $\fx_\pol$. Using this, we can always achieve that
the two embeddings are related by a local isometry of the form
\eref{iso}. There is then the following relation between the conical
coordinates $(T,R,\cdir)$, and the Minkowski coordinates $\fx_\pol$ in
$\U\cap\pol$,
\begin{equation}
  \label{ff-cone}
  \fx_\pol = (T + \atau_\pol(\cdir)) \, \gam_0 
          + R \, \gam(\cdir - \adir_\pol(\cdir) ),
\end{equation}
where $\atau_\pol$ and $\adir_\pol$ are two linear functions
satisfying
\begin{equation}
  \label{asm-atau-adir}
  \atau_\pol'(\cdir) = 4 G S , \qquad
  \adir_\pol'(\cdir) = 4 G M . 
\end{equation}
For the fall off condition \eref{asm-ee-om} to be satisfied, the field
$\gx_\pol$ is then necessarily given by
\begin{equation}
  \label{gg-cone}
  \gx_\pol = \expo{-\adir_\pol(\cdir)\gam_0/2}.  
\end{equation}
Using this, we can finally also derive the special transition
functions $\chol_{\pm\ecut}$, and the unit vectors $\dis_{\pm\ecut}$
for the external links $\ecut\in\cuts_\infty$. For this purpose, we
have to evaluate the fields $\gx_\pol$ and $\fx_\pol$ on the external
links $\pm\ecut$, in the limit $R\to\infty$. What we find is
\begin{eqnarray}
  \label{gg-ff-ecut}
  \gx_{\pol_{-\ecut}} &\to&  \expo{\bdir_\ecut^-\gam_0/2}, \qquad
  \fx_{\pol_{-\ecut}} \to 
       (T_\ecut + \atau_{\pol_{-\ecut}}(\cdir_\ecut)) \, \gam_0 
    + R \, \gam(\cdir_\ecut + \bdir_\ecut^- ),
   \nwl 
   \gx_{\pol_{\ecut}}  &\to&  \expo{\bdir_\ecut^+\gam_0/2}, \qquad
   \fx_{\pol_{\ecut}}   \to 
       (T_\ecut + \atau_{\pol_\ecut}(\cdir_\ecut)) \, \gam_0 
    + R \, \gam(\cdir_\ecut + \bdir_\ecut^+ ),
\end{eqnarray}
where we introduced the \emph{deviations} already defined in
\eref{adir-ecut}, thus
\begin{equation}
  \label{adir-ecut-emb}
  \bdir^+_\ecut = - \adir_{\pol_\ecut}(\cdir_\ecut), \qquad 
   \bdir^-_\ecut = - \adir_{\pol_{-\ecut}}(\cdir_\ecut) . 
\end{equation}
Now, consider first the vectors $\dis_{\pm\ecut}$, as defined in
\eref{dis-emb-cut}. In the limit $R\to0$, the conical coordinate $R$
asymptotically defines the proper length on a spatial half line, and
therefore we can use $s=\pm R$ as a parameter on the links $\pm\ecut$.
It then follows from \eref{gg-ff-ecut} that
\begin{equation}
  \label{ecut-dis-emb}
   \dis_\ecut = \gam(\cdir_\ecut + \bdir^+_\ecut), \qquad
   \dis_{-\ecut} = - \gam(\cdir_\ecut + \bdir^-_\ecut) , \qquad
   \ecut\in\cuts_\infty,
\end{equation}
which is the same as \eref{ecut-vec}. To find the transition functions
$\chol_{\pm\ecut}$, we have to use the defining relation
\eref{cut-con}, which can also be evaluated in the limit $R\to0$. What
we get is 
\begin{equation}
  \label{ecut-trans}
  \chol_\ecut = \gx_{\pol_{-\ecut}} \, \gx\inv_{\pol_\ecut} 
     = \expo{-(\bdir_\ecut^+ - \bdir_\ecut^-)\gam_0/2}.
\end{equation}
We can therefore make the definition 
\begin{equation}
  \label{M-emb}
  8\pi G M_\ecut = \bdir_\ecut^+ - \bdir_\ecut^- 
   \follows 
  \chol_\ecut = \expo{- 4\pi GM_\ecut\gam_0}, 
\end{equation} 
and recover the definition \eref{ecut-rot} of the energy $M_\ecut$,
and also the second equation in \eref{bdir-eqs}. The first equation in
\eref{bdir-eqs} follows directly from the definition
\eref{adir-ecut-emb}. With $\pol=\pol_{\ecut}=\pol_{-\ecut'}$ we have
\begin{equation}
  \label{angle-emb}
  \bdir^-_{\ecut'} - \bdir^+_\ecut =
    - \adir_{\pol}(\cdir_{\ecut'}) 
    + \adir_{\pol}(\cdir_\ecut) =
    - 4 GM \, (\cdir_{\ecut'} - \cdir_\ecut).  
\end{equation}
Finally, what remains is the average condition \eref{bdir-ave}. In
\sref{inter} we have seen that this is equivalent to the condition
that the integral of the functions $\adir(\cdir)$ over the circle at
infinity vanishes, hence
\begin{equation}
  \label{dev-emb}
  \inti{\infty} \dd s \, \del_s \cdir \, \adir(\cdir) = 
  \sumi{\pol\in\pols_\infty} \, \inti{\infty\cap\pol} \, 
   \dd s \, \del_s \cdir \, \adir_\pol(\cdir) = 0 .  
\end{equation}
This is the same as \eref{deviation-1}, we only changed the notation
slightly, because now we have consider this as an integral along a
curve on $\N$, and $\cdir$ is a field variable on $\U\subset\N$.  The
notation $\infty\cap\pol$ means that the integral is evaluated along
the segment of the circle at infinity which belongs to the non-compact
polygon $\pol$. 

\subsubsection*{The symplectic structure}
With these formulas at hand, we can now compute the symplectic
structure on the subspace defined by the kinematical constraints
\eref{kin-con}. The embedding fields $\gx_\pol$ and $\fx_\pol$ can be
used as coordinates on this subspace. To derive the symplectic
potential, we insert the parameterization \eref{emb-sol} into the
original expression \eref{pot-ee-om},
\begin{equation}
  \label{pot-ee-om-x}
  \pot = \frac1{8\pi G} \inti{\N} \dd^2x \, \eps^{ij} \, 
                  \Trr{\dd\om_i \, \ee_j } 
      - \frac M{2\pi} \inti\infty \dd s \, \del_s \cdir \, 
            ( \dd  T  +  4 G S \, \dd \cdir ) .
\end{equation}
For simplicity, let us first ignore all boundary terms at infinity, or
consider a closed universe where these are absent. The symplectic
potential is then given by the bulk term only. To write this as a
function of the embedding fields $\gx_\pol$ and $\fx_\pol$, we first
have to split the integral into a sum of integrals over the individual
polygons,
\begin{equation}
  \label{pot-pol-sum}
  \pot_0 = \frac1{8\pi G} \, \sumi{\pol\in\pols}  
           \, \inti{\pol} \dd^2x \, \eps^{ij} \, 
                  \Trr{\dd\om_i \, \ee_j } 
        = \frac1{8\pi G} \, \sumi{\pol\in\pols} 
            \, \inti{\pol} \dd^2x \, \eps^{ij} \,
         \del_i\Trr{ \dd \gx_\pol\, \gx\inv_\pol \, \del_j \fx_\pol}. 
\end{equation}
Here we have inserted \eref{emb-sol}, and performed some simple
algebraic manipulations. The integrand is thus a total derivative, and
we can write the result as a boundary integral,
\begin{equation}
  \label{pot-pol-bnd}
  \pot_0 = \frac1{8\pi G} \, \sumi{\pol\in\pols}
        \, \inti{\del\pol} \dd s \,
          \Trr{ \dd \gx_{\pol} \, \gx\inv_{\pol} \del_s \fx_{\pol} }.
\end{equation}
The boundary $\del\pol$ of the polygon $\pol$ is always traversed in
counter clockwise direction. It consists of the edges
$\cut\in\cuts_\pol$, and the segments of the particles boundaries
belonging to the polygon $\pol$. The particle boundaries, however, do
not contribute to the integral, because according to the point
particle condition $\fx_\pol$ is constant, and therefore the integrand
vanishes. What remains is
\begin{equation}
  \label{pot-pol-edges}
  \pot_0 = \frac1{8\pi G} \, \sumi{\pol\in\pols} 
                   \,\,\,     \sumi{\cut\in\cuts_\pol} 
        \, \inti{\cut} \dd s \,
          \Trr{ \dd \gx_{\pol} \, 
                   \gx\inv_{\pol} \del_s \fx_{\pol} }.
\end{equation}
Now, we may equally well sum over all links $\cut\in\cuts$, because
each link appears exactly once as a boundary of a polygon, 
\begin{equation}
  \label{pot-cut-sum}
  \pot_0 = \frac1{8\pi G} \, \sumi{\cut\in\cuts}  \, 
        \inti{\cut} \dd s \,
          \Trr{ \dd \gx_{\pol_\cut} \, 
                   \gx\inv_{\pol_\cut} \del_s \fx_{\pol_\cut} }.
\end{equation}
For a closed universe, all links are internal links, thus
$\cuts=\cuts_0$. Using this and the decomposition
$\cuts_0=\cuts_+\cap\cuts_-$, we can also write 
\begin{equation}
  \label{pot-cut-inv}
  \pot_0 = \frac1{8\pi G} \, \sumi{\cut\in\cuts_+} 
           \, \inti{\cut} \dd s \,
          \Trr{ \dd \gx_{\pol_\cut} \, \gx\inv_{\pol_\cut} 
                 \, \del_s \fx_{\pol_\cut} 
               -\dd \gx_{\pol_{-\cut}} \, \gx\inv_{\pol_{-\cut}} 
                 \, \del_s \fx_{\pol_{-\cut}}  }.
\end{equation}
We can then use the defining relations \eref{cut-con} of the
transition functions, to express $\gx_{\pol_{-\cut}}$ and
$\fx_{\pol_{-\cut}}$ in terms of $\gx_{\pol_{\cut}}$ and
$\fx_{\pol_{\cut}}$. This gives
\begin{equation}
  \label{pot-cut-chol}
   \pot_0 = - \frac1{8\pi G} \, \sumi{\cut\in\cuts_+} \, 
            \inti{\cut} \dd s \,
   \Trr{ \chol\inv_\cut \, \dd\chol_\cut \, \del_s \fx_{\pol_\cut}}. 
\end{equation}
Now, the transition function $\chol_\cut$ is a constant, and therefore
the integrand is again a total derivative. Using this and the
definition \eref{dis-emb-cut} of the relative position vector
$\dis_\cut$, we get
\begin{equation}
  \label{pot-0}
  \pot_0 = - \frac1{8\pi G} \, \sumi{\cut\in\cuts_+} 
   \Trr{ \chol\inv_\cut \, \dd\chol_\cut \,  \dis_\cut }.
\end{equation}
So, for a closed universe we have shown that the symplectic potential,
derived from the Einstein Hilbert action, is indeed given by the
expression \eref{pot-tot}. The terms associated with the external
links are in this case not present.

For an open universe, we perform the same calculation, but we also
have to take into account some boundary terms at infinity, and the
special properties of the external links. Let us first consider the
external links, still ignoring the boundary terms at infinity. The
total symplectic potential can then still be written as
\eref{pot-cut-sum}, but in \eref{pot-cut-inv} and consequently also in
\eref{pot-cut-chol} we have to sum over all external links as well. We
shall now show, however, that the external links do not contribute to
this sum.  Consider an external link $\ecut\in\cuts_\infty$, and the
corresponding term in \eref{pot-cut-chol}, which is given by
\begin{equation}
  \label{pot-ecut}
     - \frac1{8\pi G} \inti{\ecut} \dd s \,
   \Trr{ \chol\inv_\ecut \, \dd\chol_\ecut \, \del_s \fx_{\pol_\ecut}}. 
\end{equation}
The integrand is still a total derivative, but the integral cannot be
expressed like \eref{pot-0}, because a relative position vector
assigned to the external link $\ecut$ does not exist. However, the
integrand is actually zero. This can be seen as follows. According to
\eref{ecut-trans}, the transition function $\chol_\ecut$ is an element
of the subgroup $\grpSO(2)$, which implies that the term
$\chol\inv_\ecut\,\dd\chol_\ecut$ is proportional to $\gam_0$. On the
other hand, according to \eref{dis-emb-ecut} and \eref{ecut-dis-emb},
the derivative $\del_s\fx_{\pol_\ecut}$ is orthogonal to $\gam_0$, and
thus the integrand is zero.

So, there is no contribution to the symplectic potential from the
external links. The only additional terms for an open universe are the
boundary terms at spatial infinity. There are two types of such
boundary terms. Let us first consider the boundary term in
\eref{pot-ee-om-x}. It is an integral along the circle at infinity. We
split it into segments belonging to the external polygons
$\pol\in\pols_\infty$,
\begin{equation}
  \label{pot-inf-1}
   - \frac1{8\pi G} \sum_{\pol\in\pols_\infty}
         \inti{\infty\cap\pol} \dd s \, 4 G M \, \del_s \cdir 
          \,  (  \dd  T  + 4 G S \, \dd \cdir).
\end{equation}
Remember that the abbreviated notation $\infty\cap\pol$ means that the
integral in evaluated along the circle segment which defines the
boundary at infinity of the non-compact polygon $\pol\in\pols_\infty$.
The integrand is written in this particular form, because according to
\eref{asm-atau-adir} we have
\begin{equation}
  \label{cdir-mdir-inf}
  4 G M  \, \del_s \cdir = \del_s \adir_\pol(\cdir).
\end{equation}
Hence, \eref{pot-inf-1} is equal to 
\begin{equation}
  \label{pot-inf-2}
   -  \frac1{8\pi G} \sum_{\pol\in\pols_\infty}
         \inti{\infty\cap\pol} \dd s \,
              \del_s \adir_\pol(\cdir) \,
               (  \dd  T  + 4 G S \, \dd \cdir).
\end{equation}
There is then a second boundary term at spatial infinity, which arises
from the boundary integral \eref{pot-pol-bnd}. The boundary of each
non-compact polygon also includes a segment of the circle at infinity,
which we did not take into account so far. Hence, yet another
contribution to the symplectic potential is
\begin{equation}
  \label{pot-inf-3}
   \frac1{8\pi G} \sum_{\pol\in\pols_\infty} 
      \inti{\infty\cap\pol} \dd s \,
     \Trr{ \dd \gx_{\pol} \, \gx\inv_{\pol} \del_s \fx_{\pol} }.
\end{equation}
The path along which this integral is evaluated lies entirely within
the segment $\U\cap\pol$ of the neighbourhood of infinity. We can
therefore insert the expressions \eref{ff-cone} for $\fx_\pol$ and
\eref{gg-cone} for $\gx_\pol$. Using this, one finds that
\eref{pot-inf-3} is equal to
\begin{equation}
  \label{pot-inf-4}
   \frac1{8\pi G}  \sum_{\pol\in\pols_\infty}
         \inti{\infty\cap\pol}\dd s \,\,
     \dd\left( \adir( \cdir ) \right) \, 
        ( \del_s T + 4 G S \, \del_s \cdir ) .
\end{equation}
Let us add to this a total exterior derivative, which we are always
allowed to, as the only relevant object is the symplectic two-form
$\sym=\dd\pot$. Then we have
\begin{equation}
  \label{pot-inf-5}
   - \frac1{8\pi G}  \sum_{\pol\in\pols_\infty}
         \inti{\infty\cap\pol}\dd s \,
       \adir_\pol( \cdir ) \, [ 
        \del_s ( \dd T + 4 G S \, \dd \cdir ) 
       + 4 G \, \dd S \, \del_s \cdir ]  .
\end{equation}
The term proportional to $\dd S$ is the expression which is set to
zero in \eref{dev-emb}. It therefore vanishes. What remains is,
together with \eref{pot-inf-2}, again a total derivative,
\begin{equation}
  \label{pot-inf-6}
   \pot_\infty = - \frac1{8\pi G}  \sum_{\pol\in\pols_\infty}
         \inti{\infty\cap\pol}\dd s \, \del_s [
       \adir_\pol( \cdir ) \, 
       ( \dd T + 4 G S \, \dd \cdir ) ].
\end{equation}
The integral can be evaluated and becomes
\begin{equation}
  \label{pot-inf-7}
   - \frac1{8\pi G} \,\, \sumi{\pol\in\pols_\infty}
      [ \adir_\pol( \cdir_{\ecut'} ) \, 
            ( \dd T_{\ecut'} + 4 G S \, \dd \cdir_{\ecut'} ) 
         -  \adir_\pol( \cdir_{\ecut} ) \, 
            ( \dd T_{\ecut} + 4 G S \, \dd \cdir_{\ecut} )  ].
\end{equation}
Here we used that the segment of the circle at infinity belonging to
the polygon $\pol$ is bounded by the two external link
$\ecut\in\cuts_\infty$ and $-\ecut\in\cuts_{-\infty}$, where
$\pol=\pol_\ecut=\pol_{-\ecut'}$. And we inserted the conical
coordinates on these edges, which are given by \eref{T-cdir-cone}.
Finally, we can use the definitions \eref{adir-ecut-emb} and rearrange
the summation, which gives
\begin{equation}
  \label{pot-inf-8}
   \pot_\infty = - \frac1{8\pi G} \,  \sumi{\ecut\in\cuts_\infty}
       ( \bdir_\ecut^+ - \bdir_\ecut^- ) \, 
            ( \dd T_\ecut + 4 G S \, \dd \cdir_\ecut ) .
\end{equation}
And now, we just have to use \eref{M-emb} and make another
definition, namely
\begin{equation}
  \label{L-emb}
  L_\ecut = - 4 G S \, M_\ecut,
\end{equation}
to obtain the final result. The contribution to the symplectic
potential from the boundary at infinity is, again up to a total
derivative, 
\begin{equation}
  \label{pot-infty}
   \pot_\infty = - \sumi{\ecut\in\cuts_\infty}
       M_\ecut \, ( \dd T_\ecut + 4 G S \, \dd \cdir_\ecut ) 
   = \sumi{\ecut\in\cuts_\infty}
      ( T_\ecut \,  \dd M_\ecut + L_\ecut \, \dd \cdir_\ecut ). 
\end{equation}
This completes the technical part of the derivation. The sum
$\pot=\pot_\infty+\pot_0$ is the same as \eref{pot-tot}. Now we just
have to sort out what this equality actually means, and how it fits
together with the definitions made in \sref{phase}. First of all, it
follows that all those variations of the embedding fields $\gx_\pol$
and $\fx_\pol$ are gauge symmetries of general relativity, which
preserve the values of link variables showing up in \eref{pot-0} and
\eref{pot-infty}. In other words, the link variables are the
\emph{observables} of the particle model, in the sense that all
physical degrees of freedom are encoded in these variables. This is
what we had to assume in the beginning of \sref{phase}.

Let us consider the quotient space of all field configurations
$\gx_\pol$ and $\fx_\pol$, divided by the gauge transformations that
preserve the values of the link variables. This quotient space is
obviously the kinematical phase space $\ksp_\cuts$ defined in
\sref{phase}. To see this, we first note that by definition the
observables, hence the link variables are the coordinates on this
space. And furthermore, they are not independent but subject to the
kinematical constraints (\ref{icon}-\ref{jcon}). It is therefore not
the extended phase space $\esp_\cuts$ which is recovered here, but the
kinematical subspace $\ksp_\cuts\subset\esp_\cuts$. To check this, we
verify that the constraints defined in \sref{phase} are indeed
satisfied by the link variables introduced above. 

The constraint $\pcon_\pol\approx0$ for each compact polygon
$\pol\in\pols_0$ is an immediate consequence of the definition
\eref{dis-emb-cut}. The same applies to the constraint
$\econ_\pol\approx0$ for each non-compact polygon
$\pol\in\pols_\infty$. This follows from the definition of the clocks
in \eref{T-cdir-cone}, the fall off condition \eref{ff-cone} for
$\fx_\pol$ at infinity, and again the definition \eref{dis-emb-cut} of
the vectors $\dis_\cut$. And finally, the constraint
$\jcon_\ecut\approx0$ for each external link $\ecut\in\cuts_\infty$
simply follows from the definition \eref{L-emb} of the auxiliary
variable $L_\ecut$. So, everything fits together with the definitions
made in \sref{phase}.

\section*{Acknowledgments}
For many helpful discussions and hospitality during the long history
of this work, I would like to thank Ingemar Bengtsson, S\"oren Holst,
Gerard 't~Hooft, Jorma Louko, Hermann Nicolai, Martin Reuter and Max
Welling.

\begin{appendix}
\section*{Appendix}
\setcounter{equation}{0} 
\def\thesection{A} 
Here we summarize some facts about the spinor representation of the
three dimensional Lorentz algebra $\algsl(2)$ of traceless $2\times2$
matrices, and the associated Lie group $\grpSL(2)$. A more
comprehensive overview, using the same notion, is given in
\cite{matwel}. As a vector space, $\algsl(2)$ is isometric to three
dimensional Minkowski space. An orthonormal basis is given by the
gamma matrices
\begin{equation}
  \label{gamma}
  \gam_0 = \pmatrix{ 0 & 1 \cr -1 & 0 } , \qquad
  \gam_1 = \pmatrix{ 0 & 1 \cr 1 & 0 }, \qquad
  \gam_2 = \pmatrix{ 1 & 0 \cr 0 & -1 }.
\end{equation}
They satisfy
\begin{equation}
  \label{gamma-alg}
  \gam_a \gam_b = \eta_{ab} \, \one - \eps_{abc} \, \gam^c, 
\end{equation}
where $a,b=0,1,2$. The metric $\eta_{ab}$ has signature $(-,+,+)$, and
for the Levi Civita symbol $\eps^{abc}$ we have $\eps^{012}=1$.
Expanding a matrix in terms of the gamma matrices, we obtain an
isomorphism of $\algsl(2)$ and Minkowski space,
\begin{equation}
  \vv = v^a \, \gam_a  \equivalent
  v^a = \ft12\Trr{\vv\gam^a}.
\end{equation}
Some useful relations are that the scalar product of two vectors is
equal to the trace norm of the corresponding matrices, and the vector
product is essentially given by the matrix commutator,
\begin{equation}
  \label{vec-mat}
   \ft12\Trr{\vv\ww} = v_a w^a , \qquad
   \ft12 [ \vv , \ww ] = - \eps^{abc} \, v_a w_b \, \gam_c.
\end{equation}
Sometimes it is useful to introduce cylindrical coordinates in Minkowski
space, writing
\begin{equation}
  \label{vec-cyl}
   \vv = \tau \, \gam_0 + \rho \, \gam(\varphi),
\end{equation}
where $\tau$ is the time, $\rho\ge0$ the radial, and $\varphi$ the
angular coordinate, which is redundant for $\rho=0$ and has a period
of $2\pi$. The unit vector $\gam(\varphi)$ defines the angular
direction $\varphi$ in Minkowski space. This vector and its derivative
$\gam'(\varphi)$ are two orthogonal unit spacelike vectors, forming
together with $\gam_0$ a rotated orthonormal basis,
\begin{equation}
  \label{gamma-rot}
  \gam(\varphi) = \cos\varphi \, \gam_1 + \sin\varphi \, \gam_2, \qquad
  \gam'(\varphi) = \cos\varphi \, \gam_2 - \sin\varphi \, \gam_1. 
\end{equation} 
Useful relations are 
\begin{equation}
  \gam_0 \gam(\varphi) = \gam'(\varphi), \qquad
  \gam_0 \gam'(\varphi) = - \gam(\varphi),
\end{equation}
and 
\begin{equation}
  \gam(\varphi_1) \, \gam(\varphi_2) = 
        \cos(\varphi_1-\varphi_2) \, \one + 
        \sin(\varphi_1-\varphi_2) \, \gam_0. 
\end{equation}
The Lie group $\grpSL(2)$ consists of matrices $\uu$ with unit
determinant. The group acts on the algebra in the adjoint
representation, so that 
\begin{equation}
  \label{adjoint-action}
  \vv \mapsto \uu^{-1} \vv \uu.
\end{equation}
This defines a proper Lorentz rotation of the vector $\vv$. The
conjugacy classes of the algebra are characterized by the invariant
length $v^av_a=\ft12\Trr{\vv^2}$. For timelike and lightlike vectors
$\vv$ (with $\ft12\Trr{\vv^2}\le0$) we distinguish between positive
($v^0=\ft12\Trr{\vv\gam^0}>0$) and negative
($v^0=\ft12\Trr{\vv\gam^0}<0$) timelike and lightlike vectors,
respectively. Special group elements are those representing spatial
rotations about the $\gam_0$-axis. The matrix
\begin{equation}
  \label{rot-mat}
  \uu = \expo{-\emul\gam_0/2} = 
         \cos(\emul/2) \, \one - \sin(\emul/2) \, \gam_0, 
\end{equation}
represents a rotation by $\emul$ in counter clockwise direction, 
\begin{equation}
  \label{rotation}
  \gam_0 \mapsto \uu^{-1} \gam_0 \, \uu = \gam_0, \qquad
  \gam(\varphi) \mapsto  \uu^{-1} \gam(\varphi) \, \uu = 
                   \gam(\varphi+\emul). 
\end{equation}
A generic element $\uu\in\grpSL(2)$ can be expanded in terms of the
unit and the gamma matrices, defining a scalar $u$ and a vector
$\pp\in\algsl(2)$,
\begin{equation}
  \label{project}
  \uu = u \, \one + p^a \, \gam_a \follows
   \pp = p^a \, \gam_a.
\end{equation}
The scalar $u$ is half of the trace of $\uu$, and the vector $\pp$ is
called the \emph{projection} of $\uu$. The determinant condition
implies that
\begin{equation}
  \label{det-con}
  u^2 = p^a p_a + 1. 
\end{equation}
According to the property of the vector $\pp$, we distinguish between
timelike, lightlike and spacelike group elements $\uu$. For a timelike
element we have $-1<u<1$, and it represents a rotation about some
timelike axis, which is specified by the vector $\pp$. The angle of
rotation $\emul$ obeys $u=\cos(\emul/2)$. Lightlike elements are null
rotations, and spacelike elements are boosts.

Finally, a parity transformation, which acts on Minkowski space as a
reflection of space, can be defined as follows, 
\begin{equation}
  \label{reflection}
  \vv \mapsto - \gam_2 \, \vv \, \gam_2 .
\end{equation}
It represents a reflection of the $\gam_2$-axis,
\begin{equation}
  \gam_0 \mapsto \gam_0 , \qquad
  \gam_1 \mapsto \gam_1 , \qquad
  \gam_2 \mapsto - \gam_2. 
\end{equation}
Note that this is not a proper Lorentz rotation, because $\gam_2$ is
not an element of the group $\grpSL(2)$. 

\end{appendix}

\end{document}